\documentclass[prd,twocolumn,nofootinbib,showpacs]{revtex4-1}

\usepackage{amsfonts}
\usepackage{amsmath}
\usepackage{amssymb}
\usepackage{bm}
\usepackage{dcolumn}
\usepackage{graphicx}
\usepackage{graphics}
\usepackage[latin1]{inputenc}
\usepackage{latexsym}
\usepackage{rotating}
\usepackage{hyperref}
\usepackage[all]{hypcap} 
\usepackage{xspace} 
\usepackage[usenames,dvipsnames]{color}
\usepackage{mathrsfs}
\usepackage{subfigure}
\usepackage{aas_macros}
\usepackage{amsfonts}
\usepackage{booktabs}
\usepackage{siunitx}

\usepackage{ulem}
\usepackage{outlines}
\usepackage{enumitem}
\setenumerate[1]{label=\Roman*.}
\setenumerate[2]{label=\Alph*.}
\setenumerate[3]{label=\roman*.}
\setenumerate[4]{label=\alph*.}

\definecolor {darkgreen}{rgb}{0.2,0.7,0.2}

\newcommand\be{\begin{equation}}
\newcommand\ba{\begin{eqnarray}}
\newcommand\ee{\end{equation}}
\newcommand\ea{\end{eqnarray}}

\newcommand\bw{\begin{widetext}}
\newcommand\ew{\end{widetext}}

\begin{document}
\title{A 3PN Fourier Domain Waveform for Non-Spinning Binaries with Moderate Eccentricity}

\author{Blake Moore}
\affiliation{eXtreme Gravity Institute, Department of Physics, Montana State University, Bozeman, MT 59717, USA.}

\author{Nicol\'as Yunes}
\affiliation{eXtreme Gravity Institute, Department of Physics, Montana State University, Bozeman, MT 59717, USA.}

\date{\today}

\begin{abstract} 
While current gravitational wave observations with ground based detectors have been consistent with compact binaries in quasi-circular orbits, eccentric binaries may be detectable by ground-based and space-based instruments in the near future. 
Eccentricity significantly complicates the gravitational wave signal, and we currently lack fast and accurate models that are valid in the moderate to large eccentricity range.
In a previous paper, we built a Fourier domain, eccentric waveform model at leading order in the post-Newtonian approximation, i.e.~as an expansion in small velocities and weak fields.
Here we extend this model to 3rd post-Newtonian order, incorporating the effects of periastron precession and higher post-Newtonian order effects that qualitatively change the waveform behavior. 
Our 3PN model combines the stationary phase approximation, a truncated sum of harmonics of combinations of two orbital frequencies (an azimuthal one and a radial one), and a bivariate expansion in the orbital separation and the eccentricity. 
We validate the model through comparisons with a fully-numerical, time domain post-Newtonian model, and find good agreement (matches between $97\% - 99\%$) in much of the parameter space.
We estimate in what regions of parameter space eccentric effects are important by exploring the signal to noise ratio of eccentric corrections.
We also examine the effects of higher post-Newtonian order terms in the waveform amplitude, and the agreement between different PN-consistent, numerical, time-domain models.
In an effort to guide future improvements to the model, we gauge the error in our 3PN model incurred by the different analytic approximations used to construct it.
This model is useful for preliminary data analysis investigations and it could allow for a phenomenological hybrid that incorporates eccentricity into an inspiral-merger-ringdown model.  
\end{abstract}



\maketitle

\section{Introduction}
\label{intro}

With the multiple detections of gravitational waves (GWs) from binary black hole and neutron star systems by ground based detectors \cite{2017PhRvL.119p1101A, 2017PhRvL.119n1101A, 2017ApJ...851L..35A, 2017PhRvL.118v1101A, 2016PhRvL.116x1103A, 2016PhRvL.116f1102A} and the imminent launch of space based detectors \cite{lisa}, we are now in a position to perform gravitational wave astronomy. With this comes the ability to answer important questions about the physical characteristics of the sources producing those waves, the environments in which they are formed, the validity of General Relativity, and the nature of the universe at large. To answer these questions, we rely heavily on our ability to model these sources -- modeling which initially was focused on binaries in quasi-circular orbits. In more recent years, there has been much work to model the GWs incorporating the effects of orbital eccentricity. 

This work has been motivated by astrophysical studies which suggest that, while quasi-circular binaries dominate potential ground based detector sources, there are a number of scenarios that can produce binaries in eccentric orbits while emitting GWs in the sensitivity band of these detectors. Recently, Rodriguez et al. incorporated PN effects into globular cluster simulations and as a result predicted a merger rate for LIGO/Virgo of 0.5 Gpc$^{-3}$ yr$^{-1}$ for binaries with detectable eccentricity above $0.1$ (with an overall merger rate between 10 and 213 Gpc$^{-3}$ yr$^{-1}$) \cite{2018PhRvD..98l3005R}. Other promising eccentric formation scenarios include field triples \cite{2017ApJ...841...77A}, GW capture and triples near supermassive BHs \cite{2009MNRAS.395.2127O, 2014ApJ...781...45A}. As a result, the detection of GWs from multiple eccentric sources could help us identify the environment in which they form \cite{2017MNRAS.465.4375N}.

Neglecting the effects of eccentricity in source modeling can lead to a loss in detection rate and parameter bias. Several studies have investigated the ability of quasi-circular template banks to recover eccentric signals and found significant loss in detection rate for binaries with eccentricities greater than about 0.1 \cite{Huerta:2013qb, 2010PhRvD..81b4007B, 2009CQGra..26d5013C}. However, even for eccentricities smaller than 0.1,  parameter biases can arise when the eccentricity is neglected in the model and the signal comes from a binary with eccentricity as small as about $10^{-3}$ \cite{2013arXiv1310.8288F}.

Clearly then, if we wish to recover and perform data analysis on eccentric signals we require models which incorporate the effects of orbital eccentricity, particularly efficient Fourier domain waveforms. The Post-Circular (PC) formalism provides a fully analytic (at Newtonian order) prescription to compute a frequency-domain waveform, while incorporating eccentric effects \cite{PhysRevD.80.084001}. This was extended to 2PN order in \cite{Tanay:2016zog} and recently to 3PN order~\cite{Gopu-in-prep}. In \cite{Moore:2016qxz}, the authors used a circular GW amplitude and incorporated the leading-order eccentric effects in the binary dynamics while working at 3PN order.  These models expand all relevant quantities in small eccentricity and relate the Fourier frequency to the orbital frequency in order to map the model to the Fourier domain. As a result, these models are computationally efficient, but likely only valid in the small eccentricity regime ($e_0 \leq 0.3$). 

In Reference \cite{2018CQGra..35w5006M} (hereafter referred to as paper 1), we built a Fourier domain, eccentric waveform model at leading order in the post-Newtonian (PN) approximation~\footnote{The PN approximation is one in which the field equations are solved assuming small velocities and weak gravitational fields in an expansion in powers of $(\frac{v}{c})$, where $v$ is the orbital velocity and $c$ is the speed of light \cite{blanchet-review}. By $n$PN order we mean an expansion to order $({v}/{c})^{2 n}$.}. This Newtonian order model schematically took the form 
\begin{equation}
\tilde{h}_{+,\times} \sim \sum_{j=1}^{N}A^{(j)}_{+,\times}(f) \; e^{i\psi_j(f)} \,,
\end{equation} 
where $A^{(j)}_{+, \times}$ are slowly-varying amplitudes and the $\psi_j$ are rapidly-varying Fourier phases. The multiple harmonics in the Fourier domain arise from the fact that an eccentric signal can be decomposed as a sum over harmonics of the mean orbital frequency in the time domain. We validated this waveform for large eccentricities ($e_0 \leq 0.9$) using the match (a measure of agreement between two models) for advanced LIGO at design sensitivity against a fully-numerical, time-domain model. The key to our Newtonian Fourier-domain model's accuracy and distinction from PC models is an analytic prescription in terms of the orbital eccentricity $e$ for the Fourier phases and the amplitudes, with $e$ then related to the fourier frequency though the stationary phase condition. A similar model to 1PN order is presented in \cite{2015PhRvD..92d4038M}.

We here extend the Newtonian model of paper 1 to 3PN order (developing a model similar to TaylorF2~\cite{2009PhRvD..80h4043B} but for eccentric binaries), and we validate it in a large region of initial-separation/initial-eccentricity parameter space. We follow the same procedure as in paper 1, with the most major complication arising from the inclusion of periastron precession due to PN effects. This gives rise to harmonics that are combinations of the two orbital frequencies in the eccentric problem -- one related to the radial period and the other related to the azimuthal period. In addition, the Fourier phases do not admit exact solutions in terms of $e$ as was the case in paper 1, where those phases could be expressed exactly in terms of hypergeometric functions. However, in paper 1 we found that the small eccentricity expansion of those hypergeometric functions were still highly accurate, and so we here employ that same expansion. 

\begin{figure}[htp]
\includegraphics[clip=true,angle=0,width=0.4\textwidth]{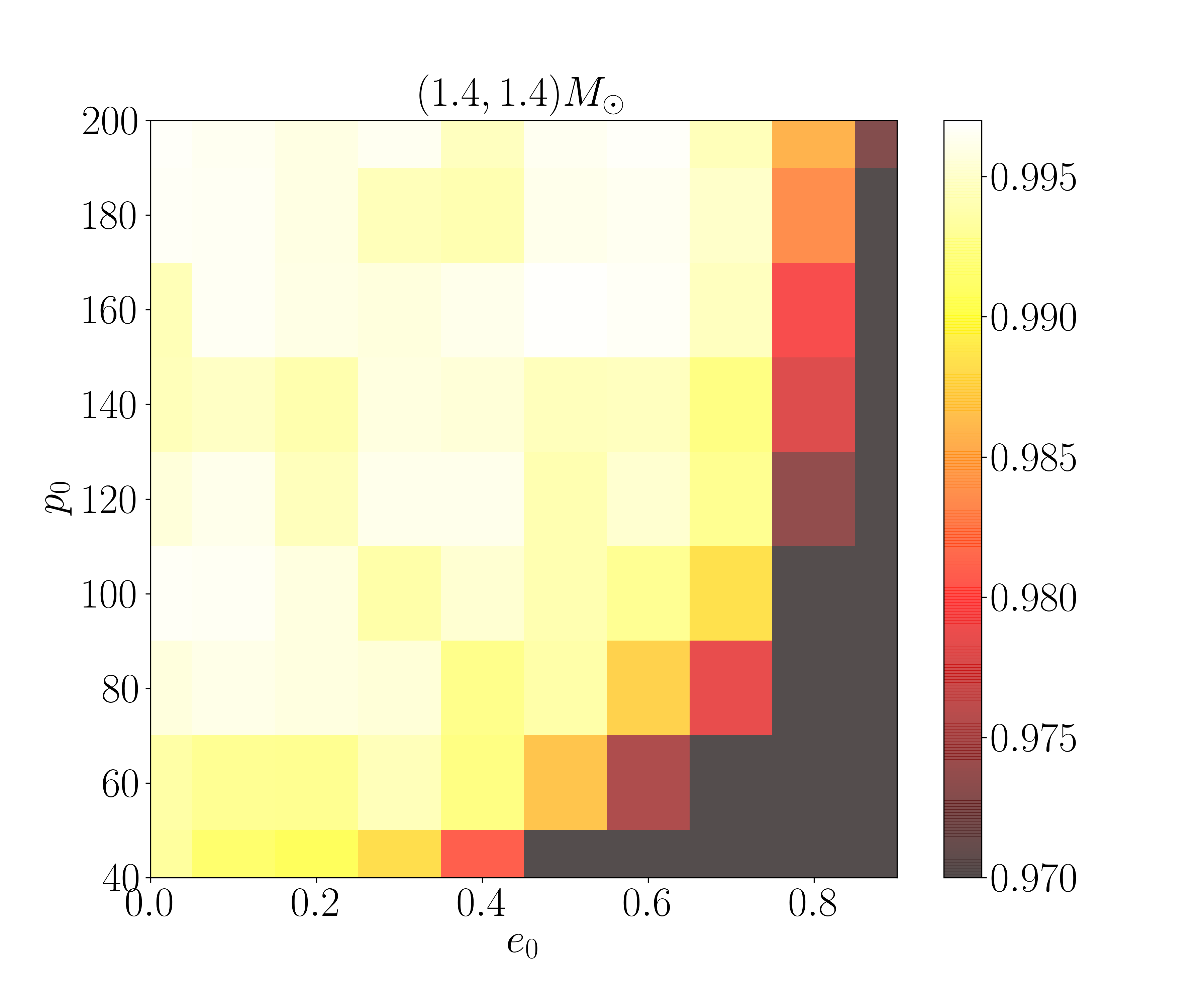}
\caption{\label{fig:match_intro} Match between our 3PN accurate Fourier domain waveform and a fully numerical PN time domain waveform as a function of the initial eccentricity and initial dimensionless semi-latus rectum for a $(1.4,1.4)M_{\odot}$ system. The match is greater than 97\% for much of parameter space explored.}
\end{figure}

Figure \ref{fig:match_intro} shows the match between our 3PN Fourier domain model and a fully numerical time-domain PN waveform for a $(1.4, 1.4)M_{\odot}$ binary inspiral for a number of initial separations and initial eccentricities. We see that our model is very faithful to the numerical one in a large region of the parameter space explored. The Fourier and time-domain models begin to disagree for very eccentric binaries that start at very small initial separation, since then their pericenter velocity becomes quite large. The results presented in this figure are representative of systems with other comparable masses. The match, however, deteriorates more rapidly for systems with disparate masses, suggesting that higher PN order models may be needed for these cases. 

In order to develop our model and to better understand the features of the model, we produce several byproducts. First, we obtain a 3PN accurate decomposition of the time-domain waveform into an infinite sum of harmonics and we truncate it at a finite order. With this at hand, we then use numerical solutions to the orbital dynamics to validate the truncated model against another model that keeps all harmonics, so as to determine the loss of faithfulness due to harmonic truncation. Second, we investigate how different PN-consistent ways of solving the orbital dynamics lead to a loss of faithfulness in numerically-solved, time domain models due only to the truncation of the PN series. Third, we produce two Fourier domain waveforms, one with more terms kept in the eccentricity expansions and one with less terms kept, and perform a faithfulness analysis on each in order to understand the error incurred by our expansions in small eccentricity. Lastly, we swap the analytic components of our Fourier domain waveform with numerical solutions in order to get a sense of which of these analytic approximations leads to a dominant loss in faithfulness. Since each of these results isolate a particular approximation made in generating our Fourier domain model, we are able to pinpoint which approximation causes the most error, which can guide future advances aimed at increasing the accuracy of our model.

The organization of the paper is as follows. In Section \ref{sec:orb_dyn} we review the PN dynamics of an eccentric orbit, and in Section \ref{sec:GW_time} the time domain GW signal and its decomposition into harmonics of the two orbital frequencies. Section \ref{sec:F2} presents the GW signal in the Fourier domain, and provides the formalism that we employ to obtain our model. Section \ref{sec:imp_val} provides details on the implementation of the Fourier domain model, and carries out a faithfulness study. In Section \ref{sec:conc} we conclude and point to future work. Throughout this work we use geometric units (c = G = 1). 
\section{Review of Eccentric Orbital Dynamics}
\label{sec:orb_dyn}
In this section we review the orbital dynamics of an eccentric binary. We first begin by reviewing the dynamics at Newtonian order. Then we present the dynamics including effects up to 3PN order in the quasi-Keplerian (QK) formalism presented in \cite{Damour:2004bz}. The most notable changes that occur when including higher PN effects is periastron precession (entering at 1PN order), which introduces a second frequency (and thus a second phase angle) into the dynamics, and higher order corrections to Kepler's equation.

\subsection{Newtonian Order}
\label{subsec:newt_orb_dyn}
At Newtonian order the dynamics of an elliptical orbit restricted to a plane are described by 
\begin{subequations}
\label{eq:newt_orb}
\begin{align}
r &= a (1 - e\cos u) \, , \\
\phi &= \lambda + W = \lambda + v - u + e \sin u  \, , \\
\lambda &= n(t - t_0) + c_{\lambda} \, , \\
\label{eq:Kepler}
l &= n (t - t_0) + c_l = u - e \sin u \, , \\
v &= 2\arctan \left[\left( \frac{1+e}{1-e}\right)^{1/2} \tan\left(\frac{u}{2}\right)  \right] \, .
\end{align}
\end{subequations}
Here $r$ is the magnitude of the relative separation vector, which we choose to be on the $x-y$ plane $\vec{r} = (r \cos \phi, r \sin \phi, 0)$. The angle $\phi$ is the orbital phase, $e$ the orbital eccentricity, and the angle $l$ is the mean anomaly. The mean motion $n$ is related to the orbital period $P$ by $n = 2 \pi/P$, and the semi-major axis via Kepler's third law $n^2a^3 = m$ with $m$ the total mass of the system. We have written $\phi$ in terms of a linearly increasing phase angle $\lambda$, and an oscillatory phase angle $W$, which accounts for the fact that the rate of change of the orbital phase $\dot{\phi}$ is not constant throughout one orbit. In the absence of radiation reaction, the conservation of energy and angular momentum demands that the orbital eccentricity and semi-major axis remain fixed. To obtain $r(t)$ and $\phi(t)$, a separation and an orbital eccentricity must be specified and then these are used to solve Kepler's equation [Eq.~\eqref{eq:Kepler}] for $u(t)$ -- which can only be exactly solved numerically. This points to a clear complication of eccentric binary modeling -- even in the Newtonian treatment and in the absence of radiation reaction, the explicit time domain dynamics require a numerical solution. 

In GR, GWs carry away energy and angular momentum from the binary, and in response the orbital eccentricity and mean motion (or alternatively the separations associated with the ellipse) vary in time. Equations \eqref{eq:newt_orb} remain the same with the exception of $l$, which obeys
\begin{equation}
l = \int^t n(t')dt' \, .
\end{equation}
The time evolution of the eccentricity and mean motion were first derived by Peters and Matthew \cite{PetersMathews} and is given by
\begin{subequations}
\begin{align}
\frac{dn}{dt} &= \frac{\eta}{m}(mn)^{11/3}\left(\frac{96+292e^2+37e^4}{5(1-e^2)^{7/2}}\right) \, , \\
\frac{de}{dt} &= -\frac{\eta}{m}(mn)^{8/3}e \left(\frac{304 + 121e^2}{15(1-e^2)^{5/2}}\right) \, .
\end{align}
\end{subequations}
The appearance of the negative sign in $\dot{e}$ implies the well known result that radiation reaction causes a binary to circularize. In paper 1, we solved this coupled set of differential equations exactly for $n(e)$, which in turn allowed us to obtain $t(e)$ and $l(e)$ exactly in terms of hypergeometric functions and the AppelF1 function. We now turn to treat the effects of higher order PN effects in the orbital dynamics.

\subsection{Inclusion of PN Effects}
\label{subsec:pn_orb_dyn}
The quasi-Keplerian formalism (QK) of~\cite{Damour:2004bz} allows one to describe elliptic motion while including higher order PN effects. Equations~\eqref{eq:newt_orb} keep a similar form, with the notable difference being that the orbital angle $\lambda$ is related to an azimuthal frequency and $l$ is related to a radial one, a distinction caused by periastron precession. Including PN effects causes Eqs.~\eqref{eq:newt_orb} to take the form
\allowdisplaybreaks[4]
\begin{subequations}
\label{eq:pn_orb}
\begin{align}
r &= a_r(1 - e_r \cos u) \, , \\
\phi &= \lambda + W \, , \\
\label{eq:WPN}
W &= (1+k)(v - l) + \left(f_{4\phi} + f_{6\phi}\right)\sin 2\nu \nonumber
\\ & + \left(g_{4\phi} + g_{6\phi}\right)\sin 3\nu + i_{6\phi} \sin 4\nu + h_{6\phi} \sin 5\nu \, , \\
\lambda &= (1+k)n(t-t_0) + c_{\lambda} \, , \\
l &= n(t - t_0) + c_{l} \nonumber \\
& = u - e \sin u + \left( g_{4t} + g_{6t}\right) (\nu - u) \nonumber \\
& + \left(f_{4t} + f_{6t}\right) \sin \nu + i_{6t} \sin 2\nu + h_{6t} \sin 3\nu \, , 
\label{eq:3PNKepler}
\\
\label{eq:vPN}
v &= 2\arctan \left[\left( \frac{1+e_{\phi}}{1-e_{\phi}}\right)^{1/2} \tan\left(\frac{u}{2}\right)  \right] \, .
\end{align}
\end{subequations} 
Note that now there are three different eccentricities: the time eccentricity $e_t$, the radial eccentricity $e_r$, and the phi-eccentricity $e_{\phi}$. The three eccentricities can be related to one another through PN accurate relations, which can be found in \cite{Memmesheimer:2004cv, Damour:2004bz}.  Throughout this work we parameterize the problem in terms of the time eccentricity, $e_t$, and therefore we will drop the $t$ subscript from hereon out. The functions $g_{n\phi}, g_{nt}, f_{n\phi}, f_{nt}, i_{n\phi}, i_{nt}, h_{n\phi}, h_{nt}$ are $n/2$ PN coefficients that depend on the orbital energy, angular momentum and the masses of the binaries~\cite{Memmesheimer:2004cv}.  

We adopt a PN parameter $y$ defined by
\begin{equation}
y = \frac{(m(1+k)n)^{1/3}}{\sqrt{1-e^2}} = \mathcal{O}(v) .
\end{equation}
Note that in the zero eccentricity limit, $y$ corresponds exactly to the orbital speed $v$. At Newtonian order, the parameter $y$ is related to the dimensionless semilatus rectum, $p$, by $y = 1/\sqrt{p}$. This PN parameter is well-behaved even for highly eccentric orbits, since one can think of it as effectively requiring the semi-latus rectum to be large instead of requiring the semi-major axis to be large, which could result in large pericenter velocities.

The effect of periastron precession is encoded in the parameter $k = \delta \Phi / (2\pi)$, where $\delta \Phi$ is the advance of periastron over a period. In terms of $y$ at 3PN order 
\begin{align}
k &=  3 y^2
+ \left[ 54  - 28 \eta + ( 51 - 26 \eta ) e^2 \right]\frac{ y^4 }{ 4 } 
\nonumber \\
&+ \bigg\{
6720 - (20000 - 492 \pi^2) \eta + 896 \eta ^2
\nonumber \\
& + [ 18336 - (22848 - 123 \pi^2) \eta + 5120 \eta^2 ] e^2
\nonumber \\
&+ ( 2496 - 1760 \eta + 1040 \eta^2  ) e^4
\nonumber \\
&+ \left[ 1920 - 768 \eta + ( 3840 - 1536 \eta ) e^2 \right] \sqrt{1 - e^2}
\bigg\}\frac{ y^6 }{ 128 }.
\end{align} 
We can use this expression for $k$ to relate the radial orbital angular frequency, $n$, which corresponds to the periastron to periastron period to the azimuthal orbital angular frequency, $\omega = (1+k)n$, which is related to the period to return to the same angle in the orbit. 

The functional dependence of the time eccentricity on the PN expansion parameter $y$ depends on the choice of gauge. Two commonly employed gauges are the Arnowitt-Deser-Misner (ADM) and modified harmonic (MH) ones, which differ at 2PN order; throughout this work, we use the MH gauge. In order to express results in either gauge one needs to relate the time eccentricity in one gauge to another. These relations can be found in \cite{Moore:2016qxz}, but we also list them here:
\begin{align}
\label{eq:ADMtoMH}
e^{\rm ADM}_t &= e^{\text{MH}}_t\left\{1+\left(\frac{1}{4}+\frac{17}{4}\eta\right)y^4
\right.
\nonumber \\
&\left. +\left[\frac{1}{2}+\left(\frac{16739}{1680}-\frac{21}{16} \pi^2 \right)\eta 
-\frac{83}{24} \eta^2
\right. \right.
\nonumber \\
&\left. \left.
+\left(\frac{1}{2}+\frac{249}{16}\eta-\frac{241}{24}\eta^2\right)e_t^2\right]y^6\right\}\,.
\end{align}

Much like in the Newtonian case, the effect of the emission of GWs is to cause the orbital frequency (or alternatively $y$) and the eccentricity to vary with time. To fully solve for an eccentric orbit including PN effects, one has to solve the following set of coupled ordinary differential equations:
\begin{subequations}
\label{eq:orb_param_t}
\begin{widetext}
\begin{align}
\frac{dl}{dt} &=  n(e, y) = \frac{y^3}{m}(1-e^2)^{3/2}\left\lbrace 1 - 3y^2 - \frac{1}{4}y^4\left[18 -28\eta + e^2\left(51 - 26\eta\right)\right] 
\nonumber \right. 
\\ & \left. 
- \frac{1}{128} y^6 \left[-192 - (14624 - 492 \pi^2)\eta + 896\eta^2  + e^2 \left(5120 \eta ^2-\left(17856-123 \pi ^2\right) \eta +8544\right) 
\nonumber \right. \right. 
\\ & \left. \left.
+ e^4 \left(1040 \eta ^2-1760 \eta +2496\right) + \sqrt{1-e^2} \left(e^2 (3840-1536 \eta )-768 \eta +1920\right)\right]  \right\rbrace
\end{align}
\end{widetext}
\begin{align}
\frac{d\lambda}{dt} &= \omega(e, y) = \frac{y^3}{m}(1-e^2)^{3/2} \, , \\
m\frac{dy}{dt} &= (1-e^2)^{3/2}\eta y^9 \left(a_0 + \sum_{n=2}^6 a_n y^n\right) \, , \\
m\frac{de}{dt} &= -\frac{(1-e^2)^{3/2}}{2e}\eta y^8 \left(b_0 + \sum_{n=2}^6 b_n y^n \right) \, .
\end{align}
\end{subequations}
where the $a_n$ and $b_n$ can be found in modified harmonic gauge in Appendix C of \cite{2018arXiv180108542K} and  $\eta = m_1 m_2/ m^2$ is the symmetric mass ratio. The equations for $\dot{y}$ and $\dot{e}$ depend on certain enhancement functions $\phi_y$, $\phi_e$, $\psi_y$, $\psi_e$, $\zeta_y$, $\zeta_e$, $\kappa_y$, and $\kappa_e$~ \cite{2018arXiv180108542K}, which we have expanded to $\mathcal{O}(e^{50})$. Once the solution to this set of ordinary differential equations is found, one can insert them into Eqs.~\eqref{eq:pn_orb} to obtain the evolution of the different orbital quantities (while Kepler's equation must still be solved numerically). 

\section{Eccentric GWs in the Time Domain}
\label{sec:GW_time}
In this section we present the GW signal in the time domain. We begin by analyzing the effect of PN corrections to the amplitude of the waveform. This analysis suggests that we can faithfully model signals while restricting to leading PN order in the amplitude. We then obtain and analyze the decomposition of this signal (keeping only leading order PN corrections in the amplitude) into harmonics of the azimuthal and radial orbital frequencies. We validate this decomposition by computing the overlap (a measure of the agreement between two models) between the waveform decomposed into a number of harmonics and one which has not been decomposed into harmonics. Lastly, we discuss different PN consistent ways of solving Eq.~\eqref{eq:orb_param_t} and the implications this has for modeling eccentric systems. Throughout we provide details on how we obtain the fully numerical model with which we will compare and validate our frequency domain model.

\subsection{PN Effects in the Amplitude}
\label{subsec:PN_amp_effects}
To 2PN order in the amplitude, the GW polarizations read: 
\begin{align}
\label{eq:pn_amp}
h_{+,\times} &= \frac{m \eta}{R}y^2(1-e^2)\left(H^{(0)}_{+,\times } + y H^{(1/2)}_{+,\times } 
\right. 
\nonumber \\
& \left. +  y^2 H^{(1)}_{+,\times } +
y^3 H^{(3/2)}_{+,\times } + y^4 H^{(2)}_{+,\times }\right)
\end{align}
The lengthy expressions for $H^{(n)}_{+,\times}$ can be found in \cite{2002PhRvD..65h4011G} in ADM gauge and using a PN expansion parameter related to $mn$. The corrections needed to express them in MH and in terms of $y$ can be found in Appendix \ref{app:MH_polarizations}. For reference we list the 0PN polarizations here (which are unaffected by gauge differences or the change of PN parameter between $mn$ and $y$):
\begin{subequations}
\label{eq:0pn_polar}
\begin{align}
H^{(0)}_{+} &= \left(P^{0}_{C2C2} \cos 2W + P^0_{S2C2} \sin 2W\right) \cos 2\bar{\lambda} 
\nonumber \\
&+ \left(P^0_{C2S2} \cos 2W+ P^0_{S2S2} \sin 2W\right) \sin 2 \bar{\lambda} + P^0 \, , 
\\
H^{(0)}_{\times} &= \left(X^{0}_{C2C2} \cos 2W + X^0_{S2C2} \sin 2W\right) \cos 2\bar{\lambda} 
\nonumber \\
&+ \left(X^0_{C2S2} \cos 2W+ X^0_{S2S2} \sin 2W\right) \sin 2 \bar{\lambda} \, , 
\end{align}
\end{subequations}
where we have defined
\begin{subequations}
\label{eq:P_X}
\begin{align}
P^{0}_{C2C2} &= \frac{(1+C^2)}{(1-e\cos{u})^2}  \left[ (e\cos{u})^2 - e\cos{u} - 2 e^2 +2 \right]\,,
\\
P^{0}_{C2S2} &= \frac{2 (1+C^2)}{(1-e\cos{u})^2}  \sqrt{1-e^2}e \sin{u} \, ,
\\
P^0 &= \frac{S^2}{(1-e\cos{u})}e\cos{u} \, ,
\\
X^{0}_{C2C2} &= -4 C \frac{\sqrt{1-e^2}}{(1-e\cos{u})^2}e\sin{u} \, ,
\\
X^{0}_{C2S2} &= \frac{2C}{(1-e\cos{u})^2} \lbrace (e\cos{u})^2 - e\cos{u} - 2 e^2 +2 \rbrace \, .
\end{align}
\end{subequations}
and where the different orbital parameters entering the polarizations are described in Sec.~\ref{sec:orb_dyn}, while 
\begin{align}
C &= \cos i\,, \qquad S = \sin i\,,\qquad \bar{\lambda} = \lambda - \beta\,.
\end{align}
with angles $\beta$ and $i$ polar angles that describe the polarization axes. 

Before we can attempt to obtain a Fourier domain model for the GW signal, we must first decide on how many PN corrections to keep in the amplitude of the polarization in Eq.~\eqref{eq:pn_amp}. In order to do so, we will make use of the overlap, i.e.~the normalized inner product in Fourier space, weighted by the spectral noise of the detector. Between two signals $h_1$ and $h_2$, the overlap is defined by 
\begin{align}
\label{eq:olap}
\mathcal{O}(h_1,h_2) = \frac{(h_1|h_2)}{\sqrt{(h_1|h_1)(h_2|h_2)}} \, ,
\end{align}
where the inner product is defined as
\begin{align}
(h_1|h_2) = 4 \text{Re}\left[\int_{f_1}^{f_2} \frac{\tilde{h}_1^{\ast}\tilde{h}_2}{S_n(f)} df \right] \, .
\end{align}
and $S_n(f)$ is the spectral noise density of the detector. An overlap of unity between two models indicates perfect agreement, while an overlap of -1 indicates models perfectly out of phase, with a vanishing overlap indicating perfect disagreement. 

In order to gauge the importance of inclusion of higher order PN corrections to the amplitude we calculate the overlap between two models. One model consists of numerically solving for the orbital parameters (i.e. $u,~W,~l,~\lambda,~e,~\text{and }~y$) using their evolution equations at 3PN order, and keeping all corrections up to 2PN order in the amplitude. The other consists of again numerically solving for the orbital parameters using their evolution equations at 3PN order, but varying the PN order to which the amplitude of the polarizations in Eq.~\eqref{eq:pn_amp} is kept. We also explore the effect of solving for $u(t)$ using the Newtonian version of Kepler's equation.

Before presenting the results of the overlap, let us provide some technical details of how these models are constructed. We begin by numerically solving (using the Burlisch-Stoer method implemented in \texttt{C++}) the coupled set of differential equations $({\dot{\lambda}, \dot{l}, \dot{e}, \dot{y}})$ provided by Eqs~\eqref{eq:orb_param_t}. We stop the evolution when $y > (3(1+e))^{-1}$, a choice that is explained in detail in Section \ref{subsec:TaylorF2e}, but we see that in the circular limit this corresponds to an orbital velocity of a third the speed of light. With the time evolution of these parameters in hand, we must then solve the 3PN Kepler's equation. This is done by adapting Mikkola's method \cite{Mikkola1987}, a fast and accurate numerical scheme to solve Kepler's equation at Newtonian (0PN) order. To solve Kepler's equation at higher PN order [Eq.~\eqref{eq:3PNKepler}], we first use Mikkola's method on the Newtonian equation, and then use the result as an initial guess to solve Eq.~\eqref{eq:3PNKepler} using Mikkola's method again (see \cite{2007MNRAS.374..721T, Tanay:2016zog} for more details). To obtain $W$, we plug in the solution for $u$, $y$, and $e$ into Eq.~\eqref{eq:vPN} to obtain the true anomaly $v$, and then, all of these are substituted into Eq.~\eqref{eq:WPN} to obtain $W$. 

\begin{figure*}[htp]
\includegraphics[clip=true,angle=0,width=0.475\textwidth]{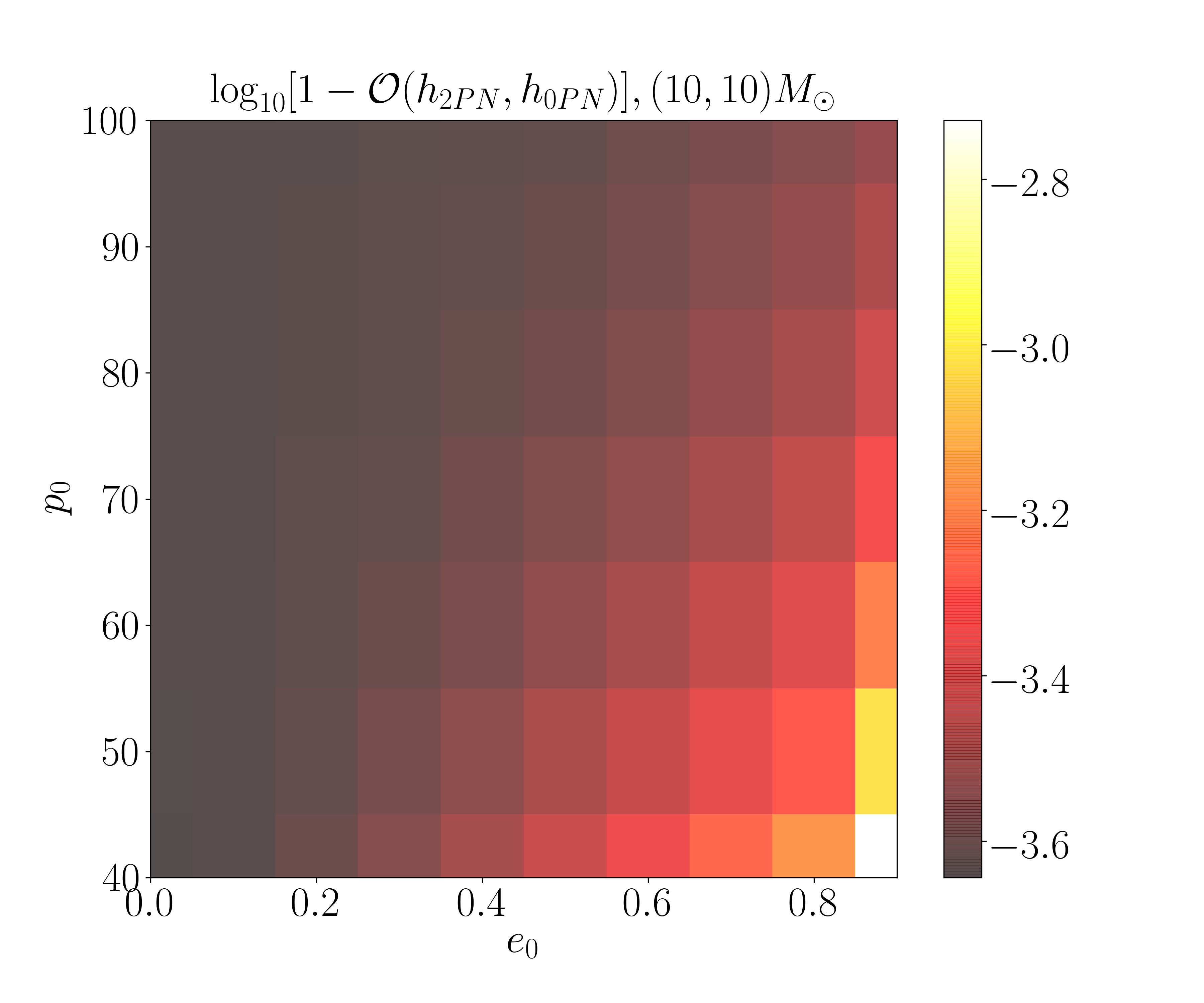}
\includegraphics[clip=true,angle=0,width=0.475\textwidth]{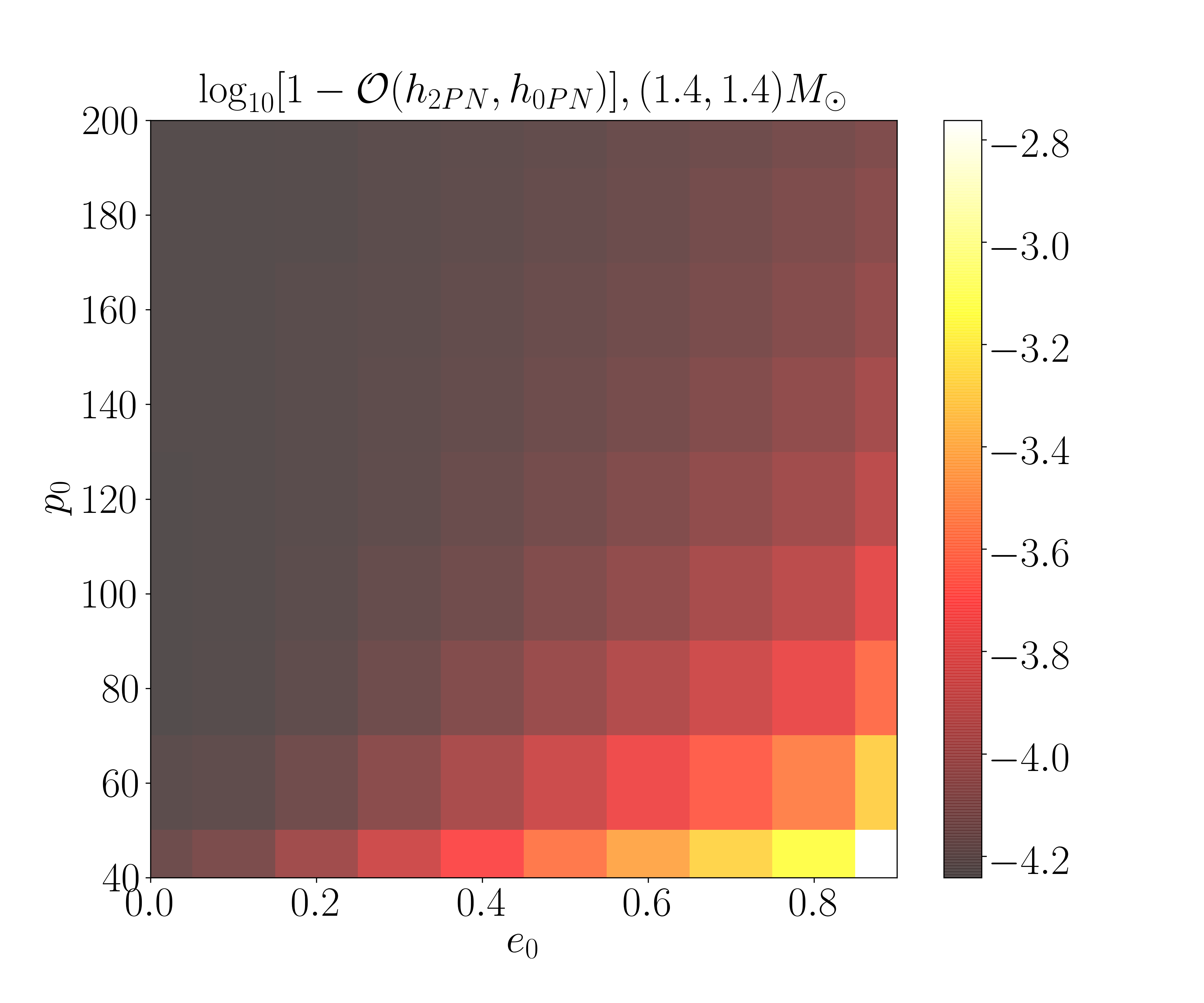}
\includegraphics[clip=true,angle=0,width=0.475\textwidth]{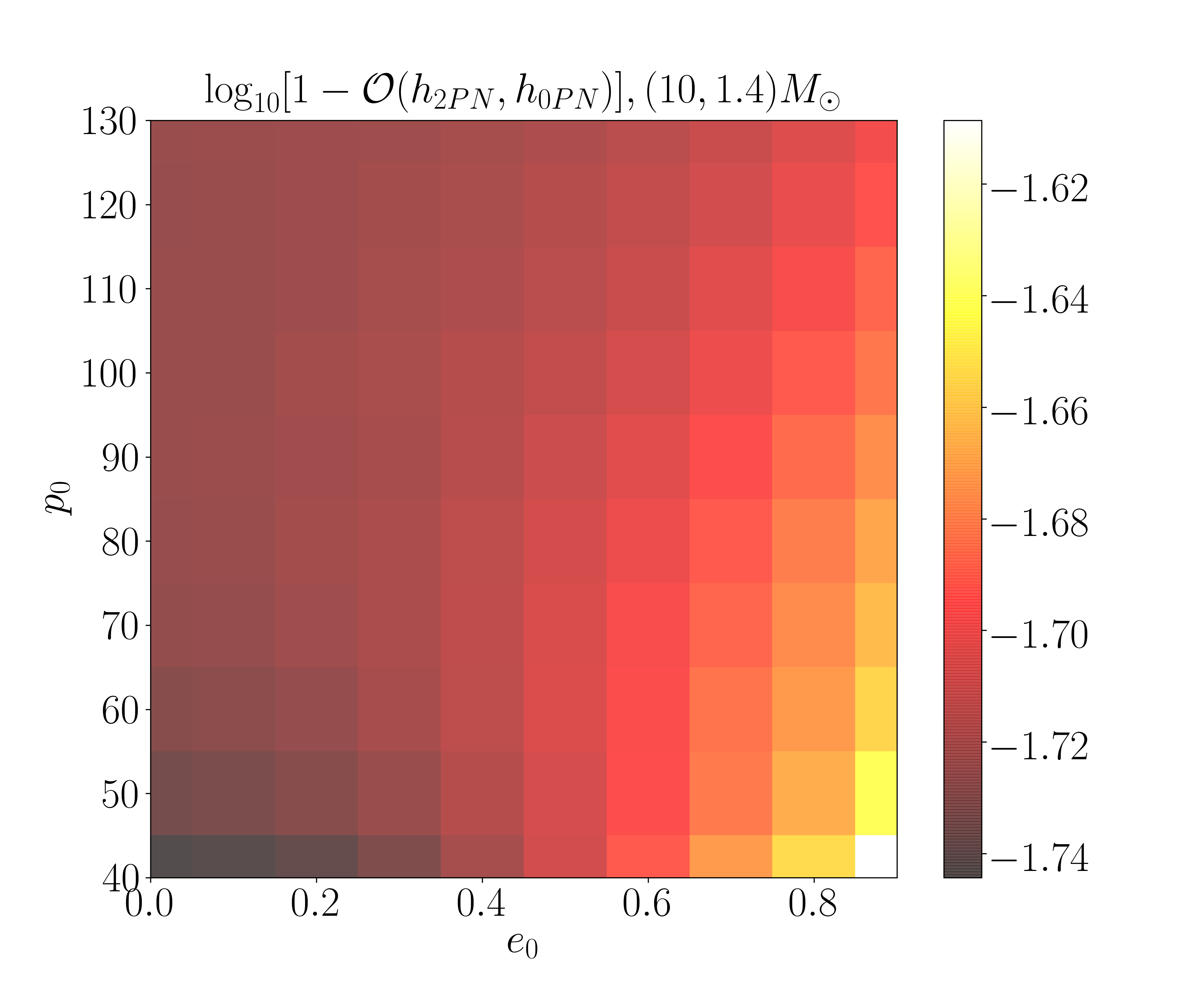}
\includegraphics[clip=true,angle=0,width=0.475\textwidth]{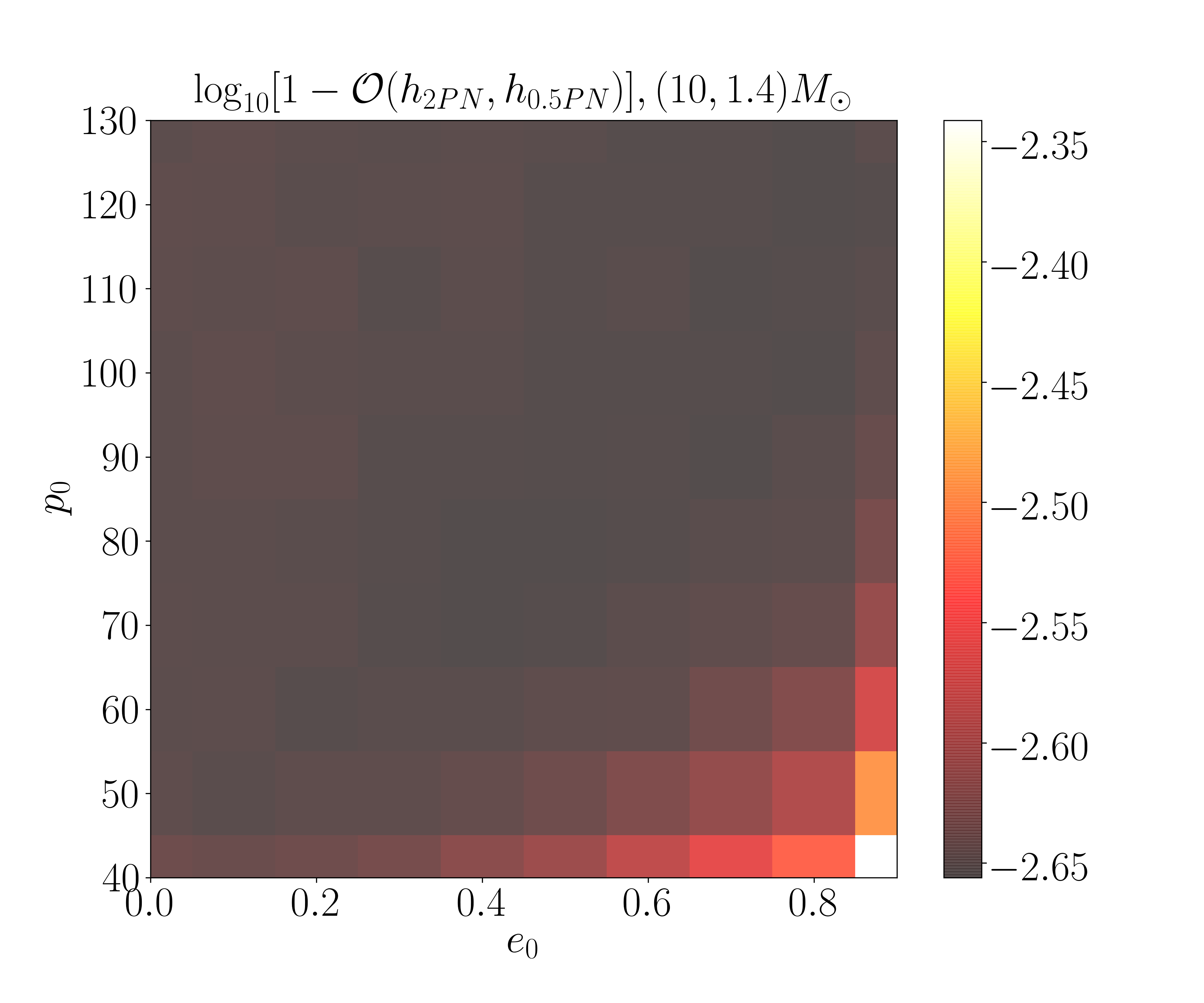}
\caption{\label{fig:PN_amp_study} $\log_{10}$ of $1-\mathcal{O}$ between a waveform keeping only the leading order corrections and one keeping 2PN corrections in the amplitudes for a $(10, 10)M_{\odot}$ (top left), $(1.4, 1.4)M_{\odot}$ (top right), and $(10, 1.4)M_{\odot}$ (bottom left) system. On the bottom right panel we show the same but between a waveform keeping 0.5PN corrections in the amplitude and one keeping 2PN corrections for the unequal mass case. There is good agreement in the equal mass case, but significant loss in the unequal mass case if we keep only leading order corrections in the amplitude.}
\end{figure*}

The numerical solutions for the orbital elements are plugged into the GW polarization to form $h(t)$ (where we choose to sample all time domain quantities at 8192 Hz). This time series is then zero padded on either side such that its length is the nearest integer power of 2, and we take the discrete Fourier transform using the \texttt{FFTW} library implemented in \texttt{C++}. The overlaps are computed from $f_1 = 1$Hz to $f_2 = 4096$Hz. We use an interpolation of a set of values of the noise curve corresponding to stationary Gaussian noise in aLIGO at design sensitivity \cite{2017CQGra..34d4001A}. The integrals appearing in the overlap are evaluated as sums with frequency resolution determined by the frequency resolution of the discrete Fourier transform, which depends on the length of signal, and as a result varies between different initial conditions. 

Figure \ref{fig:PN_amp_study} presents density contours for the value of the overlap between various templates as a function of the initial eccentricity and dimensionless semi-latus rectum. In the equal mass case, the overlaps are not affected past the third decimal place if we keep the amplitude to Newtonian order. We also find that keeping Kepler's equation at Newtonian order when solving for $u(t)$ only affects the overlaps in the fifth decimal place. This suggests we can keep the amplitude and Kepler's equation to Newtonian order, which considerably simplifies the decomposition of $h(t)$ into a sum over harmonics of $l$ and $\lambda$, which in turn is crucial to analytically approximate the Fourier transform. However, half order contributions vanish in the equal mass limit and thus the overlap is affected in the 2nd digit for unequal mass binaries. This amount of loss in overlap may be considerable, depending on the data analysis application one has in mind, and it is not due to solving Kepler's equation at Newtonian order, but rather due to neglecting higher order PN effects in the amplitude. We will return to this loss of accuracy for non-equal mass systems later.  

\subsection{Harmonic Decomposition}
\label{subsec:harm_decomp}

Following paper 1, a crucial step in obtaining a Fourier domain waveform is to decompose the time domain signal into a sum over harmonics of the angles related to the mean orbital frequencies $n$ and $\omega$. We carry out such a decomposition here and then verify that it faithfully represents the un-decomposed signal using overlaps. The results of Section \ref{subsec:PN_amp_effects}, however, indicate that we may retain only the leading order (Newtonian) contributions in the GW amplitude and in Kepler's equation, at least for comparable mass binaries, and thus, we will restrict attention to such a model. 

The results of this section rely heavily on the work to analytically solve 3PN accurate Kepler's equation by Boetzel et. al. in \cite{2017PhRvD..96d4011B}, which provides a useful decomposition of harmonics of $W$ into an infinite sum of harmonics of $l$:
\begin{equation}
e^{imW} = \sum_{n = -\infty}^{\infty} \mathcal{P}^{mW}_{n}e^{inl}.
\end{equation}
Re-expressing the polarizations in Eq.~\eqref{eq:0pn_polar} in terms of complex exponentials and inserting the above decomposition yields
\begin{align}
h(t) &= \frac{m\eta}{R}y^2(1-e^2)(F_{+}H^{(0)}_{+} + F_{\times}H^{(0)}_{\times}) \,,
\end{align}
with
\begin{align}
\label{eq:Hp_Hc_decomp1}
H_{+}^{(0)} &= P^{0}+ \frac{P^{0}_{C2C2}+iP^{0}_{C2S2}}{2} \sum_{s = -\infty}^{\infty} \!\!\! \mathcal{P}^{2W}_{s}e^{-i(sl+2\bar{\lambda})} + {\textrm{c.c.}} \, ,
\\
H_{\times}^{(0)} &= \frac{X^{0}_{C2C2}+iX^{0}_{C2S2}}{2} \sum_{s = -\infty}^{\infty} \!\!\! \mathcal{P}^{2W}_{s}e^{-i(sl+2\bar{\lambda})} + {\textrm{c.c.}} \, ,
\end{align}
where ${\textrm{c.c.}}$ stands for the complex conjugate, and where $F_{+}$ and $F_{\times}$ are the antenna functions of the detector
\begin{align}
F_{+}(\theta,\Phi,\psi) &= \frac{1}{2}(1+\cos^{2}\theta)\cos 2\Phi \cos 2\psi
\nonumber \\ &
- \cos \theta \sin 2\Phi \sin 2\psi\,, \\
F_{\times}(\theta,\Phi,\psi) &= F_{+}(\theta,\Phi,\psi-\pi/4)\,,
\end{align}
with $\theta$ and $\Phi$ the sky angles associated with the orientation of the detector relative to the source, and $\psi$ the polarization angle that defines the rotation from the wave's polarization basis to that defined by the arms of the detector.

Inspection of Eqs.~\eqref{eq:0pn_polar} and ~\eqref{eq:P_X} reveals that harmonic decompositions of three additional expressions are required:
\begin{align}
\chi_0 &= \frac{e\cos{u}}{1-e\cos{u}}\,, 
\\
\chi_1 &= -\frac{1}{1-e\cos{u}}\left[ (e\cos{u})^2 - e\cos{u} - 2e^2 + 2 \right]\,,
\\
\chi_2 &= e\sqrt{1-e^2}\frac{\sin{u}}{(1-e\cos{u})^2}\,.
\end{align}
The decomposition of $\chi_0$ at 0PN order, which is consistent with our 0PN treatment of Kepler's equation, is very straightforward and we obtain (using Eq.(B13) of \cite{2001MNRAS.325..358P})
\begin{align}
\chi_0 &= \sum_{s = -\infty}^{\infty} G_{s}e^{isl} \, ,
\end{align}
with $G_0 = 0$ and $G_{s \neq 0} = J_s(se)$, where $J_s(x)$ are Bessel functions of the first kind. With a little more work, we further obtain 
\begin{align}
\chi_1 &= \sum_{s=-\infty}^{\infty}I_{s}e^{isl}\,,
\end{align}
with the definitions
\begin{subequations}
\begin{align}
I_0 &= (e^2 - 1) 2A_0^2\,,
\\
I_{s \neq 0} &= J_s(se) + (e^2 - 1)A_{s}^2\,,
\end{align}
\end{subequations}
where the $A_j^s$ are given by Eq.~(42b) of \cite{2017PhRvD..96d4011B}. Lastly, for $\chi_2$ we have
\begin{align}
\chi_2 &= i\sum_{s = -\infty}^{ \infty}L_{s}e^{isl}
\end{align}
where we have defined
\begin{subequations}
\begin{align}
L_s &= e\sqrt{1-e^2}\left[ J'_s(se)A_0^1+\frac{1}{2}K_{|s|}\right]\,,
\\
K_s &= \sum_{j=1}^{s-1}A^1_jJ'_{s-j}[(s-j)e] - \sum_{j = s + 1}^{\infty} A^1_jJ'_{j-s}[(j-s)e]\,,
\nonumber \\
&+ \sum_{j=1}^{\infty}A^1_jJ'_{j+s}[(j+s)e]\,.
\end{align}
\end{subequations}

We can combine all of these results to find the harmonic decomposition of the waveform. 
First, we find that
\begin{subequations}
\begin{align}
P^0_{C2C2}+iP^0_{C2S2} &= -(1+C^2)\sum_{s = -\infty}^{\infty}M_se^{isl}\,,
\\
X^0_{C2C2}+iX^0_{C2S2} &= -2iC\sum_{s = -\infty}^{\infty}M_se^{isl}\,,
\end{align}
\end{subequations}
where we have defined $M_s = I_s + 2L_s$, and we note that $M_{s} = M_{-s}$. With this, we then have 
the full decomposition of $H^{(0)}_{+,\times}(t)$ into harmonics of $l$ and $\lambda$:
\begin{subequations}
\label{eq:H_decomp_2}
\begin{align}
H^{(0)}_+ &= S^2\sum_{s=-\infty}^{\infty}G_se^{isl} 
 \nonumber \\
& - \frac{(1+C^2)}{2} \sum_{j=-\infty}^{\infty}N_{j}e^{-i(jl+2\bar{\lambda})}  + {\textrm{c.c.}} 
\\
H^{(0)}_{\times} &= -iC  \sum_{j=-\infty}^{\infty}N_{j}e^{-i(jl+2\bar{\lambda})} - {\textrm{c.c.}} \,,
\end{align}
\end{subequations}
where we have defined
\begin{align}
N_j &= \sum_{s = - \infty}^{\infty} \mathcal{P}^{2W}_{s}M_{j-s}\,.
\end{align}
The $N_j$ coefficients scale as $e^{|j|}$ to leading PN order. In this work, we represent $N_j$ as the first 20 terms in a low-eccentricity expansion past leading order and at 3PN in the contributions from $W$ (keeping only 0PN contributions from the solution to Kepler's equation) with $j=[-15,15]$. We provide each of these coefficients to 16 digits of precision in the supplemental material. 

Putting the above together to form the GW strain we then have
\begin{equation}
\label{eq:td_harm_decomp}
h(t) =\mathcal{A}\left( F_{+}S^2\sum_{s=-\infty}^{\infty} G_se^{isl} + Q\sum_{j=-\infty}^{\infty}N_je^{-i(jl+2\lambda)} + {\textrm{c.c.}} \right) 
\end{equation}
where $Q$ is the complex function of the antenna functions and sky location
\begin{subequations}
\begin{align}
Q &= - \left(F_{+}\frac{1+C^2}{2} + iCF_{\times}\right)e^{i2\beta} \, ,
\end{align}
\end{subequations}
and $\mathcal{A} = \frac{m\eta}{R} y^2(1-e^2)$. Notice that in the circular limit the first term in Eq.~\eqref{eq:td_harm_decomp} is not present,
as it scales as $G_{s} = {\cal{O}}(e^{|s|})$.

\begin{figure}[htp]
\includegraphics[clip=true,angle=0,width=\columnwidth]{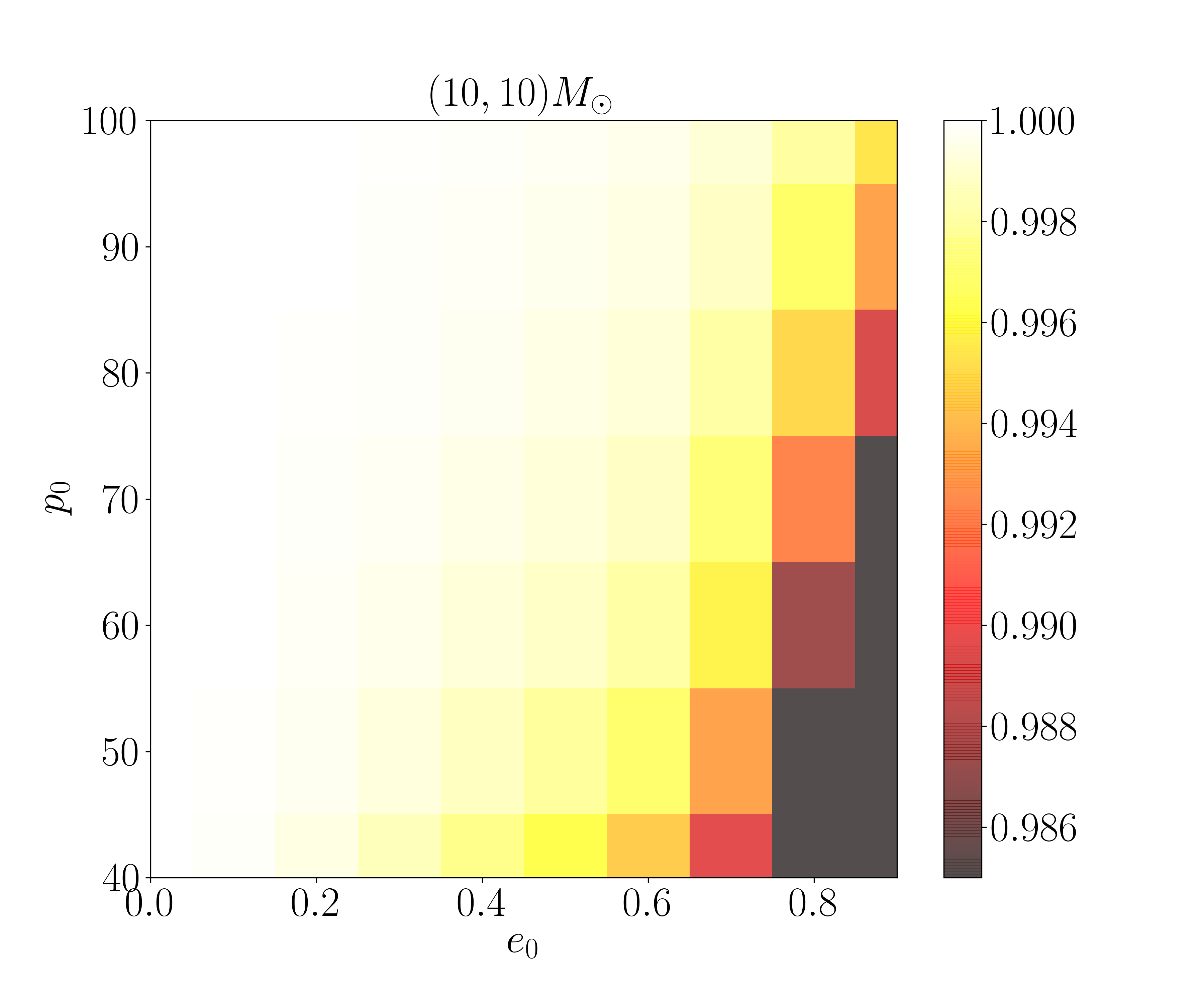} \\
\vspace{-0.3cm}
\includegraphics[clip=true,angle=0,width=\columnwidth]{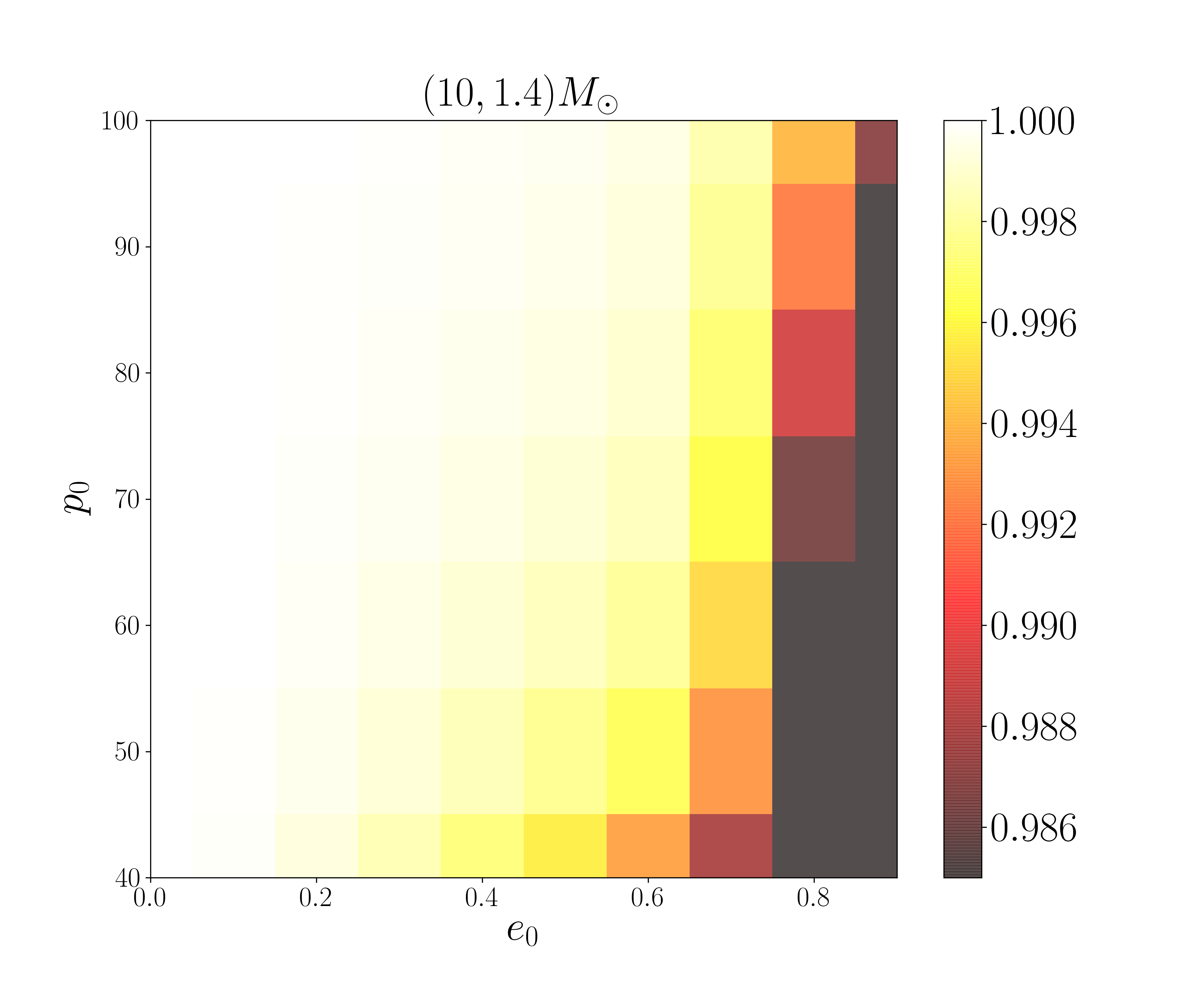} \\
\vspace{-0.3cm}
\includegraphics[clip=true,angle=0,width=\columnwidth]{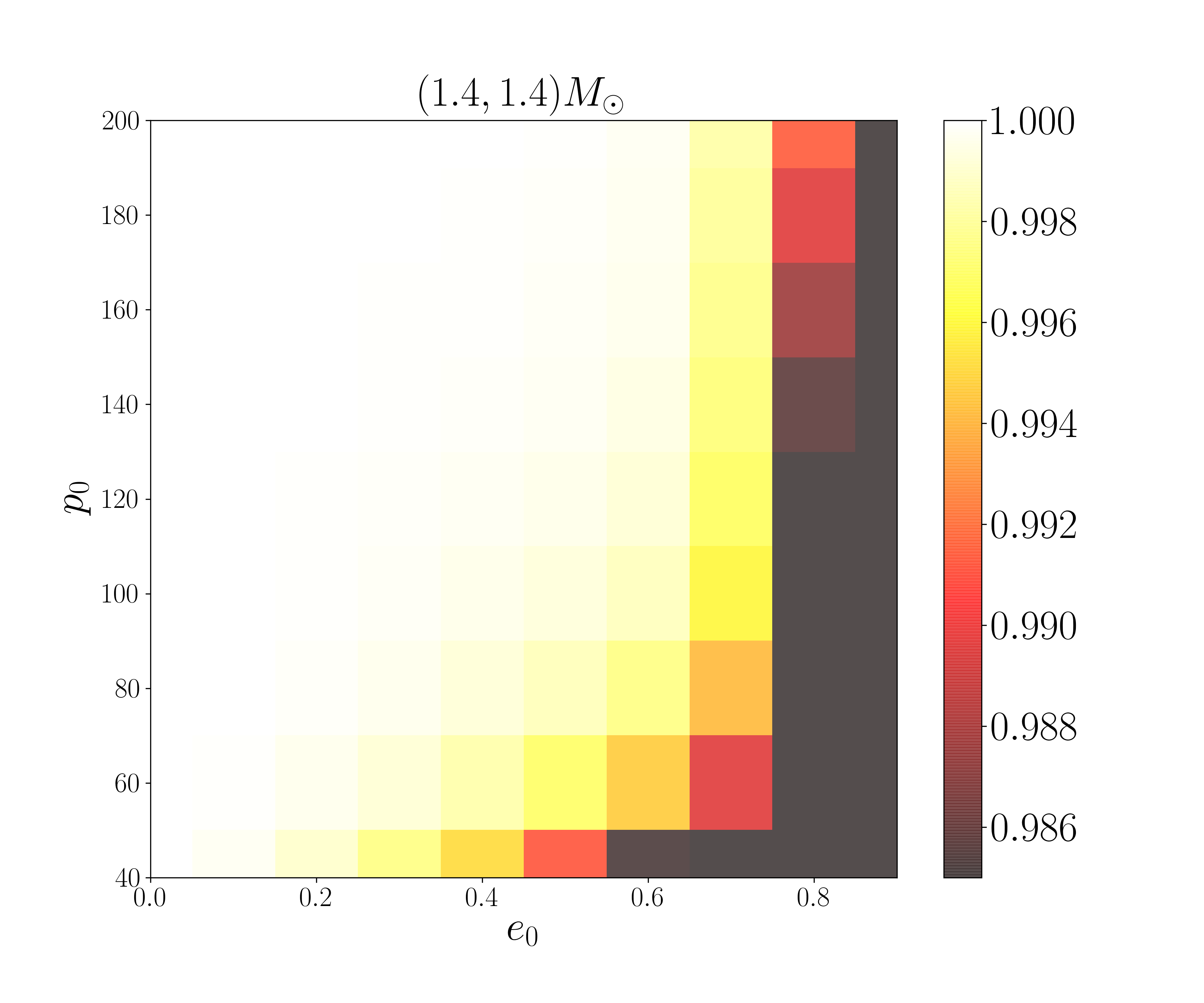}
\caption{\label{fig:harmdecomp_study} Overlap between the harmonic decomposed signal of Eq.~\eqref{eq:td_harm_decomp} with $s = [-15,15]$ and $j = [-15,15]$ harmonics and the non-decomposed signal of Eq.~\eqref{eq:0pn_polar}. The overlap is above 0.985 for much of the parameter space.}
\end{figure}

We are now in a position to check if the low eccentricity expansion of the $N_j$ and $G_s$ can be used to faithfully reconstruct the signal. In order to do so, we compute the overlap between the harmonically decomposed waveform keeping $j = [-15,15]$ and $s = [-15,15]$ harmonics in Eq.~\eqref{eq:td_harm_decomp}, and a waveform where trigonometric functions of $u$ and $W$ have not been harmonically decomposed. Figure \ref{fig:harmdecomp_study} shows these overlaps for a $(10, 10)M_{\odot}$, $(10, 1.4)M_{\odot}$, and $(1.4, 1.4)M_{\odot}$ binary. We see that for a majority of the parameter space considered, the overlap is above 0.98. 

For very high initial eccentricities ($e_{0}>0.7)$ and small dimensionless initial separations ($p_{0} < 50 $), however, the overlap decreases sharply (below the minimum value shown) to values as low as ~0.93. This is due to a combination of spectral truncation (i.e. not enough harmonics have been kept in Eq.~\eqref{eq:td_harm_decomp} to faithfully reconstruct the full signal) and inaccuracy of the harmonic amplitudes, as they have been expanded in low eccentricity. Since Bessel functions appear all throughout the harmonic amplitudes with the form $J_s(se)$, and expansions of Bessel functions are known to converge very slowly as their argument is near their index, we conclude that the faithful modeling of this small subset of the parameter space would require a considerably higher number of terms in the eccentric expansions of the amplitudes, as well as more harmonics in the sum of Eq.~\eqref{eq:td_harm_decomp}.

\subsection{Qualitative Analysis of Eccentric Signals}
Let us now develop a sense of the relative strength of the harmonics and the structure of the Fourier domain signal before moving on to explicitly deriving such a Fourier-domain waveform. We achieve this through a few different methods. First, we examine the structure of the Fourier response without including the effects of radiation reaction. This reveals the relative strength of the different harmonics and the mapping between the orbital frequencies, $n$ and $\omega$, and the Fourier frequency, $f$. Second we examine the structure of the Fourier response with radiation reaction using both the Fourier transform and a time frequency representation of the signal. Lastly we investigate the relative strength of the $N_j$ by plotting their values as function of $j$ and $e$. Knowing which amplitudes are the strongest is key if one wishes to make the model more efficient by careful truncation or selection of which harmonics to include in Eq.~\eqref{eq:td_harm_decomp}.

Let us first inspect the Fourier response of a steady state system (i.e.~in the absence of radiation reaction so that $\dot{e} = \dot{y} = 0$). As a result the orbital angles $l$ and $\lambda$ are given by $l=n(t-t_0)$ and $\lambda=\omega(t-t_0)$. The Fourier transform of the signal as written in Eq.\eqref{eq:td_harm_decomp} is then trivial: the sum of the product of the harmonic amplitude $N_j$ or $G_s$ and delta functions centered at $ s n/(2\pi)$ or $\pm (j n + 2\omega)/(2\pi)$, where $s$ and $j$ are the indices of the sums in Eq.~\eqref{eq:td_harm_decomp}. We proceed by taking the discrete Fourier transform of the numerical un-decomposed time domain signal in order to get a sense of the distribution of strength of the harmonics, as well as to verify that the result is consistent with Eq.~\eqref{eq:td_harm_decomp}.

\begin{figure*}[htp]
\includegraphics[clip=true,angle=0,width=0.475\textwidth]{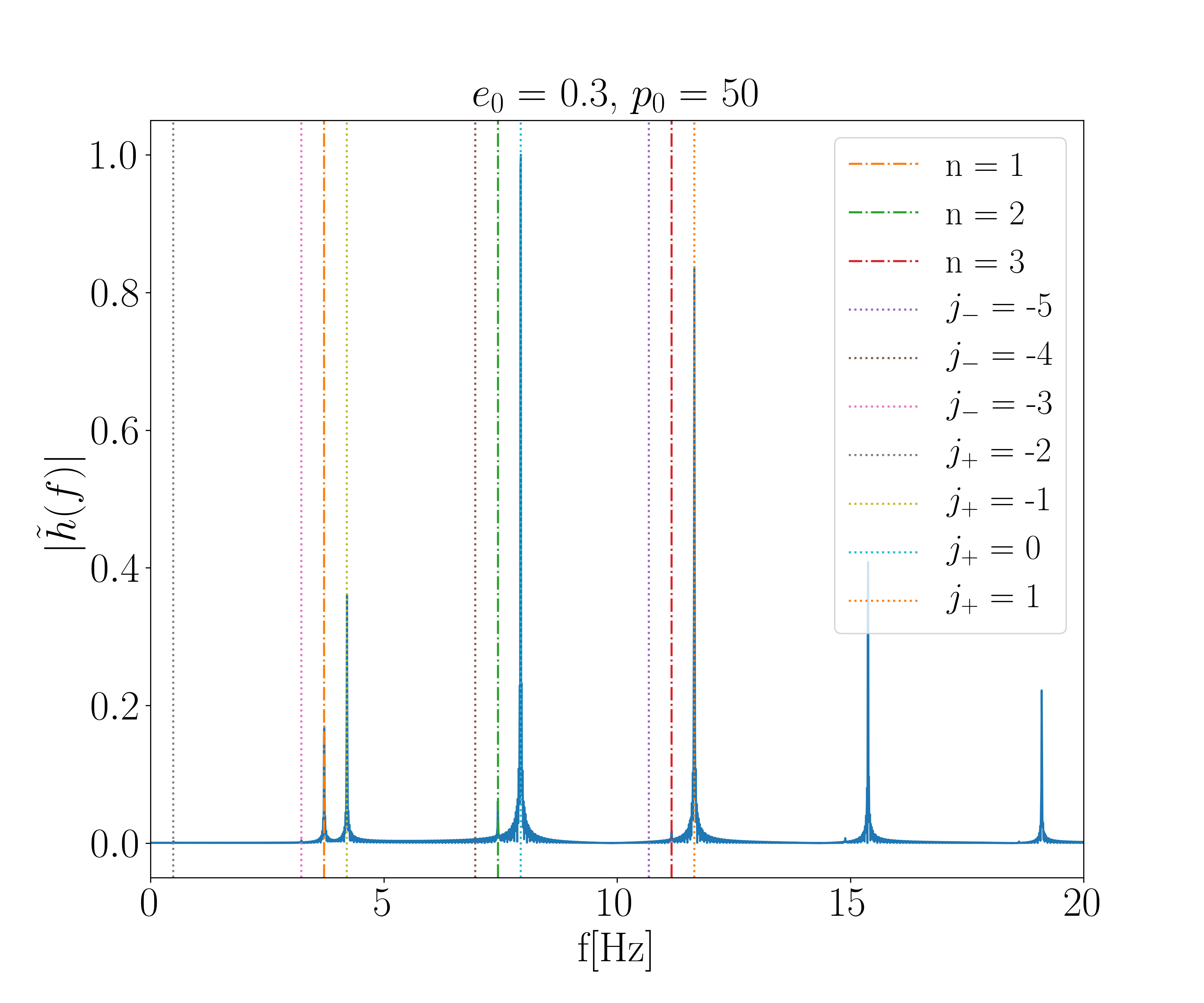}
\includegraphics[clip=true,angle=0,width=0.475\textwidth]{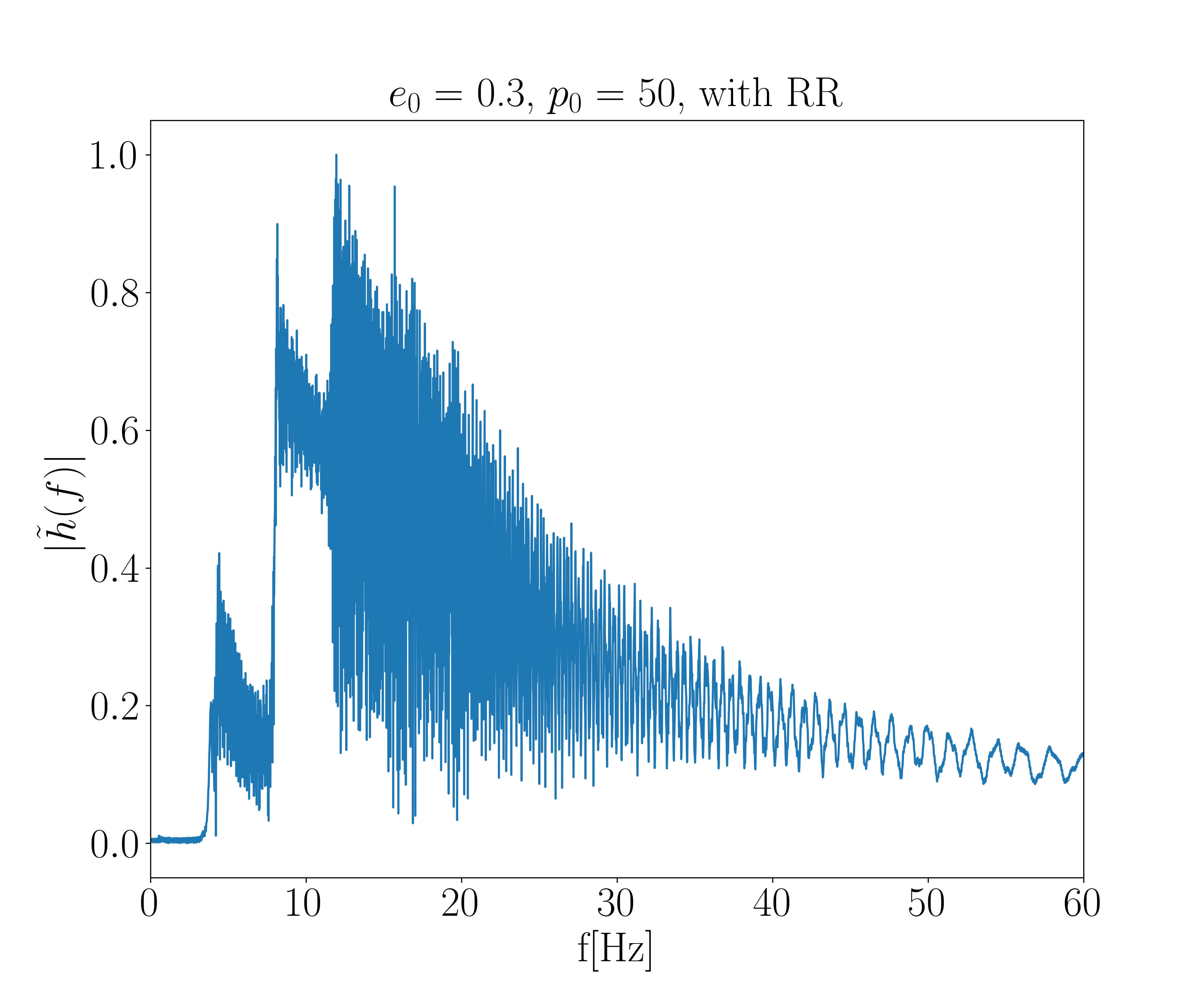}
 \\
\vspace{-0.3cm}
\includegraphics[clip=true,angle=0,width=0.475\textwidth]{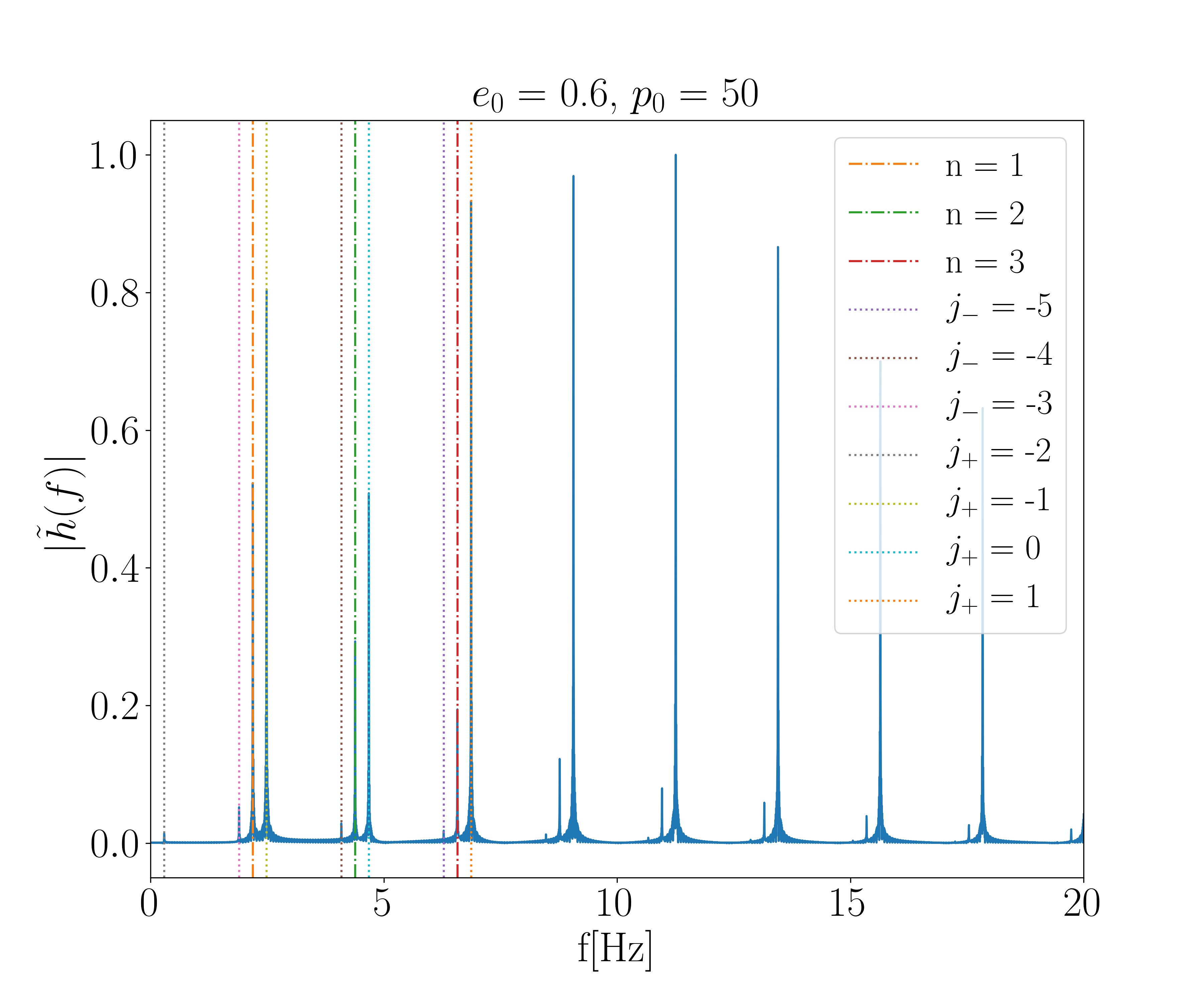} 
\includegraphics[clip=true,angle=0,width=0.475\textwidth]{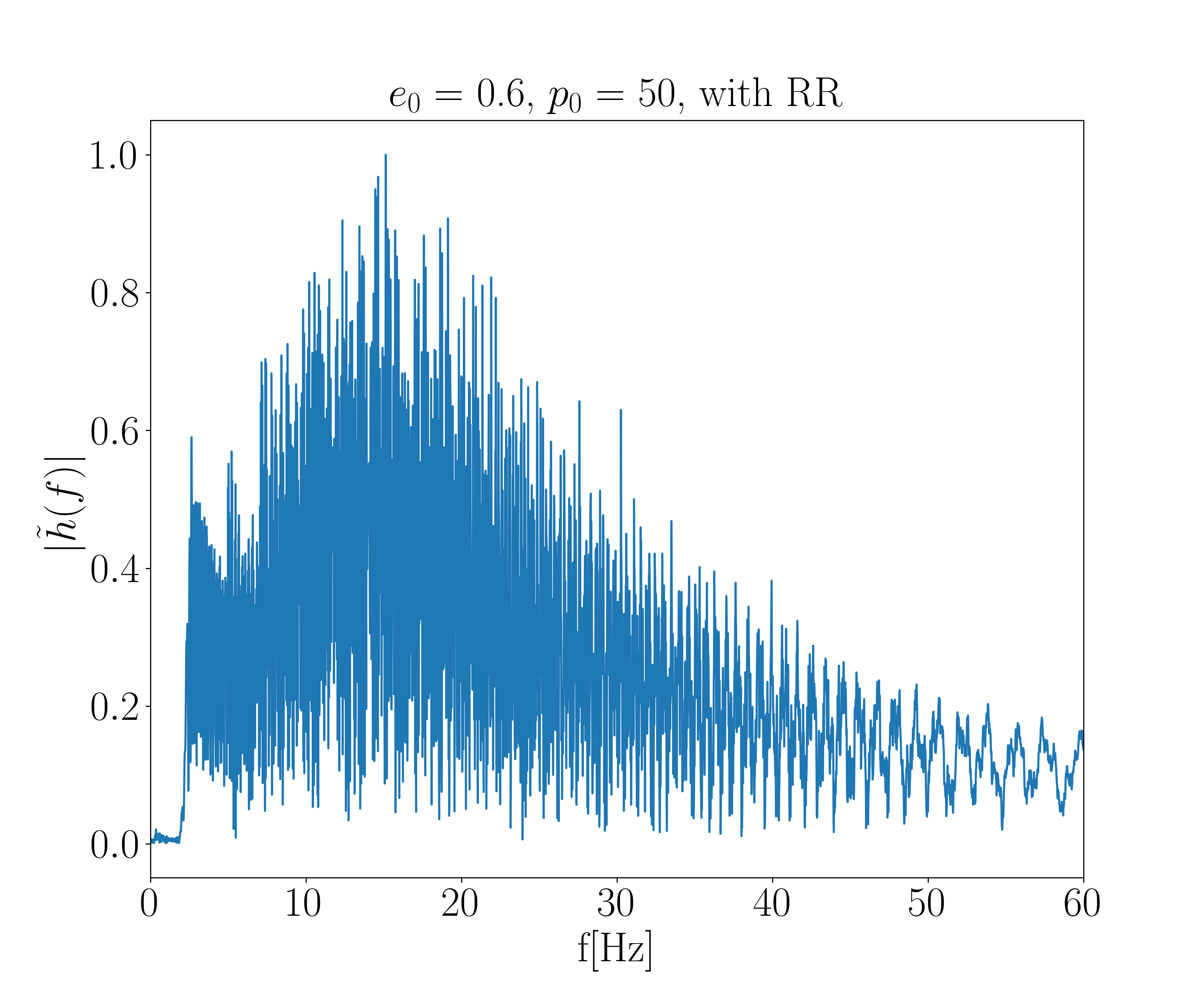}
\\
\vspace{-0.3cm}
\includegraphics[clip=true,angle=0,width=0.475\textwidth]{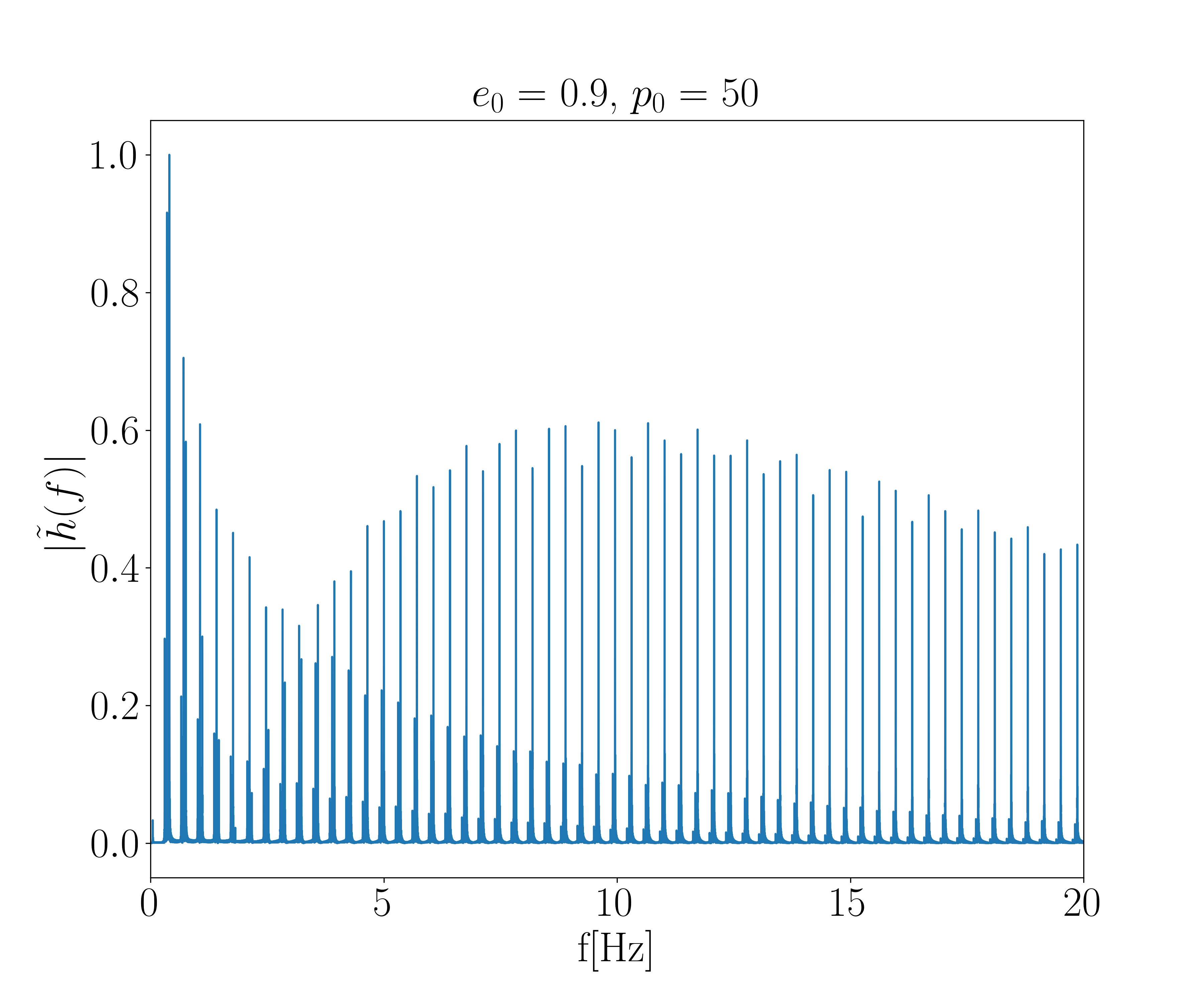}
\includegraphics[clip=true,angle=0,width=0.475\textwidth]{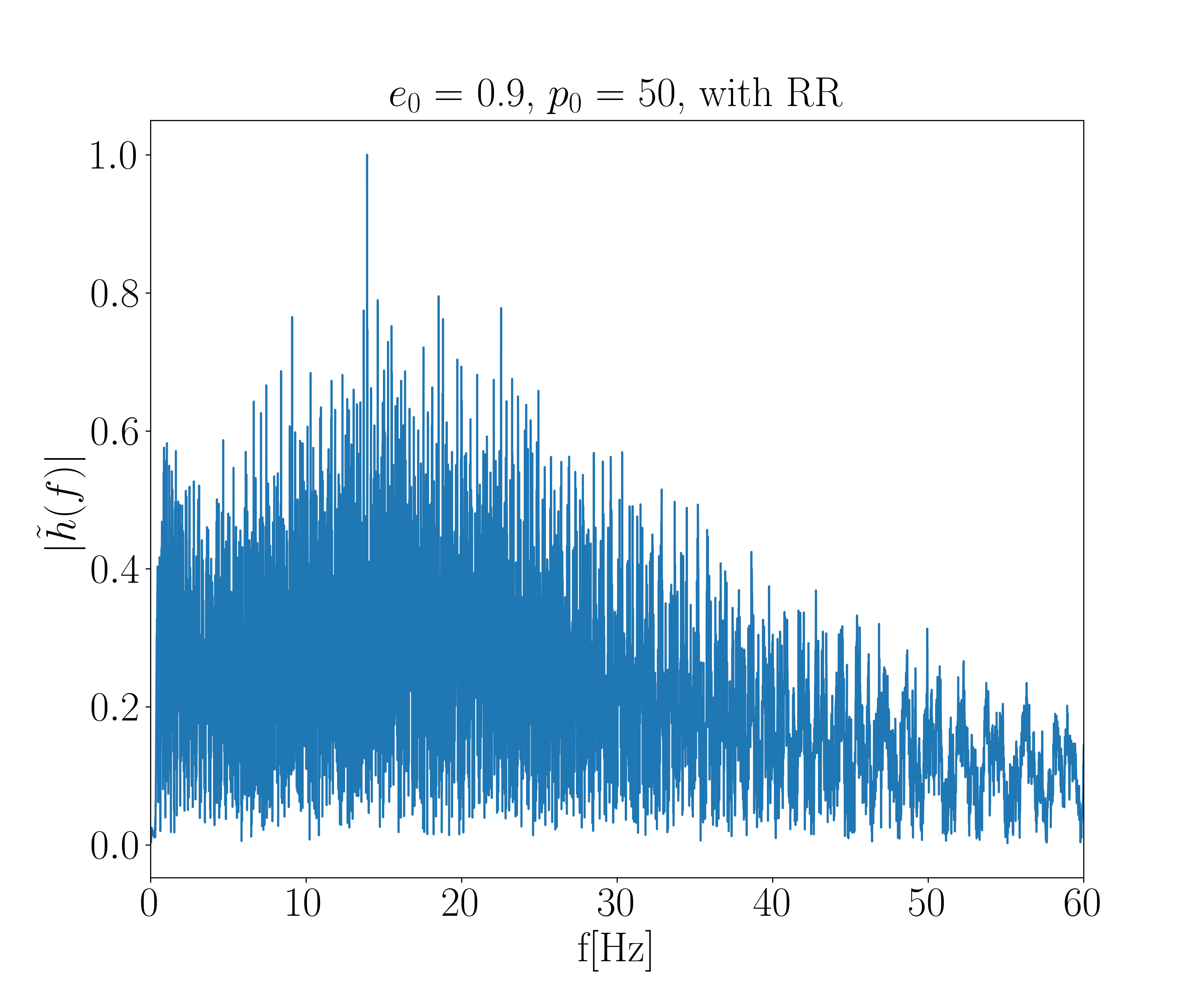}
\caption{\label{fig:noRR} DFT of the the numerically evolved time domain signal as written in Eq.~\eqref{eq:0pn_polar} without (left) and with (right) radiation-reaction effects for a BBH system with different initial conditions. When radiation-reaction is absent, the signal is split into harmonics of combinations of the orbital frequencies, with harmonic index as appearing in Eq.~\eqref{eq:td_harm_decomp} labeled. When radiation-reaction is present, the harmonics now sweep in frequency and interfere with once another producing a rapidly varying amplitude.}
\end{figure*}

The left panel of Fig.~\ref{fig:noRR} shows the amplitude of the discrete Fourier transform of the numerically evolved (at 3PN order in the dynamics, keeping only leading PN order contributions in the amplitude) time-domain signal as written in Eq.~\eqref{eq:0pn_polar}. As predicted, we see spectral lines appearing at Fourier frequencies that are combinations of the two orbital frequencies. In the case of the signals with modest initial eccentricities, we have labeled a number of frequency positions of the different harmonics. The subscript on the index $j$ indicates whether the plus or minus has been taken in the location of the $j$th harmonic given by $\pm (j n + 2\omega)/(2\pi) $. In the high eccentricity case, there are too many closely spaced harmonics to clearly indicate which is which. The harmonics of index $j_{+}$ tend to be dominant, followed by the harmonics of $s$, and the harmonics of index $j_{-}$ are clearly subdominant. We choose all sky angles to be $3\pi/7$, and have verified that the results do not qualitatively change much with the choice of sky angle (other than the fact that the harmonics of $s$ vanish in the $\iota \rightarrow 0$ limit).

When we allow the binary to inspiral due to radiation reaction, the orbital eccentricity and the two orbital frequencies change with time. As a result, the Fourier transform is not a simple sum of spectral lines any longer. Since the amplitude is slowly varying compared to the complex exponentials, we can expect the spectral lines of the left panels of Fig.~\ref{fig:noRR} to sweep with the orbital frequencies as the latter increase, forcing different harmonics to overlap and to interference. The right panel of Fig.~\ref{fig:noRR} shows the Fourier spectrum when the effects of radiation reaction are included. We see that the amplitude oscillates rapidly due to harmonic interference as expected. 

\begin{figure*}[htp]
\includegraphics[clip=true,angle=0,width=0.475\textwidth]{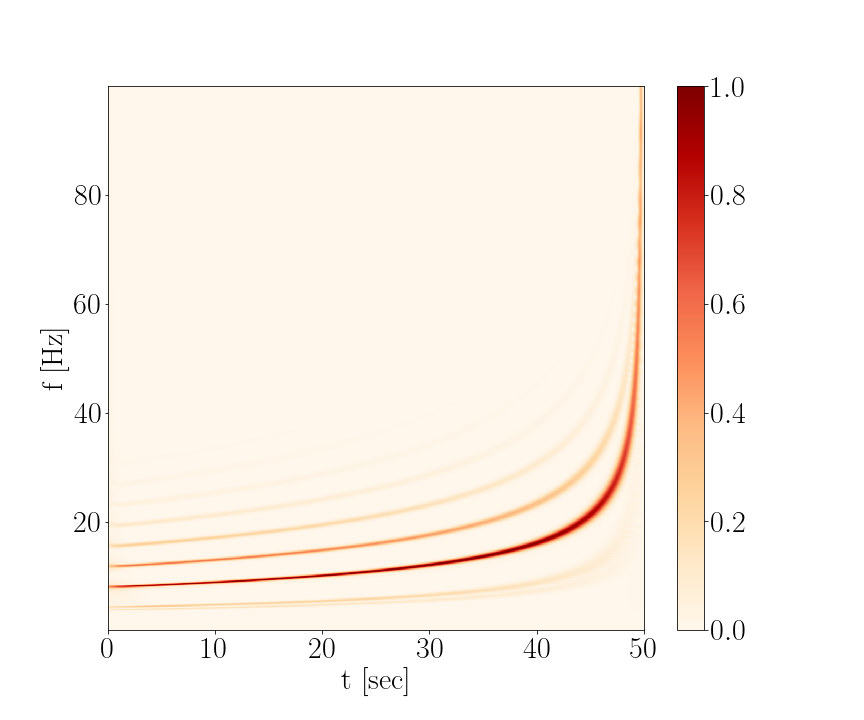}
\includegraphics[clip=true,angle=0,width=0.475\textwidth]{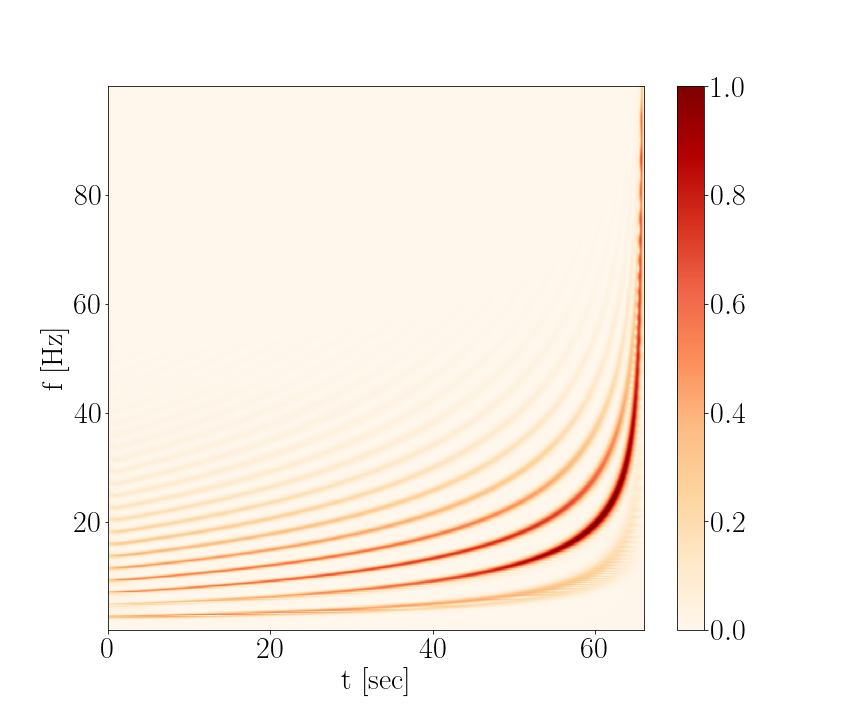}
\includegraphics[clip=true,angle=0,width=0.475\textwidth]{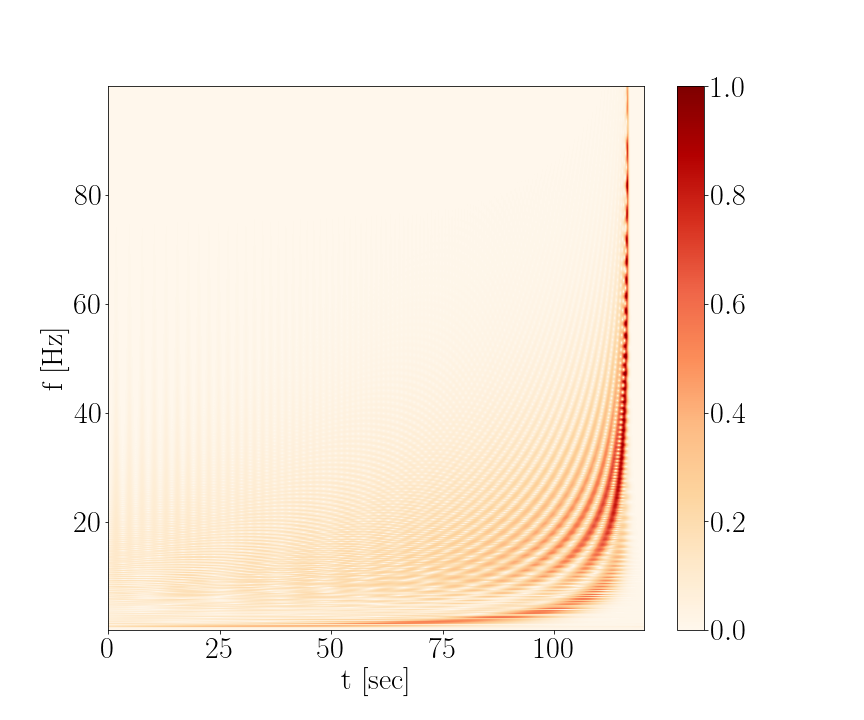}
\includegraphics[clip=true,angle=0,width=0.475\textwidth]{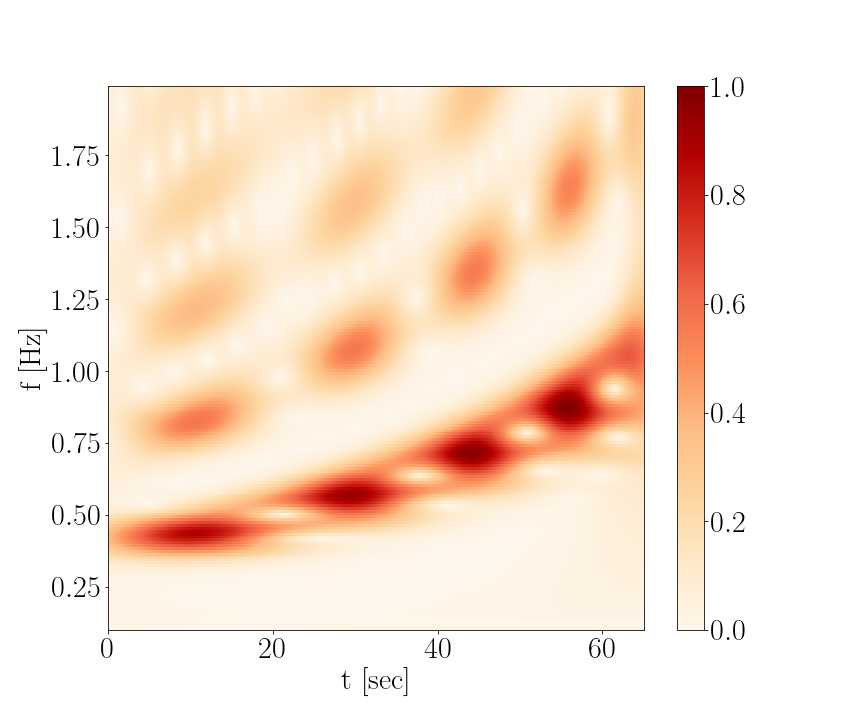}
\caption{\label{fig:timefreq} A time frequency representation of the numerically solved eccentric signal for a BBH system all with a $p_0 = 50$, $e_{0} = 0.3$ (top left), $e_{0} = 0.6$ (top right), $e_{0} = 0.9$ (bottom left). The bottom right is a zoomed in view of the high eccentricity case. We clearly see in time frequency space that the signal is composed of different harmonics which increase in frequency as time increases. The bursty nature of highly eccentric signals is encapsulated by the bottom right panel. Power rapidly decays from higher harmonics as frequency is increased, suggesting that they maybe be negligible past certain frequencies.}
\end{figure*}

Since the amplitude is so rapidly oscillating when radiation-reaction is included, it is more enlightening to look at a time-frequency representation of the GW signal. Figure \ref{fig:timefreq} shows the Q-transform of the GW signals considered in the right panel of Fig.~\ref{fig:noRR}. A Q-transform is one in which the basis functions are sine-Gaussians with quality factor Q, such that a low (high) value of Q localizes the basis more in time (frequency). With this representation of the signal, we can clearly see the behavior of the different harmonics. Each harmonic takes a distinct track in time-frequency space, as a result of the orbital frequencies increasing in time. We also see that harmonics, besides the $j = 0$ one (the only non-vanishing harmonic in the circular limit), decay with time and very rapidly with frequency. The lower left panel shows the time frequency evolution for the high eccentricity case ($e_0 = 0.9$). Although at late times the harmonic structure is clear, this is not so at early times, which is an artifact of the particular choice of Q we made in this figure (a choice of $Q=150$ does well at capturing the oscillatory nature of the late signal since it is more localized in frequency). At early times the signal consists of a series of bursts. In the lower right panel we have plotted the Q-transform of some of the early signal with $Q = 20$, to appreciate how much more localized in time the signal is at early times. 

\begin{figure*}[htp]
\includegraphics[clip=true,angle=0,width=0.475\textwidth]{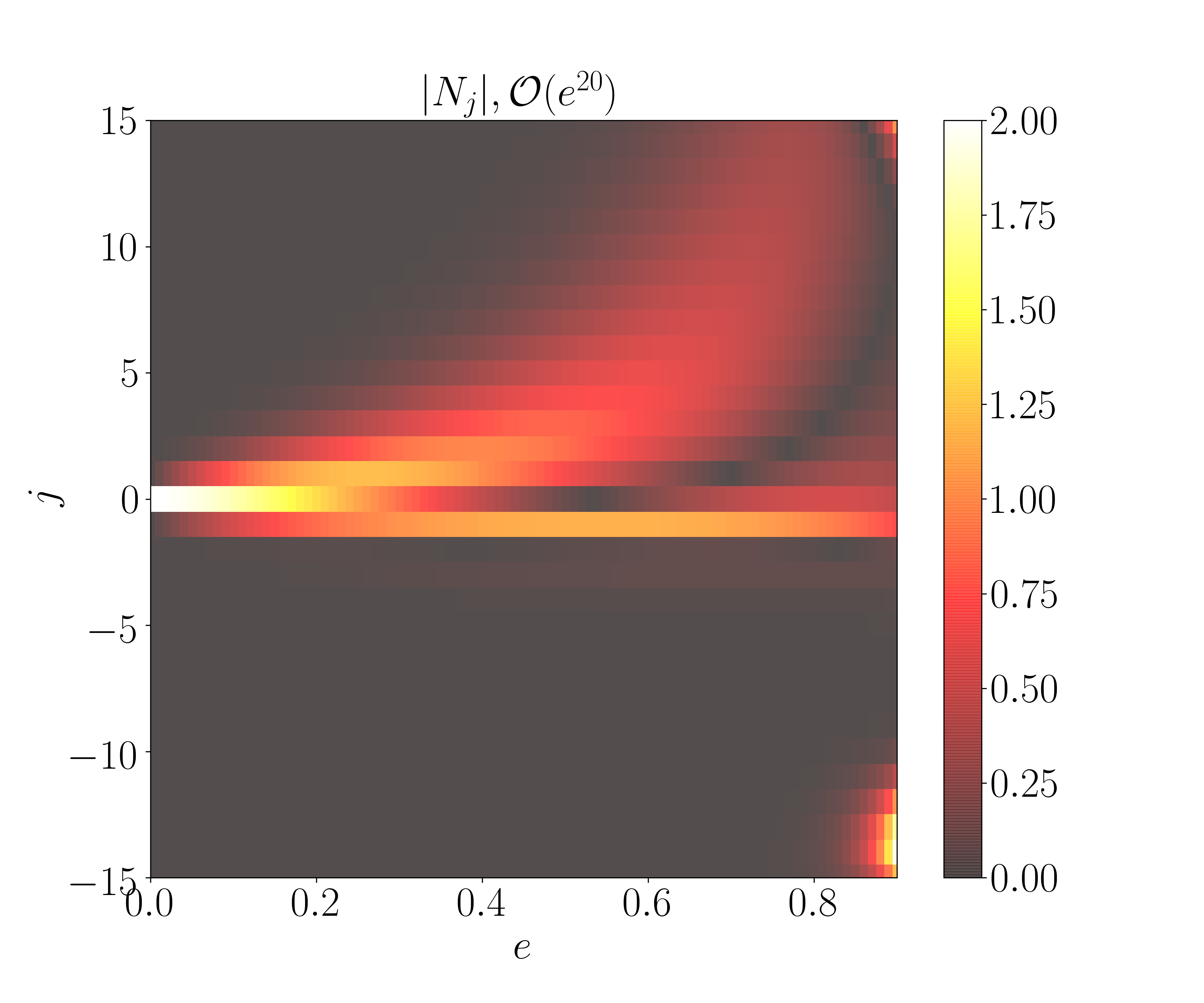}
\includegraphics[clip=true,angle=0,width=0.475\textwidth]{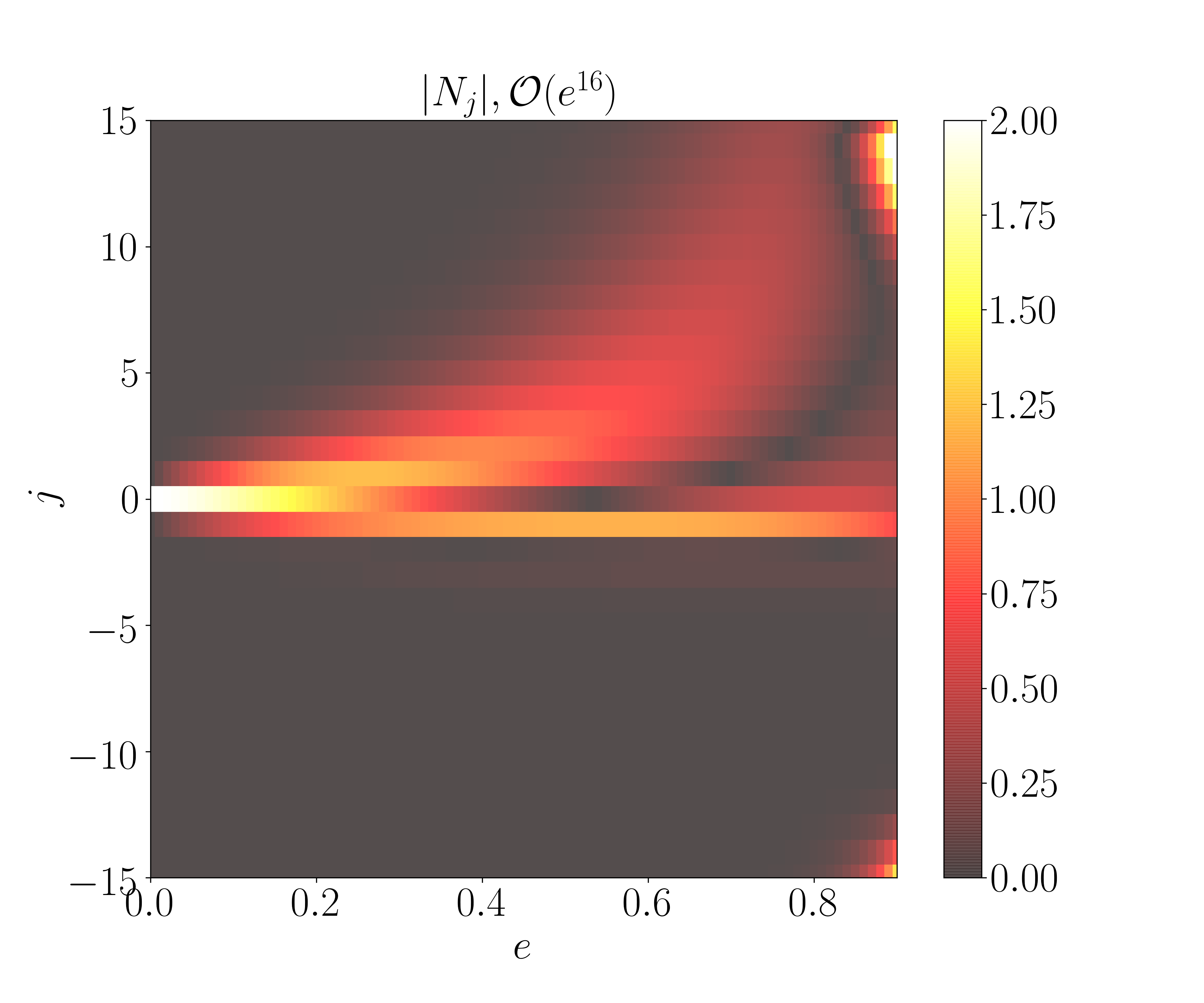}
\caption{\label{fig:Nj} Absolute value of $N_j$ with fixed $\eta = 0.25$ and $p_0 = 50$, keeping 20 terms in eccentricity past leading order (left) and 16 terms past leading (right). Most of the power is concentrated in the $j > -1$ harmonics as eccentricity is increased. For large eccentricities, we see that differences arise between the two panels suggesting that perhaps more eccentric corrections are needed to evaluate the harmonics accurately at such large eccentricities.}
\end{figure*}

Figure \ref{fig:Nj} presents the value of $N_j$ as a function of the index $j$ and the eccentricity $e$ at which they are being evaluated, keeping 20 terms in eccentricity past leading (left) and 16 terms past leading (right). We have set $\eta = 0.25$ in these results, but verified that setting $\eta = 0.1$ changes the result negligibly. We have set $y$ such that the results correspond to a dimensionless semilatus rectum of 50. In the low eccentricity limit, the $j=0$ harmonic dominates as expected as it is the only non-vanishing harmonic in the circular limit.  As the eccentricity is increased, the harmonics become more comparable in strength with the $j > -1$ harmonics clearly dominating. For large eccentricities, there is considerable strength in the larger negative $j$ harmonics. We see that at such high eccentricity there are visibly different features between the two panels suggesting that the eccentricity expansions are beginning to break down. 
\subsection{Different Time Domain Approximants}
\label{subsec:TaylorT4s}

Given the increase in the number of relevant ODEs and parameters in the eccentric problem, there are a number of different and new PN-consistent ways to solve the orbital dynamics beyond what is possible in the circular limit. In order to introduce these, let us begin by reviewing the circular TaylorT4 approximant (for a more in-depth discussion of the different circular PN approximants, see Sec.~III of~\cite{2009PhRvD..80h4043B}). In the circular TaylorT4 approximant, one numerically solves the set of ODEs:
\begin{subequations}
\begin{align}
\frac{dv}{dt} &= \frac{32}{5}\frac{\eta}{M}v^9 \left[1 - \left(\frac{743}{336} + \frac{11}{4}\eta\right) v^2 + ... + \mathcal{O}(v^7)\right]
\\
\dot{\phi} &= \frac{v^3}{M}
\end{align}
\end{subequations}
The numerical solutions to these are then substituted into a circular GW polarization and a DFT can then be taken to obtain a Fourier-domain waveform.

In the eccentric case, however, there are several different ways of solving the relevant orbital dynamics presented in Eq.\eqref{eq:orb_param_t}, and each of them is consistent in a PN sense. Given this, we introduce here 3 different time-domain approximants that extend the circular TaylorT4 approximant to eccentric inspirals: TaylorT4t, TaylorT4y, and TaylorT4e. Each of these is constructed by numerically solving a set of ODEs, with the difference being the independent variable $t$, $y$, or $e$ used to parameterized the solution.

The TaylorT4t approximant is obtained by using time $t$ as the independent variable when numerically solving the set of ODEs $( \dot{l}, \dot{\lambda}, \dot{e}, \dot{y})$ for $[\lambda(t), ~l(t), ~e(t), y(t)]$. The resulting solutions are then inserted into the appropriate expressions for the full orbital dynamics (Eq.~\eqref{eq:pn_orb}) to obtain $\phi(t)$, and the result of this is then inserted into the expression for $h(t)$ in Eq.~\eqref{eq:0pn_polar}. Finally, we take a DFT to obtain a Fourier domain waveform. 

The TaylorT4e approximant is obtained by using the eccentricity $e$ as the independent variable when solving the orbital dynamics equations. This implies we must first express the ODEs of TaylorT4t in terms of the eccentricity and then numerically solve $(d\lambda/de, dl/de, dt/de, dy/de)$ for $[\lambda(e),l(e),t(e),y(e)]$. This re-writing is obtained via the chain rule and the subsequent re-expansion of the resulting expressions to the appropriate PN order. For example, to obtain $dy/de$, we use $dy/de = \dot{y}/\dot{e}$, where the ratio is re-expanded to 3PN order in $y$. The numerical solution to these ODEs can then be expressed as functions of time by numerically inverting $t(e)$ for $e(t)$. With these substitutions done, one finds $h(t)$ through Eq.~\eqref{eq:0pn_polar} and applies a DFT to find the Fourier-domain representation. 

The TaylorT4y approximant is obtained by using the PN variable $y$ as the independent variable when numerically solving the set of ODEs $( d\lambda/dy, dl/dy, dt/dy, de/dy)$ for $[\lambda(y),l(y),t(y),e(y)]$. Again in each of these ODEs, any ratio of PN expressions arising due to the use of the chain rule is re-expanded to the appropriate PN order. With solutions for the orbital elements in terms of $y$ in hand, one then numerically inverts $t(y)$ for $y(t)$, and this is used to construct $h(t)$ via Eq.~\eqref{eq:0pn_polar} before taking a DFT to find the Fourier-domain representation. 

Each of these different ways to solve for the orbital elements is equally valid in the PN approximation, but each solution is slightly different due to the truncation of this series in the ODEs and the subsequent use of numerical solvers. Since our Fourier domain waveform will later rely on the differential equations of the TaylorT4e approximant, it is important to know how faithful TaylorT4e is relative to the TaylorT4t approximant. Such an analysis will reveal the PN limitations of our frequency domain waveform model. We choose to compare TaylorT4e to TaylorT4t because, at least in the circular limit, the latter have been shown to be the most faithful against EOB waveforms \cite{2009PhRvD..80h4043B}. 

\begin{figure}[htp]
\includegraphics[clip=true,angle=0,width=0.475\textwidth]{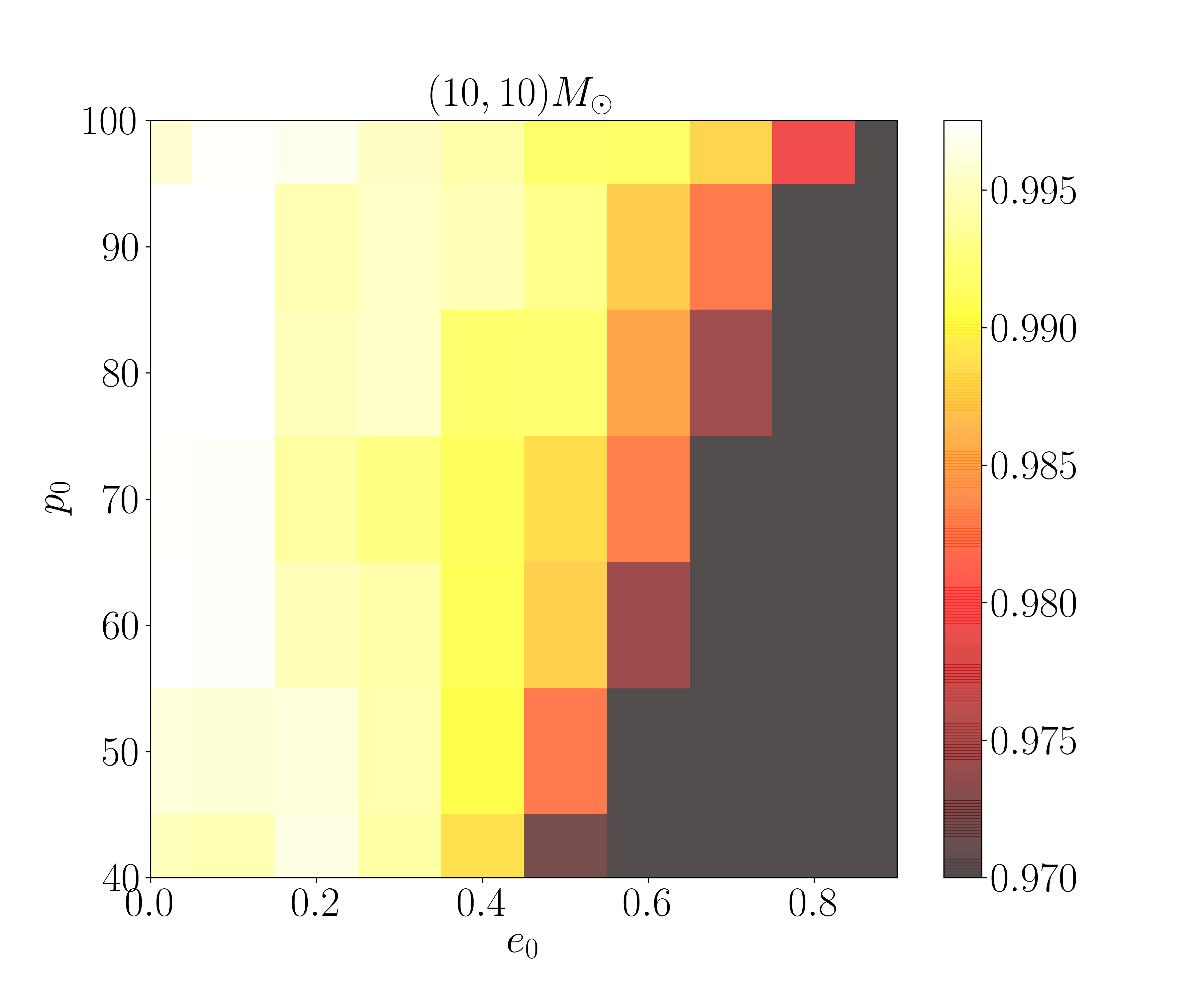}
\includegraphics[clip=true,angle=0,width=0.475\textwidth]{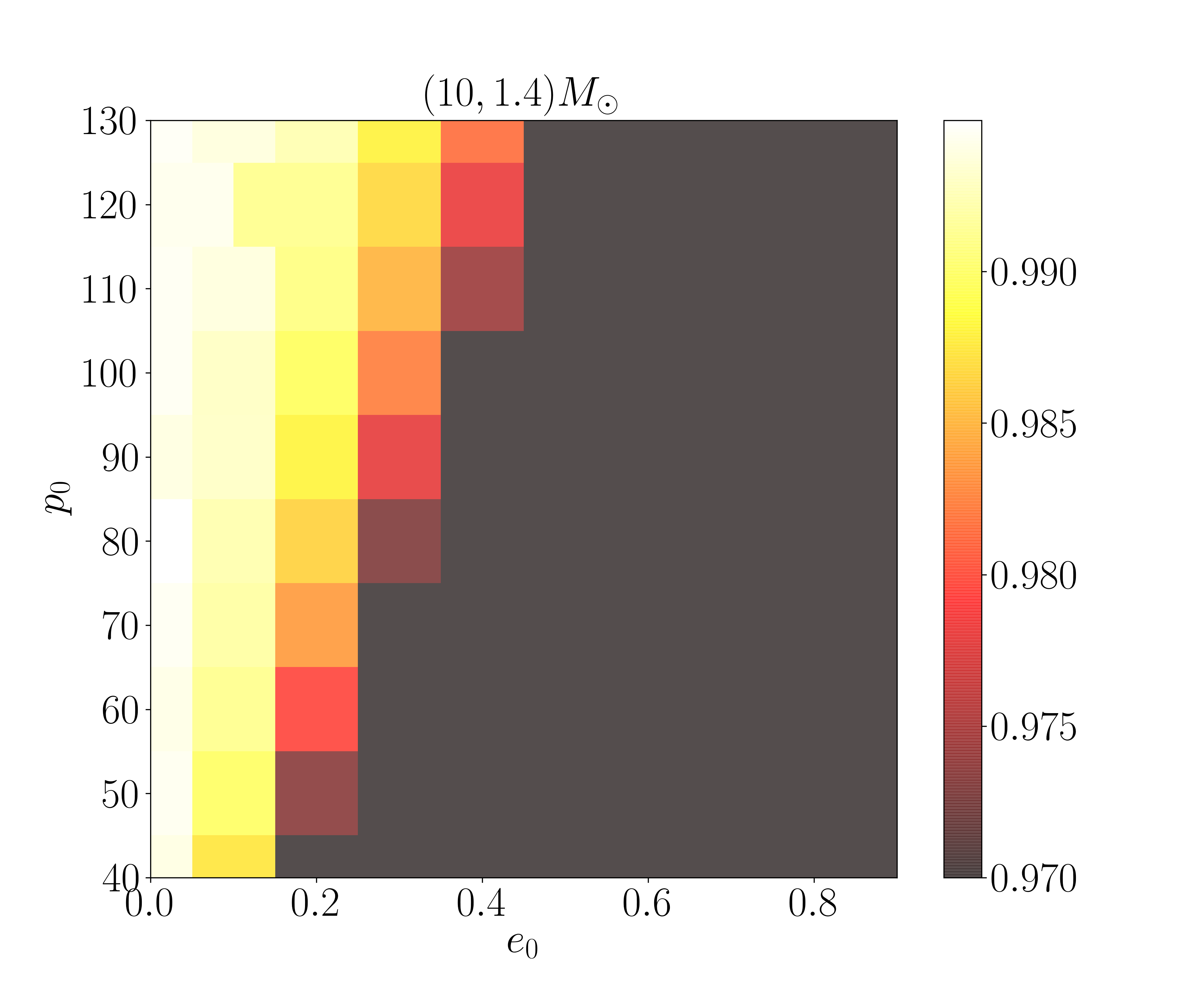}
\includegraphics[clip=true,angle=0,width=0.475\textwidth]{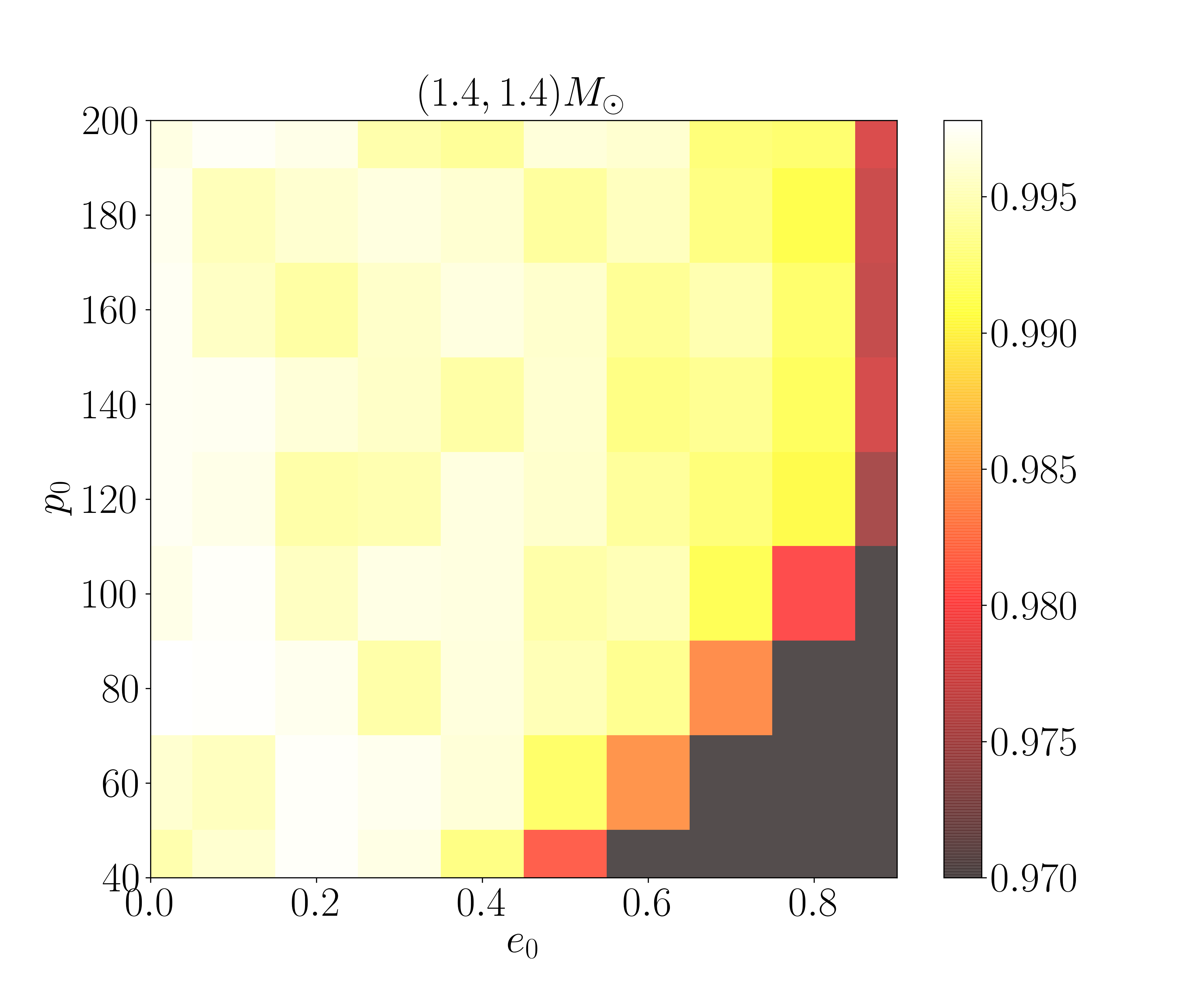}
\caption{\label{fig:T4t_T4e} Maximized overlap between the TaylorT4t and TaylorT4e approximants for three different systems with different initial conditions. In the equal-mass case, there is good agreement between the two models until the initial eccentricity is large enough or the initial separation is small enough. In the unequal-mass case, however, the disagreement becomes considerable even for small eccentricities.}
\end{figure}

Figure \ref{fig:T4t_T4e} shows the overlap maximized over a time offset and overall phase offset between the TaylorT4e and the TaylorT4t approximants for a $(10, 10)M_{\odot}$, a $(10, 1.4)M_{\odot}$, and a $(1.4, 1.4)M_{\odot}$ binary for a variety of different initial conditions ($e_0$ and $p_0$). In the equal-mass cases, we see that as the two models agree with each other in most of the initial parameter space, with disagreements only arising at sufficiently large initial eccentricity or small initial separation. To be clear, there has been no small eccentricity expansion or approximations between these two models, with the only source of disagreement coming from the truncation of the PN series when reformulating the ODEs in terms of $t$ or $e$. Both models are obtained purely numerically and without any harmonic decomposition. 

In the unequal mass case, the disagreement between the TaylorT4e and TaylorT4t approximants is much more pronounced. This is because of the well-known fact that the PN expansion is poorly behaved for unequal masses \cite{2008PhRvD..77l4006Y}, but this feature seems to become enhanced for even moderate initial eccentricities. Since the Fourier-domain model we will develop later solves for the orbital dynamics as formulated in the TaylorT4e approximants, we expect that it will also disagree with TaylorT4t to the level displayed in Fig.~\ref{fig:T4t_T4e}.

\section{Eccentric GWs in the Fourier Domain}
\label{sec:F2}
In this section we begin by reviewing the Fourier domain model that was derived in paper 1. Following the prescription laid out there, we extend the model to 3PN order in the Fourier phase. Lastly we describe how other frequency domain approximants could be derived in analogy to the different time domain approximants discussed in Section \ref{subsec:TaylorT4s}. 

\subsection{Review of Newtonian model}

In paper 1 we computed the Fourier response via the SPA of a time domain signal whose time domain harmonic decomposition took the general form: 
\begin{equation}
\label{eq:Newthoft}
h_{+, \times} = \sum^{\infty}_{j = -\infty} A_j e^{ijl} + {\textrm{c.c.}}
\end{equation} 
Here the amplitudes $A_j$ are functions of eccentricity and vary on a slower timescale (the radiation reaction timescale) than the orbital angle $l$ (which varies on the orbital timescale). This makes it appropriate for the application of the stationary phase approximation (SPA) which is described in detail in \cite{Bender}. The time domain signal in Eq.~\eqref{eq:Newthoft} is composed entirely of harmonics of $l$ because at Newtonian order $\lambda = l$, a result of periastron precession effects entering at 1PN order. Application of the SPA to the above signal yields
\begin{equation}
\tilde{h}_{+, \times} (f) = \sum^{\infty}_{j = 1} A_j(t^{\ast}_j)\sqrt{\frac{2\pi}{j\ddot{l}(t^{\ast}_j)}}e^{i\psi_j} \, ,
\end{equation}
where the Fourier phases are given by 
\begin{equation}
\psi_j = 2\pi ft^{\ast}_j - j l(t^{\ast}_j) - \pi/4 \, .
\end{equation}
The stationary phase condition relates the Fourier frequency, $f$, to time domain quantities through 
\begin{equation}
2\pi f = j \dot{l}(t^{\ast}_j) \, .
\end{equation}

Clearly then, it is necessary to express the functions $t^{\ast}_j$ and $l$ that appear in the Fourier phases in terms of some time domain quantity to evaluate the model at any given frequency. In paper 1, we chose to use the orbital eccentricity as that time domain quantity, because the orbital dynamics at Newtonian order can be solved exactly in terms of this quantity, i.e.~one is able to exactly solve the Newtonian order $dn/de$ equation to obtain $n(e)$. This solution can then be used in the integrals that define the phase functions
\begin{align}
t(e) = \int \frac{de'}{\dot{e}[n(e'),(e')]} \\ 
l(e) = \int \frac{n(e')}{\dot{e}[n(e'),(e')]} de' \,,
\end{align} 
which admit exact solutions in the form of hypergeometric functions. Evaluation of these functions, however, is computationally expensive, and thus, in paper 1 we used a simple Taylor expansion of those hypergeometric functions about small eccentricity, which we found was sufficient to faithfully model the Fourier phases, even to eccentricities as high as $\sim 0.9$. The last step to evaluate the model at a given frequency was to numerically invert the stationary phase condition to find $e(n)$. The stationary phase condition can be more suggestively written as 
\begin{equation}
2\pi f = j n[e(t^{\ast}_j)] \,,
\end{equation}
and this is a relatively cheap inversion because the solution is clearly independent of $j$ and, in the Newtonian limit, also system independent. 

The key result which allows for the model to be readily extended to 3PN order is that the Taylor expansions in orbital eccentricity of the hypergeometric functions appearing in the phase were generally sufficient to faithfully represent the signal even for large eccentricities. Only the inversion of $n(e)$ behaves very poorly for even moderate eccentricities. Thus, the route for extension to higher PN is clear: 
\begin{itemize}
\item[(i)] solve for the frequency evolution as a function of eccentricity in an expansion in both a PN parameter and the orbital eccentricity,
\item[(ii)] use this solution to express the integrands appearing in the phase functions as expansions of those same parameters and integrate the series in eccentricity,
\item[(iii)] numerically invert the stationary phase condition to relate frequency to the eccentricity.
\end{itemize}

\subsection{Eccentric 3PN Fourier Domain Model}
\label{subsec:TaylorF2e}

Let us now derive the Fourier domain model to 3PN order, which we will call TaylorF2e as it closely follows the quasi-circular TaylorF2 model, and corresponds to solving the orbital dynamics where the independent variable is $e$. 

Following the general scheme laid out in paper 1, we begin by applying the SPA to the harmonic decomposition, which was derived in Eq.~\eqref{eq:td_harm_decomp} of Sec.\ref{subsec:harm_decomp}. We reiterate the harmonic decomposition here, explicitly writing out the complex conjugate term:
\begin{align}
h(t) &= \mathcal{A}\left\lbrace F_{+}S^2\sum_{s=-\infty}^{\infty} G_se^{isl} + \left[Q\sum_{j=-\infty}^{\infty}N_je^{-i(jl+2\lambda)}
\nonumber \right. \right. \\ 
& \left. \left. - Q^{\ast}\sum_{j=-\infty}^{\infty}N_je^{i(jl+2\lambda)} \right] \right\rbrace \, .
\end{align}
Applying the SPA to each of the oscillating terms separately (see Appendix A of paper 1 for a brief description of the SPA as applied to individual harmonics of the orbital phases), we find
\begin{subequations}
\label{eq:spa_-}
\begin{align}
\mathcal{F}\left[ N_{j}e^{-i(jl+2\lambda)} \right] & \approx N_j\sqrt{\frac{2\pi}{|j\ddot{l}+2\ddot{\lambda}|}}e^{i\psi^{-}_j}\,,
\\
\psi^{-}_j &= 2\pi ft^{-}_j - (jl+2\lambda) - \frac{\pi}{4}\,,
\\
\label{eq:stat_-}
2\pi f &= j\dot{l}(t^{-}_j)+2\dot{\lambda}(t^{-}_j)\,,
\end{align}
\end{subequations}
for the first term,
\begin{subequations}
\label{eq:spa_+}
\begin{align}
\mathcal{F}\left[ N_{j}e^{i(jl+2\lambda)} \right] &\approx N_j\sqrt{\frac{2\pi}{|j\ddot{l}+2\ddot{\lambda}|}}e^{i\psi^{+}_j}\,,
\\
\psi^{+}_j &= 2\pi ft^{+}_j + (jl+2\lambda) - \frac{\pi}{4}\,,
\\
\label{eq:stat_+}
2\pi f &= -j\dot{l}(t^{+}_j)-2\dot{\lambda}(t^{+}_j)\,,
\end{align}
\end{subequations}
for the second term and
\begin{subequations}
\label{eq:spa_jl}
\begin{align}
\mathcal{F}\left[ G_{s}e^{isl} \right] &\approx G_s\sqrt{\frac{2\pi}{|s\ddot{l}|}}e^{i\psi_s}\,,
\\
\psi_s &= 2\pi ft^{\ast}_s + sl - \frac{\pi}{4}\,,
\\
2\pi f &= s\dot{l}(t^{\ast}_s)\,,
\end{align}
\end{subequations}
for the last term.  

The stationary phase condition reveals which terms in the sum contribute to the Fourier transform. In the case of harmonics of $\dot{l}$, only positive $s$ terms contribute as $\dot{l} > 0$. However, in Eqs.\eqref{eq:spa_-} and \eqref{eq:spa_+}, the appearance of $(j\ddot{l}+2\ddot{\lambda})$ in the denominator could lead to catastrophes (instances where the amplitude diverges at a subset of frequencies) in the event that this term vanishes. To determine which terms could possibly lead to these catastrophes, we re-express the term appearing in the denominator of the amplitudes as
\begin{equation}
j\ddot{l}+2\ddot{\lambda} = \dot{n}\left( j+2+2k+2n\frac{\dot{k}}{\dot{n}} \right) \, .
\end{equation}
We can rearrange the above to find the $j$ for which there are catastrophes,
\begin{equation}
j = -2\left[ 1+k(1+\chi(e)) \right]
\end{equation}
where 
\begin{equation}
\chi(e) = \frac{n}{\dot{n}}\frac{\dot{k}}{k} = -\frac{8 \left(7 e^4+e^2-8\right)}{37 e^4+292 e^2+96}.
\end{equation}

\begin{figure}[htp]
\includegraphics[clip=true,angle=0,width=0.475\textwidth]{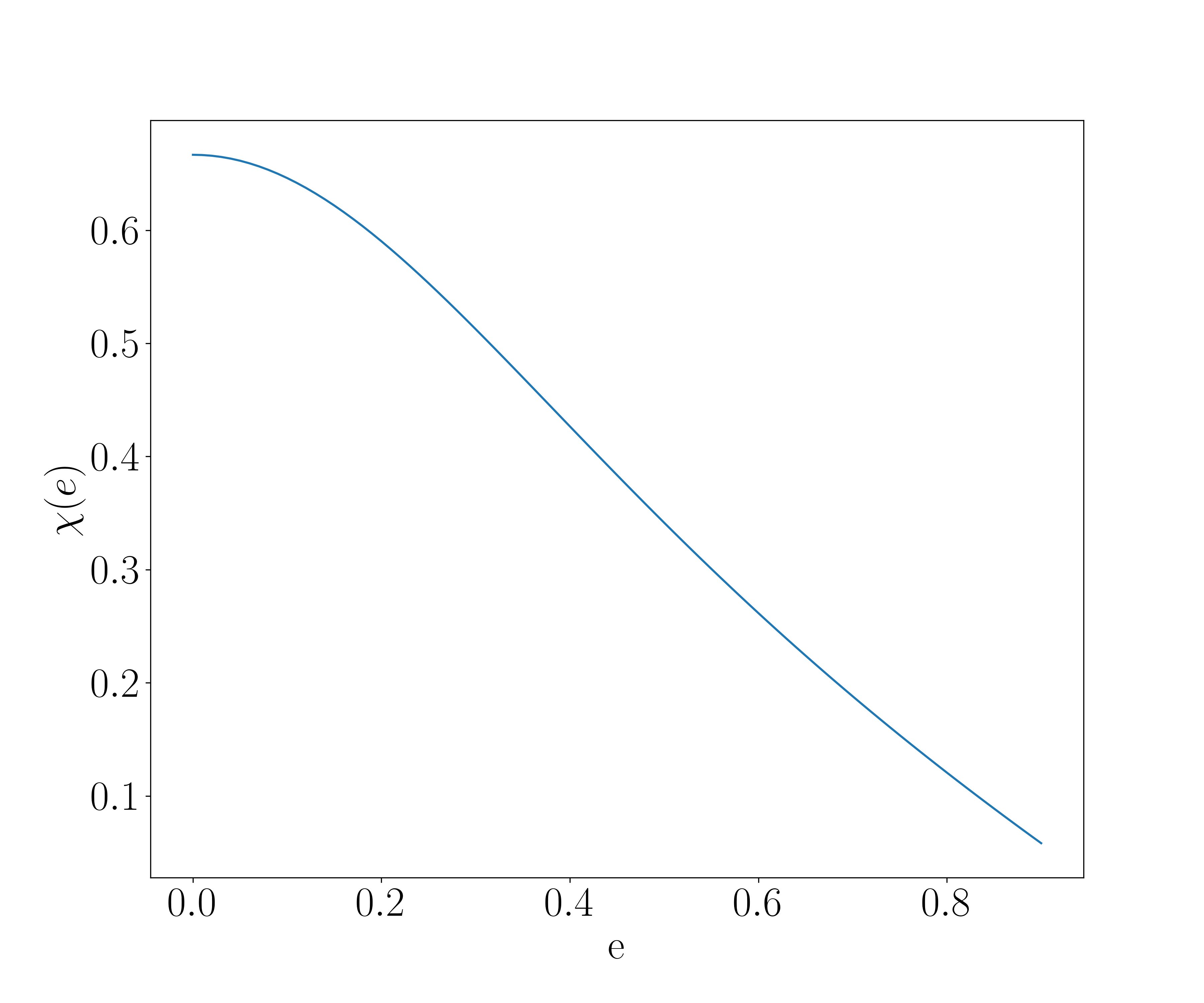}
\caption{\label{fig:G_e} The value of $\chi(e)$ as a function of $e$.}
\end{figure}

Are there catastrophes in the SPA? Figure~\ref{fig:G_e} shows $\chi(e)$, which clearly varies from $2/3$ at $e = 0$ to about $0.05$ at $e = 0.9$. If we bound the 1PN pericenter precession quantity $k$ to $0 < k < 1/2$, we see that it is possible to have a catastrophe in the $j=-3$ case when the binary is sufficiently close (i.e.~when $k \sim 1/2$) and circular ($\chi(e) \sim 2/3$). However, since accuracy in the amplitude is less important than accuracy in the phase for GW data analysis, we proceed by keeping only the leading PN order corrections in $\ddot{l}$ and $\ddot{\lambda}$. At leading PN order, $\ddot{l} = \ddot{\lambda}$, and clearly, there are only catastrophes when $j = -2$. But this harmonic corresponds to a low-frequency contribution that has a negligible amplitude contribution, as we can see from the stationary phase condition and Figure \ref{fig:Nj}. In fact, neglecting this harmonic altogether affects the matches shown in Fig. \ref{fig:harmdecomp_study} negligibly, so henceforth we neglect the $j = -2$ harmonic, which ensures our model never hits catastrophes.  

But exactly which $j$ indices contribute to the Fourier transform? Inspection of the stationary phase conditions (Eqs.~\eqref{eq:stat_+} and \eqref{eq:stat_-}) does not clearly reveal the answer to this question. In order to find the $j$ indices that do contribute, let us begin by rearranging the stationary phase conditions to set a condition on $j$ for which $f$ is positive:
\begin{subequations}
\label{eq:jineq}
\begin{align}
j > -2(1+k(t^{-}_j)),
\\
j < -2(1+k(t^{+}_j)),
\end{align}
\end{subequations}
where we have used $\dot{l} = n$ and $\dot{\lambda} = n(1+k)$. We see that as long as $k \leq 1/2$ then the stationary point corresponding to the ``-" superscript is satisfied when $j > -2$ and that corresponding to the ``+" superscript when $j < -3$.

Clearly then, in order to understand which indices $j$ contribute to the Fourier transform, we must first determine which values of $k$ are reasonable within the PN approximation. To do so, let us compare two constraints that we might impose on our waveforms: 
\begin{itemize}
\item[(i)] that the periastron velocity be small $v_p < 1/3$, 
\item[(ii)] that the rate of periastron advance be slow $k<1/2$. 
\end{itemize}
Each of these constrains the final separation of the binary, but since our results are typically quoted in terms of initial dimensionless semi-latus rectum, $p$, it is natural to consider what lower limit on $p$ each of the above constrains implies. Constraining the periastron velocity ($v_p < 1/3$) leads to $p > 9(1+e)^2$ [or equivalently $y < (3(1+e))^{-1}$], while constraining the advance of periastron ($k < 1/2$) leads to $p > 6$. Since the periastron velocity constraint is more stringent, henceforth we use it to stop our waveform evolution. Semi-latus recta below this number would incur into a regime were the PN approximation may be inaccurate \cite{2008PhRvD..77l4006Y}.

Restricting $v_p < 1/3$ ensures that $k < 1/2$, and thus, we see that the stationary phase condition of Eq.~\eqref{eq:stat_-} is satisfied only when $j > 2$ , while that of  Eq.~\eqref{eq:stat_+} is satisfied when $j < -2$ (with the understanding that we have dropped the $j = -2$ harmonic). In practice, we have found that the stationary phase condition for the $j=-3$ harmonic can become double valued, instead of adding more complication by treating this harmonic with more care, we simply neglect it as we have verified that it does not affect the results of Fig.~\ref{fig:harmdecomp_study}. Putting everything together then we have a frequency response in the SPA given by:
\begin{align}
\label{eq:spa}
\tilde{h}_{+, \times}(f) &= \mathcal{A}\left[ F_{+}S^2\sum_{s=1}^{\infty} G_s\sqrt{\frac{2\pi}{|s\ddot{l}|}}e^{i\psi_s} 
\nonumber \right.  \\ 
&+ Q\sum_{j=-1}^{\infty}N_j\sqrt{\frac{2\pi}{|j+2|\ddot{l}}}e^{i\psi^{-}_j}
\nonumber  \\ 
& \left. + Q^{\ast}\sum_{j=-\infty}^{-4}N_j\sqrt{\frac{2\pi}{|j+2|\ddot{l}}}e^{i\psi^{+}_j} \right] \,
\end{align}
where 
\begin{align}
\ddot{l} &= \frac{\eta}{5m^2}(1-e^2)^2(96+292e^2+37e^4)y^{11}\,.
\end{align}
Following paper 1 and with this in hand, we now seek solutions to express the functions appearing in the phase ($t, l, \lambda$) in terms of the orbital eccentricity.

Let us begin by seeking a relation between the orbital frequency and the eccentricity. Since either of the two orbital frequencies ($n$ or $\omega$) can be expressed in terms of the PN parameter $y$, this amounts to seeking the mapping $y(e)$. Composing $dy/de$ using the chain rule (i.e. $dy/de = \dot{y}/\dot{e}$) and expanding to 3PN order in $y$ yields
\begin{equation}
\label{eq:dyde}
\frac{dy}{de} = y \left[c_0(e) + \sum_{n = 2}^{6} c_n(e) y^n\right] \,,
\end{equation}
whose PN solution is 
\begin{equation}
\label{eq:y_pet}
y(e) = y_0 \left[d_0(e) + \sum_{n = 2}^{6} d_n(e) y_0^n \right] \,,
\end{equation}
for certain functions $d_{n}(e)$. Inserting Eq.~\eqref{eq:y_pet} into Eq.~\eqref{eq:dyde} and expanding to 3PN order (now an expansion in the parameter $y_0$) gives a set of coupled ordinary differential equations for the $d_n(e)$ functions, namely
\begin{subequations} 
\label{eq:y_e_diff_sys}
\begin{align}
d_0'(e) &= c_0 d_0  \, , \\
d_2'(e) &= c_0 d_2  + c_{2} d_{0}^3  \, , \\
d_3'(e) &= c_0 d_{3}  + c_{3} d_{0}^4  \, ,\\
d_4'(e) &= c_0 d_{4}  + c_{4} d_{0}^5  + 3c_{2} d_0^2 d_2  \, , \\
d_5'(e) &= c_0 d_{5}  + c_{5} d_{0}^6  + 3c_{2} d_0^2 d_{3}  + 4c_{3} d_0^3 d_2  \, , \\
d_6'(e) &= c_0 d_{6}  + c_{6} d_{0}^7  + 3c_{2} (d_0^2 d_4  + d_0 d_2^2 ) + 4c_{3} d_0^2 d_4 
\nonumber. \\ &+ 5c_4 d_0^4 d_2  \, , \\ 
d_{6l}'(e) &= c_0 d_{6l}  + c_{6l} d_0^7  \,,
\end{align}
\end{subequations}
where we have split the 3PN contribution into a component that is proportional to the natural logarithm of $y_0$ and one that is not proportional to it, such that the 3PN part of the solution to $y(e)$ takes the form  $(d_6 + d_{6l}\ln{y_0})y_0^6$. 

Given that 
\begin{equation}
c_0 = \frac{12}{304} \frac{8 + 7e^2}{e(1+\frac{121}{304}e^2)} \, , 
\end{equation}
we see that solving any of Eqs.~\eqref{eq:y_e_diff_sys} amounts to solving a differential equation of the form  
\begin{equation}
\frac{df}{de} = \frac{12}{304} \frac{8 + 7e^2}{e(1+\frac{121}{304}e^2)} f(e) + g(e) \,,
\end{equation}
for some functions $g(e)$.  The above differential equation has the solution 
\begin{align}
\label{eq:gen_ode_sol}
f(e) &= \frac{1}{e^{6/19}(1+\frac{121}{304}e^2)^{435/2299}} \nonumber \\
&\times \left( C + \int g(e)e^{6/19}(1+\frac{121}{304}e^2)^{435/2299} de \right) \,,
\end{align}
for some constant $C$.  The constants arising from solving for any of the coefficients $d_n$ are determined by requiring that at each PN order $y(e_0) = y_0$. Clearly, in solving for $d_0$, $g(e) = 0$ and the solution is simply
\begin{equation}
d_0(e) = \frac{e_0^{6/19}(1+\frac{121}{304}e_0^2)^{435/2299}}{e^{6/19}(1+\frac{121}{304}e^2)^{435/2299}} \, .
\end{equation}
The solutions for the higher PN coefficients $d_n$ are not so simple as the form that $g(e)$ takes becomes a more complicated function of eccentricity. In order to solve the equations, we then expand Eq.~\eqref{eq:gen_ode_sol} in eccentricity, which allows us to integrate the series analytically. The $d_n(e)$ functions and the constants of integration $C_{n}$ to order ${\cal{O}}(e_{0}^{40})$ past leading are provided in the supplemental material. 

With the solution of $y(e)$ in hand, we are now in a position to solve for the phase functions $t(e), \lambda(e),$ and $l(e)$. We begin by substituting the solution above for $y(e)$ into the integrands for each of the phase functions
\begin{align}
\label{eq:phase_fun_integ}
t(e) - t_c = \int^{e} \frac{de'}{\dot{e}[y(e'),e']} \, , \\
l(e) - l_c = \int^{e} \frac{n[y(e'),e']}{\dot{e}[y(e'),e']} de' \, , \\
\lambda (e) - \lambda_c = \int^{e} \frac{\omega[y(e'),e']}{\dot{e}[y(e'),e']} de' \, .
\end{align}
In order to carry out the integration above we expand the integrand to 3PN in $y_0$. We then expand the coefficients at each order in $y_0$ to a given order in $e$. Note that there is no restriction on the order to which we expand any PN coefficient (coefficients of $y_0$ in each of the integrands) in eccentricity other than the order to which we have obtained $y(e)$. We could, for example, keep $\mathcal{O}(e^{100})$ terms in each coefficient, but with the inclusion of so many terms the model would become very computationally inefficient to evaluate. On the other hand, if we keep too few eccentric corrections, the model becomes inaccurate. In Section \ref{subsec:imp} we introduce a scheme to pick the order in eccentricity to which we keep the expansions of each $y_0$ coefficient, and then show in Section \ref{subsec:val} how the eccentricity expansion affects the validity of the waveform for systems with different initial eccentricities. 

After carrying out the integration, we have the phase functions in terms of the orbital eccentricity, 
\begin{subequations}
\label{eq:schem_phase_fact}
\begin{align}
\lambda(e) &= \frac{1}{y_0^5 \eta} [\lambda^{(0)}(e) + y_0^2 \lambda^{(1)}(e, \eta) + y_0^3 \lambda^{(1.5)}(e, \eta)
\nonumber \\ 
& + y_0^4 \lambda^{(2)}(e, \eta) + y_0^5 \lambda^{(2.5)}(e, \eta) + y_0^6\lambda^{(3)}(e, \eta, \ln{y_0}) ] \, , \\
l(e) &= \frac{1}{y_0^5 \eta} [l^{(0)}(e) + y_0^2 l^{(1)}(e, \eta) + y_0^3 l^{(1.5)}(e, \eta)
\nonumber \\ 
& + y_0^4 l^{(2)}(e, \eta) + y_0^5 l^{(2.5)}(e, \eta) + y_0^6l^{(3)}(e, \eta, \ln{y_0}) ] \, ,
\\
t(e) &= \frac{m}{y_0^8 \eta} [t^{(0)}(e) + y_0^2 t^{(1)}(e, \eta) + y_0^3 t^{(1.5)}(e, \eta)
\nonumber \\ 
& + y_0^4 t^{(2)}(e, \eta) + y_0^5 t^{(2.5)}(e, \eta) + y_0^6t^{(3)}(e, \eta, \ln{y_0}) ] \,,
\end{align}
\end{subequations} 
where the functions $t^{(n)}, l^{(n)}$ and $\lambda^{(n)}$ are series in eccentricity (and also depend on the constants of integration appearing in $y(e)$), which are provided in the supplemental material. With these solutions in hand, the phases in Eq.~\eqref{eq:spa} are functions of only the orbital eccentricity. The final step then is to numerically invert the appropriate stationary phase condition to relate the Fourier frequency to the orbital eccentricity. The specifics of the evaluation of TaylorF2e are described in Sec.~\ref{subsec:imp}. Section \ref{subsec:val} shows that this waveform model is as faithful to TaylorT4t as the TaylorT4e model is, giving good agreement for even moderate eccentricities.

\subsection{Alternative Frequency Domain Models}

In Section \ref{subsec:TaylorT4s} we laid out different possible time domain models based on the independent variable appearing in the set of ODEs that one solves numerically to specify the orbital dynamics of the system. In the frequency domain model derived in the previous section, we solved the set of ODEs with the eccentricity $e$ as the independent variable, and thus, the Fourier domain TaylorF2e model corresponds to the SPA version of the TaylorT4e time domain approximant. In this section, we wish to explore schematically how one would go about obtaining other frequency domain approximants related to the other two time domain approximants TaylorT4t and TaylorT4y.

TaylorF2y would be obtained by first solving the set of ODEs $(dl/dy, d\lambda/dy, dt/dy, de/dy)$ analytically within the PN approximation (and possibly in a low-eccentricity expansion). With this at hand, one would then invert $y(e)$ as given in Eq.~\eqref{eq:y_pet} to obtain $e(y)$, which would be a series in both $y$ and $y_0$. This solution would then be plugged into the integrands that define the phase functions and integrated to give $l(y)$, $\lambda(y)$, and $t(y)$. The different stationary phase conditions could be numerically inverted to relate $y$ to the Fourier frequency $f$; one could attempt to invert the stationary phase condition perturbatively as well, but the error incurred would have to be investigated with the waveform in hand. We expect, however, that the resulting TaylorF2y approximant would have only limited validity in $e$, because even at Newtonian order the inversion of $y(e)$ is transcendental and very poorly behaved even for eccentricities as low as $0.3$ (see Appendix C of paper 1 where we investigated this inversion at length at Newtonian order). 
 
TaylorF2t would be obtained by first solving the set of ODEs $(\dot{l}, \dot{\lambda}, \dot{e}, \dot{y})$ analytically within the PN approximation (and possibly in a low-eccentricity expansion) to find both $e(t)$ and $y(t)$. With this at hand, one would express the integrals for $l$ and $\lambda$ in terms of only the $t$ variable, but to obtain both, one would first have to invert the solutions found in TaylorF2e and TaylorF2y to obtain $t(y)$ and $t(e)$, or else come up with some scheme to solve $\dot{e}$ and $\dot{y}$ analytically. Upon substitution of these solutions into $\dot{l}$ and $\dot{\lambda}$, one would integrate to obtain $l(t)$ and $\lambda(t)$ analytically. Again, the stationary phase condition would be numerically solved to find $t(f)$ for any given $j$ or $s$ harmonic index, or alternatively one could try to perturbatively invert the stationary phase condition, but this  could incur too much error and would have to be investigated with the approximant in hand.  

\section{Implementation and Validation}
\label{sec:imp_val}

In this section we validate the TaylorF2e model introduced in in Sec.~\ref{sec:F2} against the TaylorT4e model. To quantify the agreement between two models $h_1$ and $h_2$ it is customary to introduce the match
\begin{equation}
\label{eq:match}
M = \max \limits_{t_c,l_c,\lambda_c}\mathcal{O}(h_1,h_2) \,,
\end{equation}
which is nothing but the overlap between two waveforms maximized over the extrinsic parameters $t_c$, $l_c$, and $\lambda_c$. Note that this differs from the match in the quasicircular case, which is only maximized over $\phi_c$ and $t_c$, because in the eccentric case there is an extra phase angle associated with pericenter precession. We employ the implicit maximization scheme developed in Section 3 of Paper 1 in order to maximize over these parameters.
  
\subsection{Implementation}
\label{subsec:imp}

Let us begin by clearly describing how theTaylorF2e model is generated and implemented. As explained in Sec.~\ref{sec:F2}, the TaylorF2e model can be deployed in slightly different ways, which in essence depend on the following choices:
\begin{itemize}
\item[(i)] The order of the eccentricity expansion of each of the coefficients appearing in the phase functions of Eqs.~\eqref{eq:schem_phase_fact}.
\item[(ii)] The technique by which we invert the stationary phase condition to relate the orbital eccentricity to the Fourier frequency.
\item[(iii)] The method of obtaining the eccentricity dependence on the orbital frequency. 
\end{itemize}

Let us begin by addressing item (i). The expansion of the phase functions in small eccentricity introduces an error in our waveform model, and thus, we wish to determine the optimal expansion order such that the resulting waveform is accurate enough to lead to sufficiently high matches against a TaylorT4e model. Let us then begin by introducing the following notation for our ODEs  $(dl/de, d\lambda/de, dt/de)$ that lead to Eqs.~\eqref{eq:schem_phase_fact}:
\begin{subequations}
\label{eq:schem_ode}
\begin{align}
\frac{d\lambda}{de} &= \frac{1}{y_0^5}\left[\lambda'_{0}(e) + \lambda'_{2}(e)y_0^2 + ... \lambda'_{6}(e)y_0^6  \right]\,,
\\ 
\frac{dl}{de} &= \frac{1}{y_0^5}\left[l'_{0}(e) + l'_{2}(e)y_0^2 + ... l'_{6}(e)y_0^6  \right]\,, 
\\
\frac{dt}{de} &= \frac{M}{y_0^8}\left[t'_{0}(e) + t'_{2}(e)y_0^2 + ... t'_{6}(e)y_0^6  \right]\,.
\end{align}
\end{subequations}
In each of the PN coefficients appearing above $[\lambda_{i}'(e),l_{i}'(e),t_{i}'(e)]$, we have substituted in our solution for $y(e)$, which is valid to order $\mathcal{O} (e^{40})$ past leading, but we have not yet expanded these coefficients in eccentricity. We now wish to determine the order in eccentricity to which we should expand each of these coefficients. 

Let us then introduce the following measure of error in each of the PN coefficients  
\begin{subequations}
\label{eq:gamm_err}
\begin{align}
\gamma^{\lambda}_n(e) &= \text{Abs}\left(1 - \frac{\lambda'_n(e)}{\bar{\lambda}'_n(e)}\right)\,, \\
\gamma^{l}_n(e) &= \text{Abs}\left(1 - \frac{l'_n(e)}{\bar{l}'_n(e)}\right)\,, \\ 
\gamma^{t}_n(e) &= \text{Abs}\left(1 - \frac{t'_n(e)}{\bar{t}'_n(e)}\right)\,,
\end{align}
\end{subequations}
where the overhead bar stands for the Taylor expansion in small eccentricity to a given order. This is analogous to the Newtonian-order measure of error introduced in paper I 
\begin{equation}
\epsilon = 1 - \frac{\psi_{\rm exact}}{\psi_{\rm err}} \, ,
\end{equation}
where $\psi_{\rm exact}$ was our exact (in eccentricity) phase and $\psi_{\rm err}$ had a fixed amount of error introduced through $\epsilon$ via the above equation. We found that for moderately eccentric systems, the match began to drop for values of $\epsilon \sim 10^{-5}$, as shown in Table III of paper I.

Returning to our measures of error in Eq.~\eqref{eq:gamm_err}, we must now choose a tolerance for $\gamma_n^{\lambda,l,t}$ that is acceptable for our model, i.e.~that leads to a sufficiently small loss of match. Guided by the error investigation of paper 1, one may be tempted to set a uniform tolerance of $10^{-5}$. One may expect, however, that more error could be tolerated in the coefficients appearing at higher PN order, since they are multiplied by the small parameter $y_0$. Thus, at each PN order we impose a maximum error tolerance $\gamma_{n}^{\lambda,l,t}(e) \leq \epsilon_{n}$, where
\begin{equation}
\label{eq:err_tol}
\epsilon_n = \epsilon \; (y_0)^{-n} \,,
\end{equation}
for some constant $\epsilon$. The factor of $(y_0)$ allows for more error to be accumulated by coefficients at higher PN order, since their contribution to the phase is scaled by the small parameter $y_0$. 

\begin{figure}[htp]
\includegraphics[clip=true,angle=0,width=0.475\textwidth]{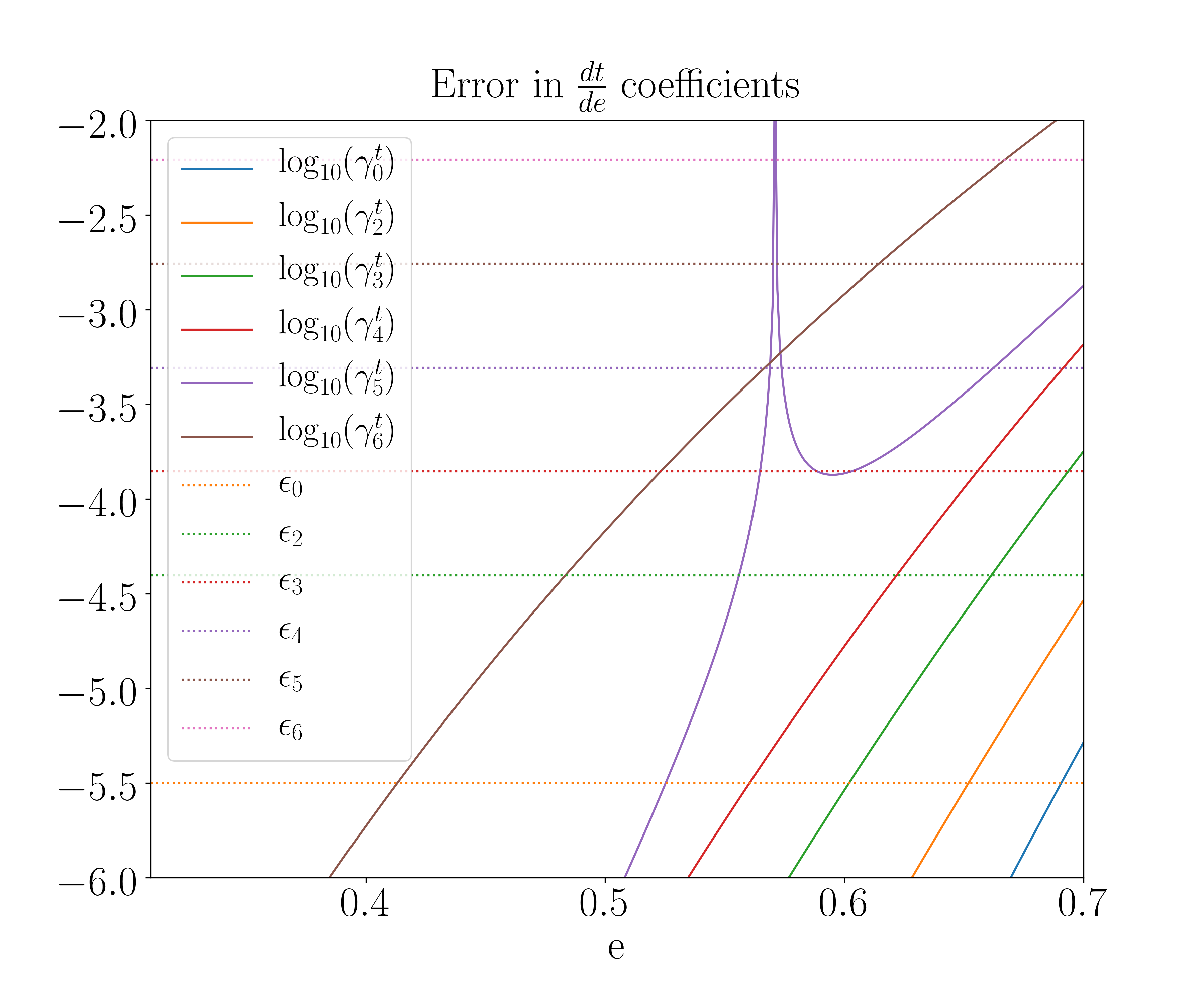} \\
\vspace{-0.3cm}
\includegraphics[clip=true,angle=0,width=0.475\textwidth]{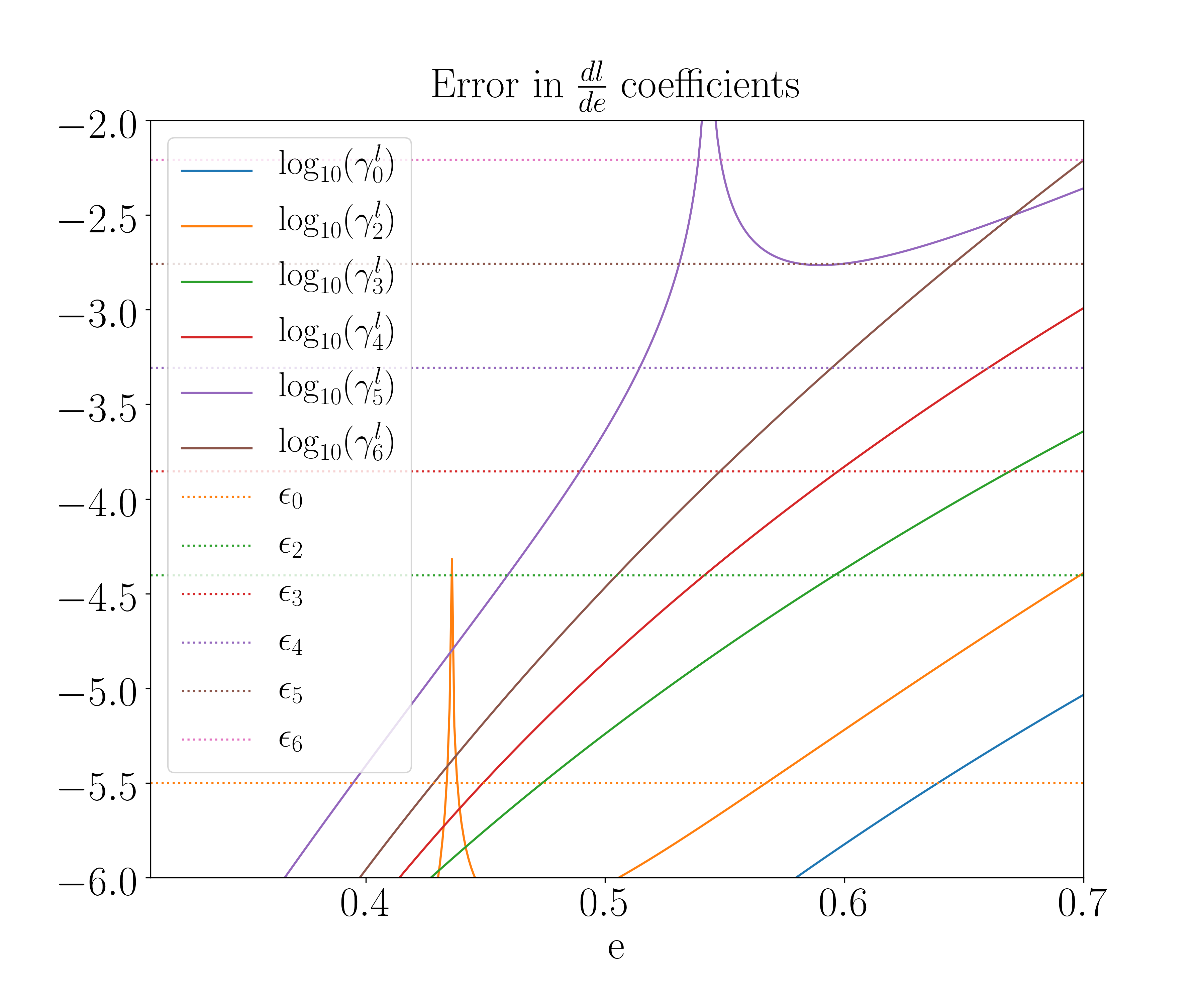} \\
\vspace{-0.3cm}
\includegraphics[clip=true,angle=0,width=0.475\textwidth]{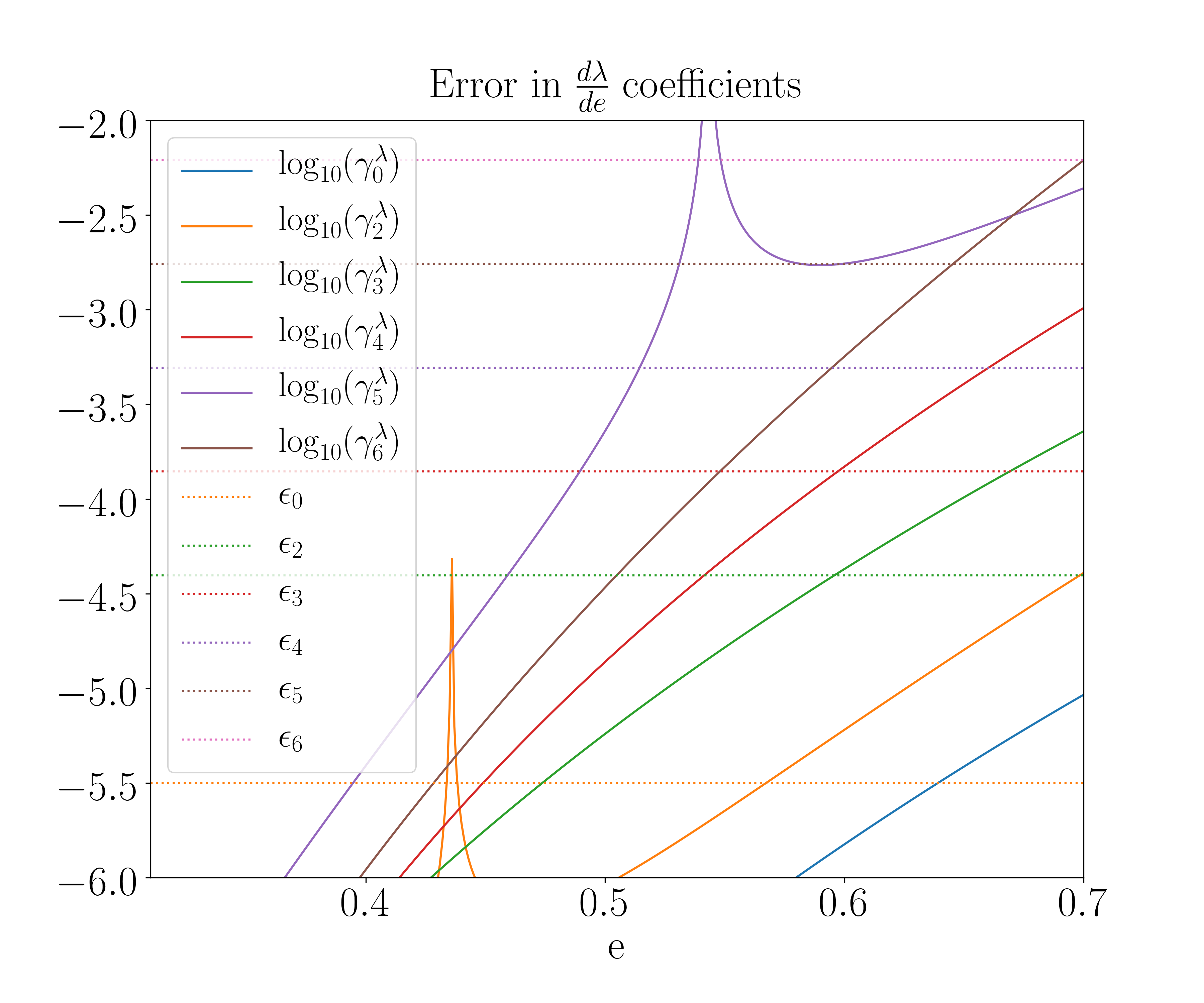}
\caption{\label{fig:phase_err} Error in the coefficients appearing in Eq.~\eqref{eq:schem_ode} given by Eq.~\eqref{eq:gamm_err}. The horizontal dashed lines are the rough error tolerances that we impose on the coefficients to select the different orders of eccentricity expansion that occur in the phase functions. }
\end{figure}

The tolerance condition described above clearly depends on $y_0$ (or $p_0$), $\eta$, $e_0$, and the constant $\epsilon$. We have investigated two different constraints, one in which we set $\epsilon = 10^{-5.5}$, $\eta = 0.25$, $p_0 = 12$, and $e_0 =  0.7$, and one in which we set $\epsilon = 10^{-5}$, $\eta = 0.25$, $p_0 = 18$, and $e_0 =  0.4$. The  first set of conditions forces us to keep more terms in eccentricity than the second set, which we may then consider a conservative choice. With these choices, we then generate two waveforms, one with more terms kept in eccentricity (TaylorF2e+) and with less terms kept (TaylorF2e-), and then we investigate the effects in the match. Figure \ref{fig:phase_err} shows $\gamma_{n}^{\lambda,l,t}$ as a function of eccentricity for the conservative choice. We see that $\gamma^{\lambda,l,t}_{n}$ is at least nearly bounded by the threshold imposed by Eq.~\eqref{eq:err_tol} (the spikes correspond to zero-crossings on a logarithmic scale). In Table \ref{tab:coeff_ord} we list the maximum order past leading to which each of our coefficients are expanded in eccentricity for TaylorF2e+ and TaylorF2e-. Note that while many terms have been kept in each of the phase functions, one can collect the resulting expression in like powers of eccentricity. For example, after such a collection, there are only 26 terms in eccentricity to be evaluated for $\lambda(e)$ in the TaylorF2e- case. 

\begin{table}
\begin{tabular}{||c|c|c||}
\hline
$n$ & $\lambda_n'(e)$ & $t_n'(e)$  \\
\hline
0 & 10 (4) & 36 (12) \\  \hline
2 & 240/19 (8) 				& 620/19 (202/19)\\  \hline
3 & 208/19 (4) 				& 512/19 (170/19)\\  \hline
4 & 278/19 (126/19) & 26 (164/19)\\  \hline
5 & 246/19 (132/19) & 468/19 (182/19)\\  \hline
6 & 14 (94/19) 				& 328/19 (100/19)\\  \hline
\end{tabular}
\caption{\label{tab:coeff_ord} Eccentricity order to which each of the coefficients in Eq.~\eqref{eq:schem_ode} is expanded past leading order for the TaylorF2e+ model and the TaylorF2e- model in parenthesis. We keep the same number of terms in $l_n'(e)$ as in $\lambda_n'(e)$}
\end{table}

To address item (ii), let us first recall the different stationary point inversions that we must achieve to relate $f$ to $e$. The equations that must be inverted are
\begin{subequations}
\begin{align}
2\pi f = j \dot{l} \pm 2 \dot{\lambda} = j n \pm 2 \omega \\
2\pi f = s \dot{l} =  s n(e)
\end{align}
\end{subequations}
where the sign in the first equation depends on the index $j$ as detailed in Sec.~\ref{sec:F2}. The second equation is particularly simple to solve since it only requires the inversion of $n(e)$. That is, defining the inverse function $\kappa$ such that $\kappa[n(e)] = e$, one can relate $f$ to $e$ for the $s$ indices via 
\begin{equation}
e_b(f) = \kappa\left[\frac{2\pi f}{s} \right] \, .
\end{equation}
Of course, the inverse function $\kappa$ must be obtained numerically, but this formulation has the advantage that once $\kappa$ is obtained, it can be used for any index $s$. The inversion of the equation for the $j$ harmonics is more complicated. First, let us rewrite the stationary condition as 
\begin{equation}
\label{eq:stat_point_j}
2\pi f = jn \pm 2(1+k)n = (j \pm 2)n \pm kn \, .
\end{equation}
We then employ the secant method \cite{Press:2007:NRE:1403886} to numerically solve Eq.~\eqref{eq:stat_point_j}, where we recognize that the term $kn$ is of higher PN order than $(j \pm 2)n$ and we leverage this to specify the approximate initial guess $e_{j}(f)^{\rm guess} = \kappa[2\pi f/(j \pm 2)]$. Indeed, this initial guess makes the secant method typically converge after 4 iterations to an accuracy of $10^{-8}$. We expect that this error tolerance could be relaxed.

Let us conclude this subsection by discussing item (iii), which in any case is required to invert the stationary phase conditions, as it needs a solution for $y(e)$. We obtain this solution by solving $\dot{y}$ and $\dot{e}$ and interpolating these solutions to compose $n(e)$ and $\omega(e)$. Given that we have obtained $y(e)$ numerically, this is then in turn used in the various harmonic amplitudes, which enter the SPA (in $\mathcal{A}$, $N_j$, and $\ddot{l}$). While somewhat computationally costly, we estimate that there are considerable speedups to be gained through numerical techniques and careful truncation of the sum appearing in Eq.~\eqref{eq:spa}. We leave a thorough investigation of computational cost to future work. Alternatively we could use the solution found in Eq.~\eqref{eq:y_pet}, which is purely analytical. However, in the interest of being consistent with paper 1, we chose here to solve $\dot{y}$ and $\dot{e}$ numerically instead.

\subsection{Validation}
\label{subsec:val}

\begin{figure*}
\includegraphics[clip=true,angle=0,width=0.4\textwidth]{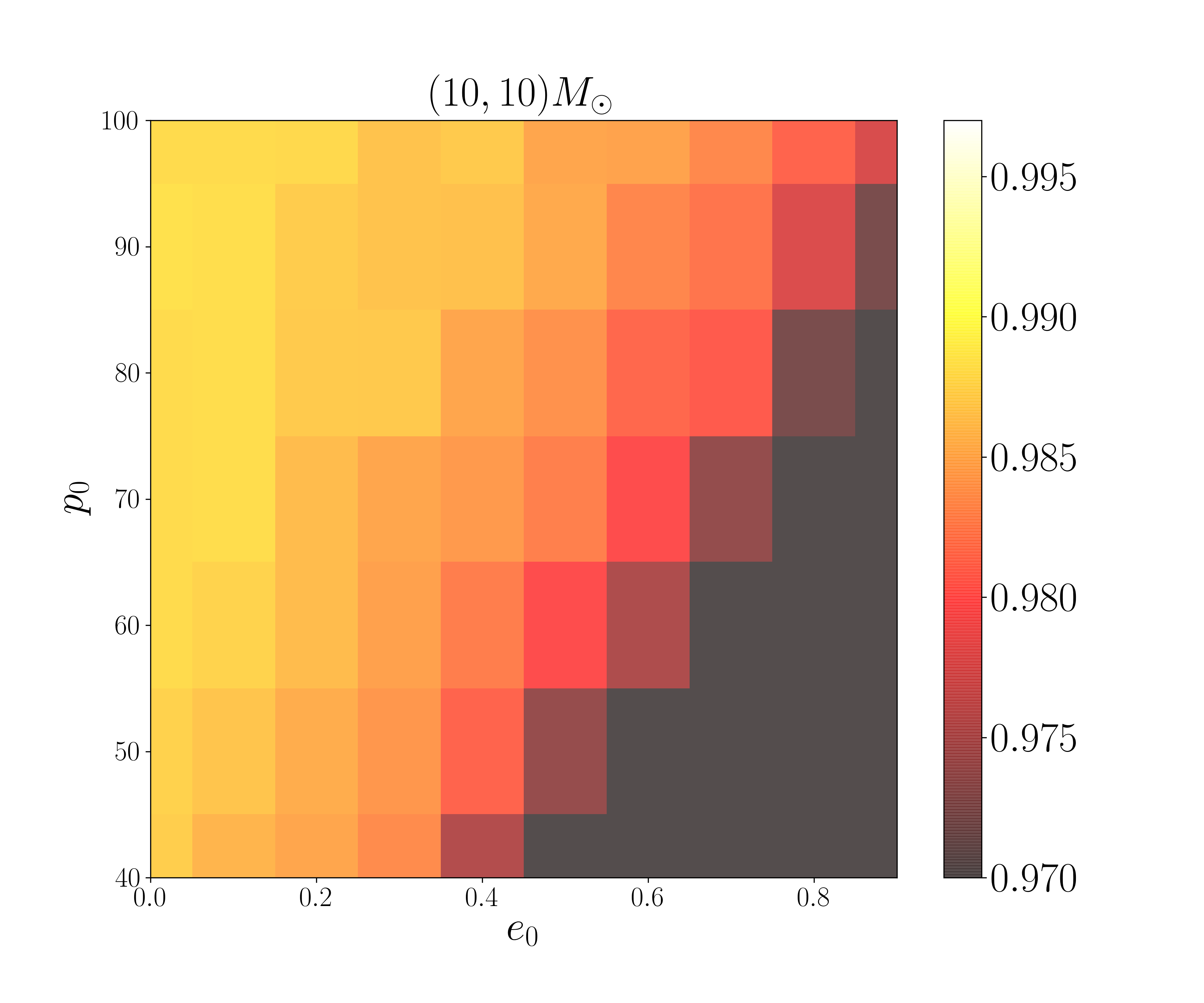}
\includegraphics[clip=true,angle=0,width=0.4\textwidth]{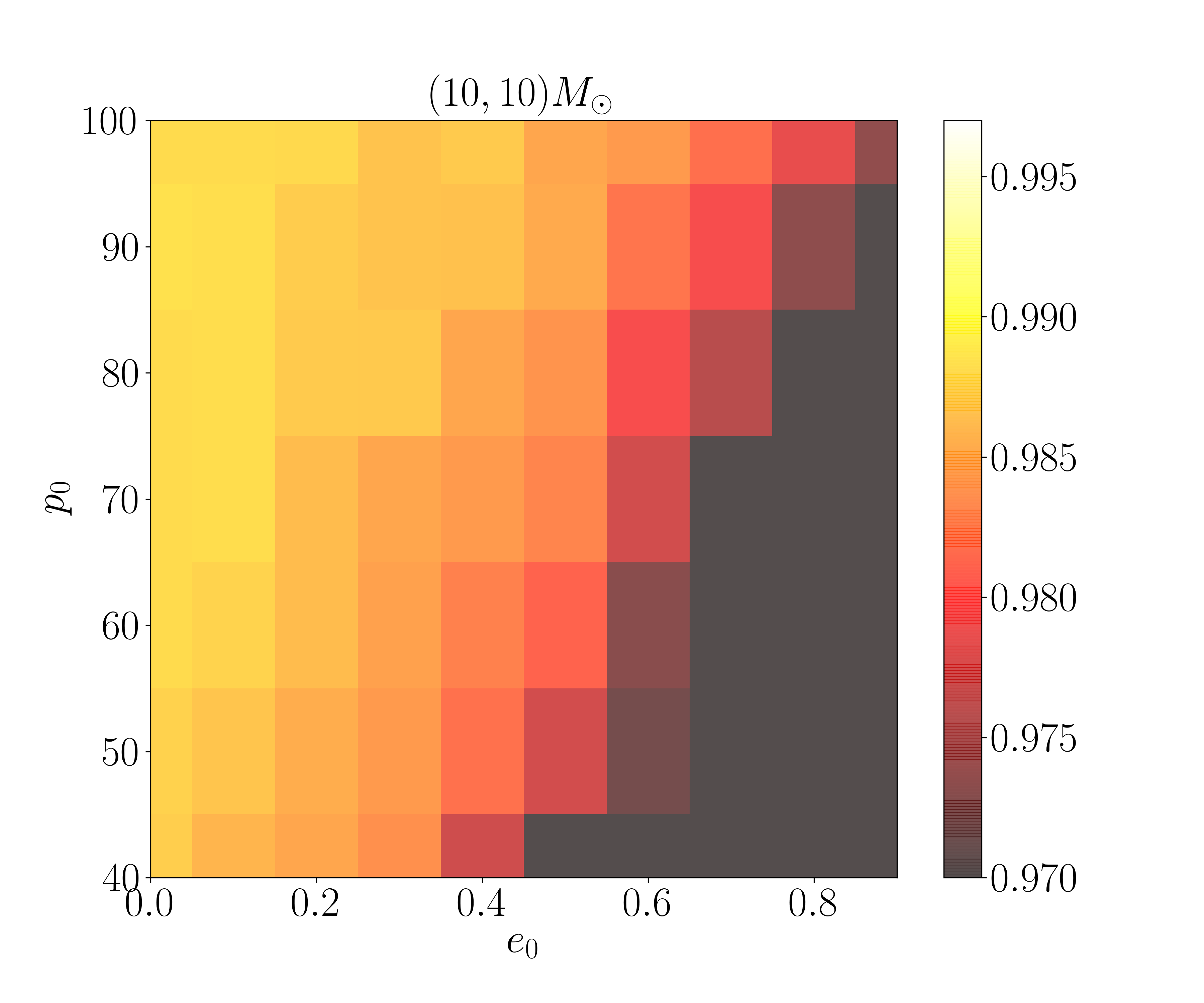}
\includegraphics[clip=true,angle=0,width=0.4\textwidth]{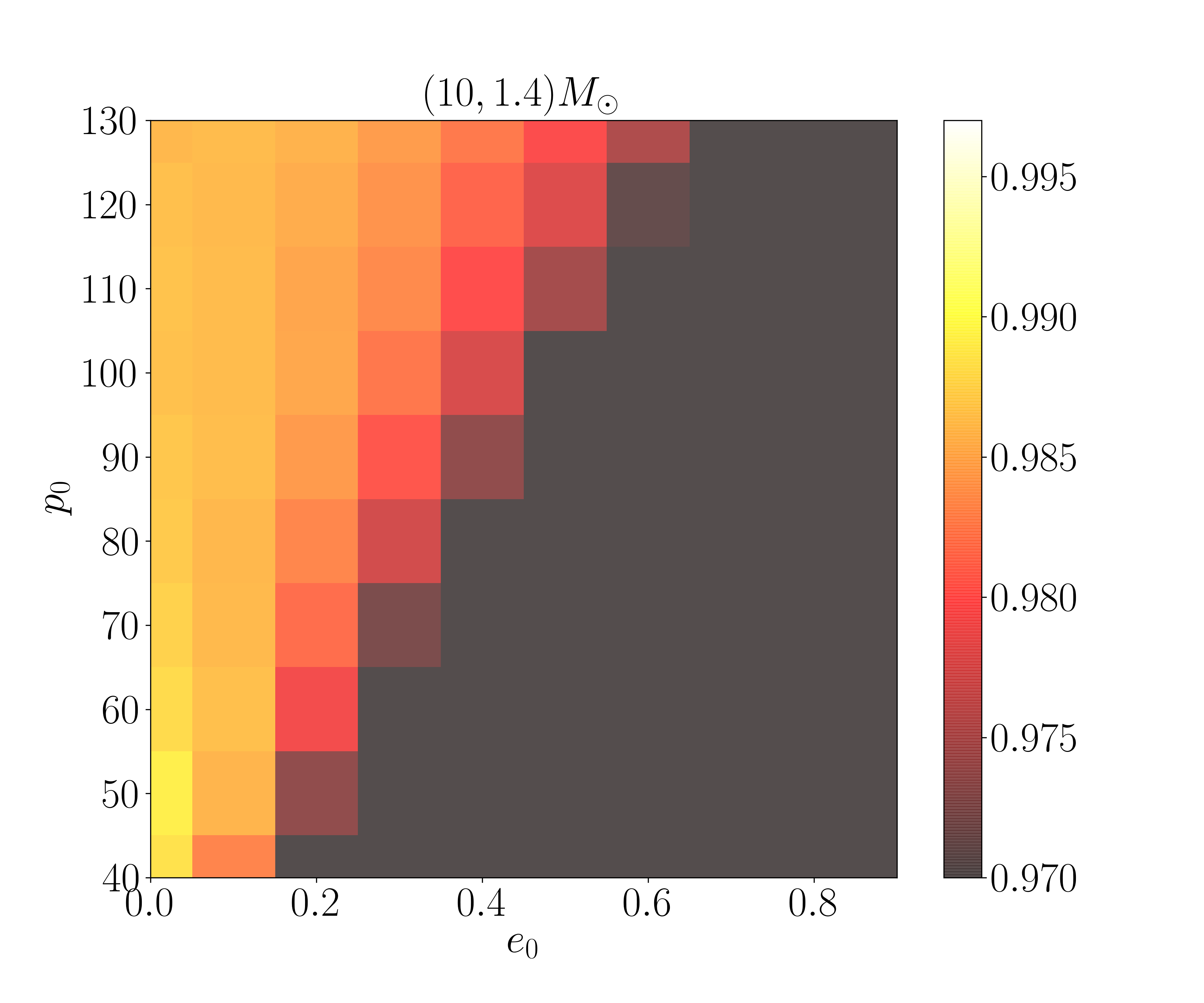}
\includegraphics[clip=true,angle=0,width=0.4\textwidth]{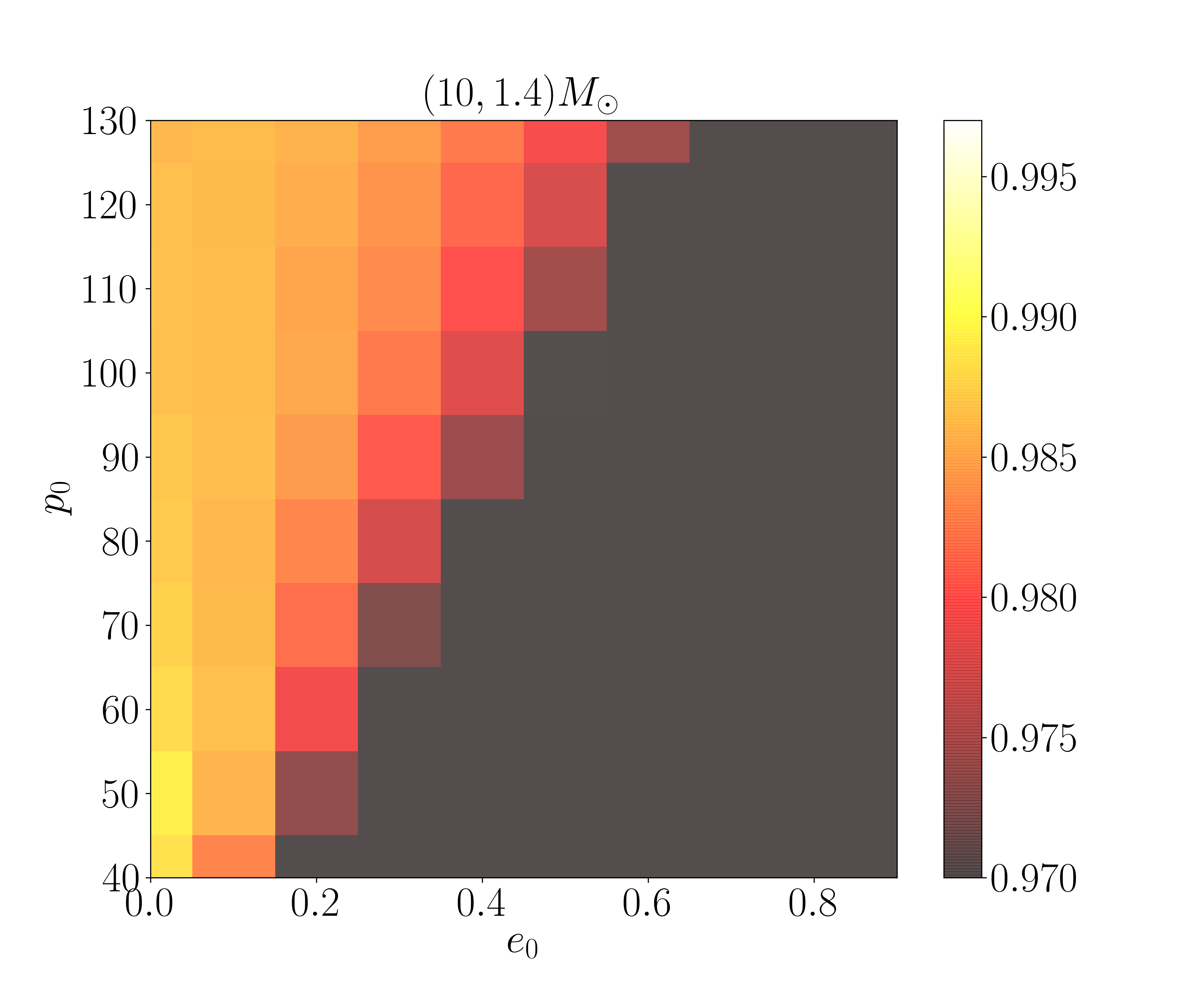}
\includegraphics[clip=true,angle=0,width=0.4\textwidth]{BNS_F2_T4t_high.png}
\includegraphics[clip=true,angle=0,width=0.4\textwidth]{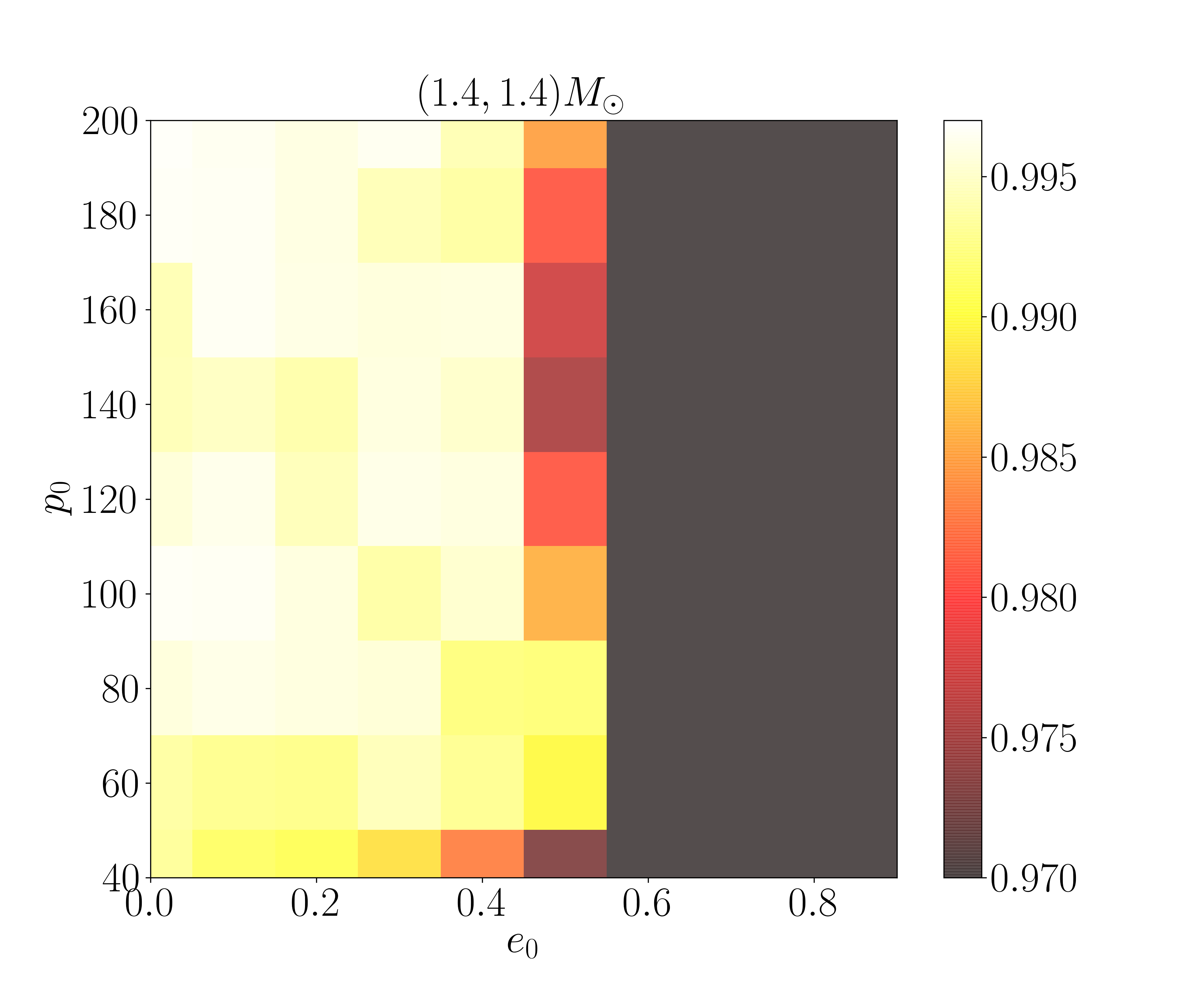}
\caption{\label{fig:match} Match between TaylorF2e+ (left) or TaylorF2e- (right) against TaylorT4t for a $(10, 10)M_{\odot}$ (top), $(10, 1.4)M_{\odot}$ (middle), and a $(1.4, 1.4)M_{\odot}$ (bottom) binary.}
\end{figure*}

In this subsection we validate the TaylorF2e model by computing the match between it and the TaylorT4t model, which was introduced in Sec.~\ref{sec:GW_time}. Figure \ref{fig:match} shows the match for a $(10, 10)M_{\odot}$, $(10, 1.4)M_{\odot}$, and a $(1.4, 1.4)M_{\odot}$ binary, with TaylorF2e+ on the left and TaylorF2e- on the right. In these match calculations there are three main sources of loss in match: 
\begin{itemize}
\item[(i)] the truncation of the summations in Eq.~\eqref{eq:spa} (explored in Fig.~\ref{fig:harmdecomp_study}), 
\item[(ii)] the PN disagreement between models obtained through solving different PN valid sets of ODEs (explored in Fig.~\ref{fig:T4t_T4e}), and 
\item[(iii)] the truncation of the eccentricity expansions in the phase functions.
\end{itemize}
Let us discuss each of these potential sources of error in turn. 

In order to address item (i), let us return to Fig.~~\ref{fig:harmdecomp_study}, which shows the overlap between two numerically calculated time-domain models (both TaylorT4t) with one model kept to all harmonics and the other restricted to $j = [-15,15]$ and $s = [1, 15]$. Comparison of that figure to Fig.~\ref{fig:match} reveals that spectral truncation is not the main source of loss in match between TaylorT4t and TaylorF2e. The loss in match due to spectral truncation (as shown in Fig.~\ref{fig:harmdecomp_study}) occurs for much greater initial eccentricities and smaller initial semi-latus recta than the loss in match that is seen between TaylorF2e and TaylorT4t (as shown in Fig.~\ref{fig:match}). 

When we compute the match between TaylorF2e+ and TaylorT4t we nearly exactly recover the results of taking the match between TaylorT4t and TaylorT4e. This is not a surprising result as TaylorF2e is built by solving the ODEs formulated with $e$ as the independent variable, as is done in TaylorT4e. This suggests that, at least when keeping many terms in eccentricity, the dominant source of loss in match is due to the PN differences between the ODEs governing the orbital dynamics, when re-expressing them in terms of a different independent variable. This result suggests that a more faithful model would require PN ODEs at higher PN order for the orbital dynamics. 

\begin{figure*}
\includegraphics[clip=true,angle=0,width=0.4\textwidth]{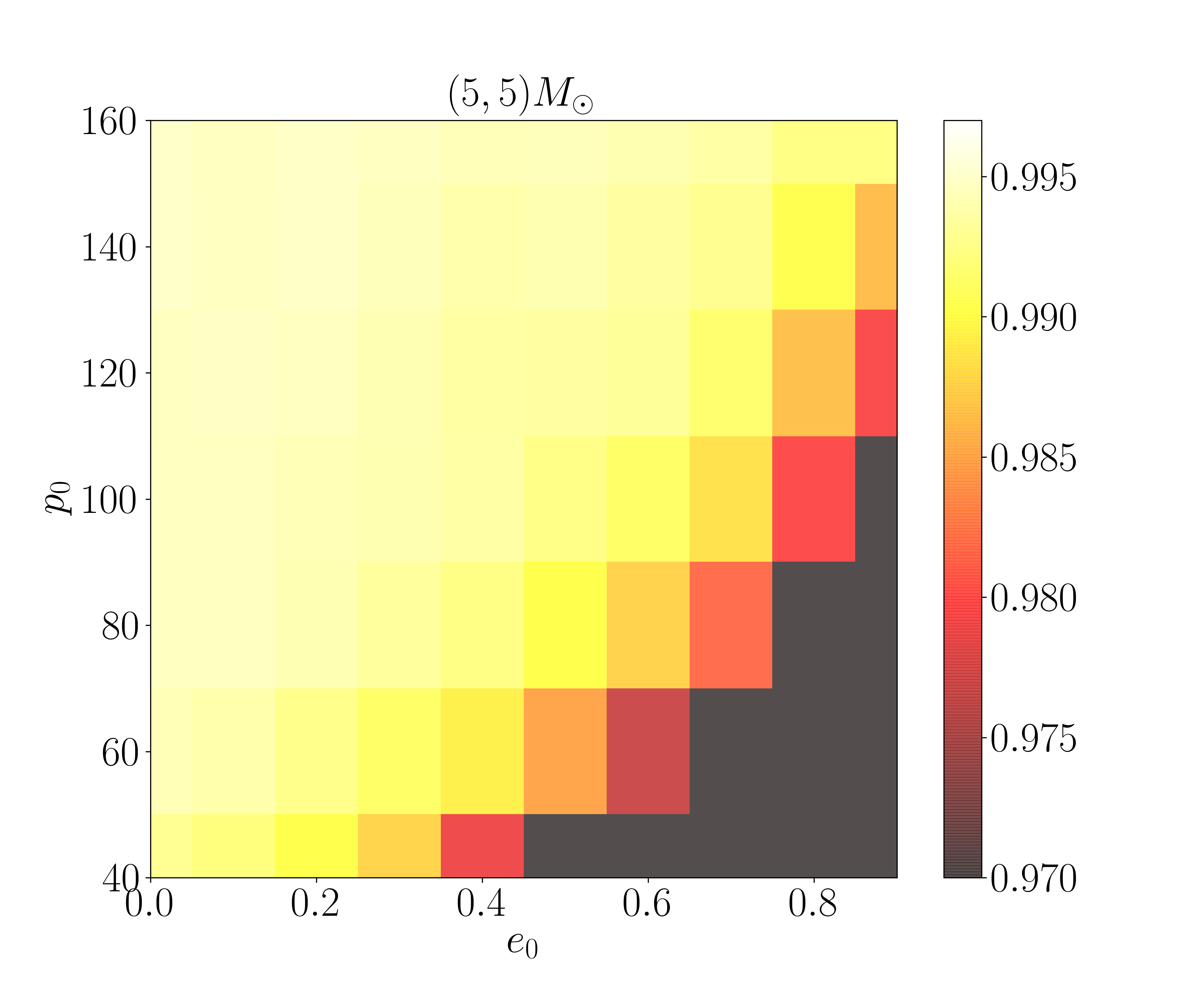}
\includegraphics[clip=true,angle=0,width=0.4\textwidth]{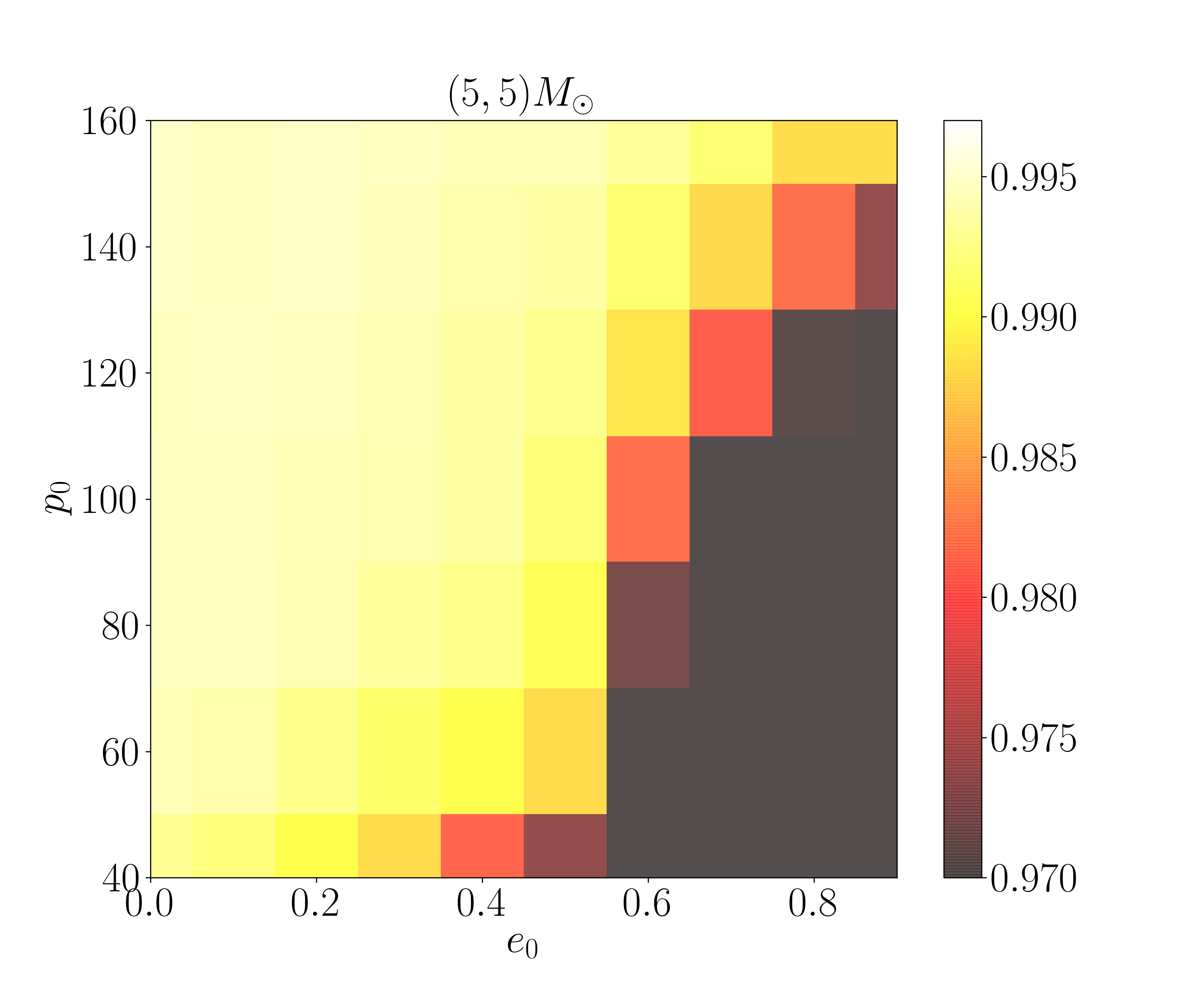}
\caption{\label{fig:bbh5_match} Match for a $(5, 5)M_{}\odot$ system, with more terms kept in eccentricity (left), and less terms kept (right). The loss in match due to the truncation of the eccentricity expansions appearing in the phase functions suggests that more terms must be kept for less massive systems.}
\end{figure*}

Let us now return to item (iii).  When comparing the left and right panels of Fig.~\ref{fig:match}, we see that for the $(10, 10)M_{\odot}$ and $(1.4, 10)M_{\odot}$ systems, there is nearly no change in the match whether we use TaylorF2e+ or TaylorF2e-. However, in the $(1.4, 1.4) M_{\odot}$ case, there is significant loss in match as the initial eccentricity exceeds $0.5$. This suggests that for less massive systems, it is important to keep more terms in the eccentricity expansions. This conclusion is also supported by Fig.~\ref{fig:bbh5_match}, which shows again the match between TaylorF2e+/- (left/right) and TaylorT4t but this time for a $(5, 5) M_{\odot}$ system. In this case, the loss in match due to truncation of the eccentricity expansions becomes considerable for close initial separations and initial eccentricities around $0.6$. 

\begin{figure*}[htp]
\includegraphics[clip=true,angle=0,width=0.4\textwidth]{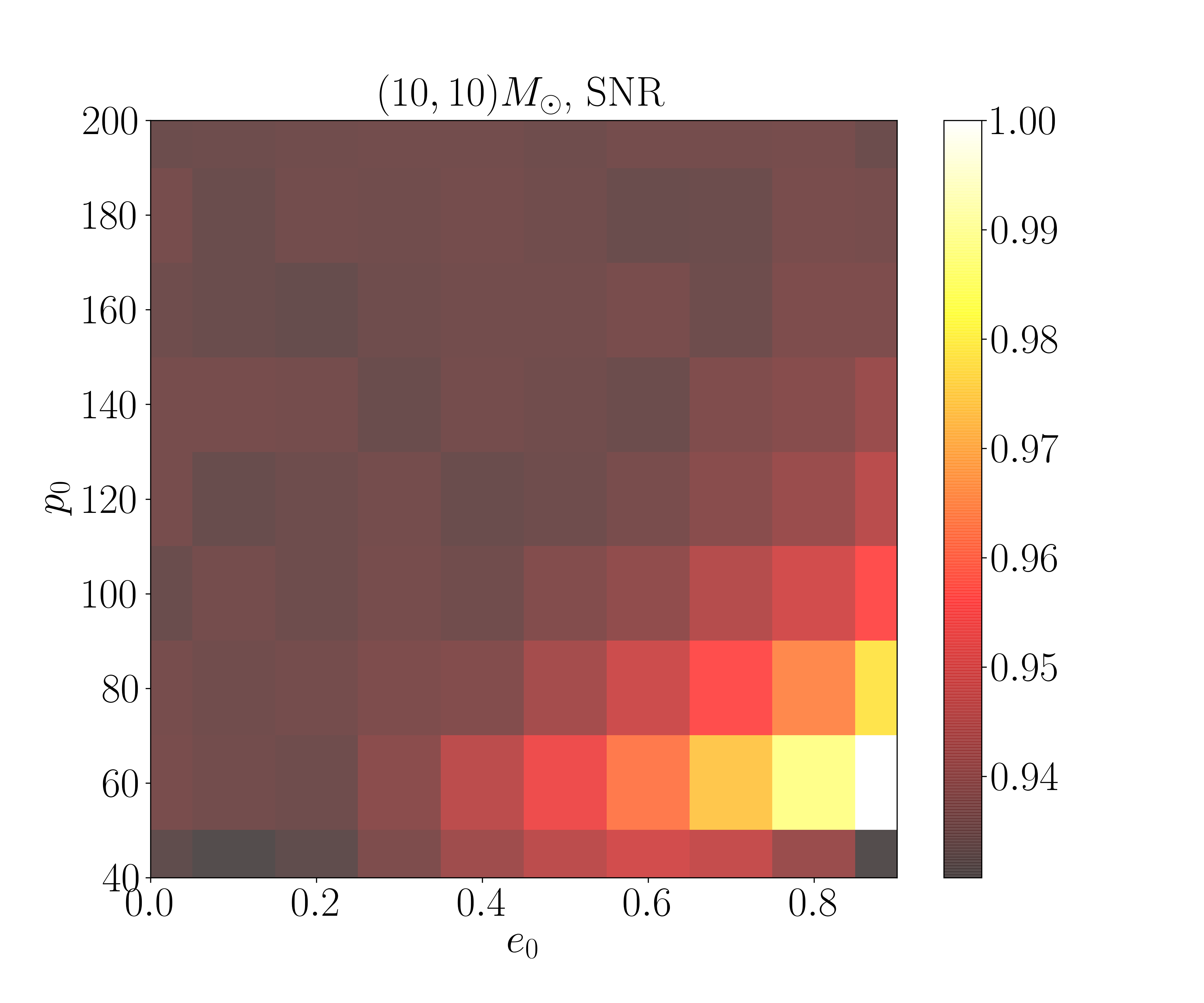}
\includegraphics[clip=true,angle=0,width=0.4\textwidth]{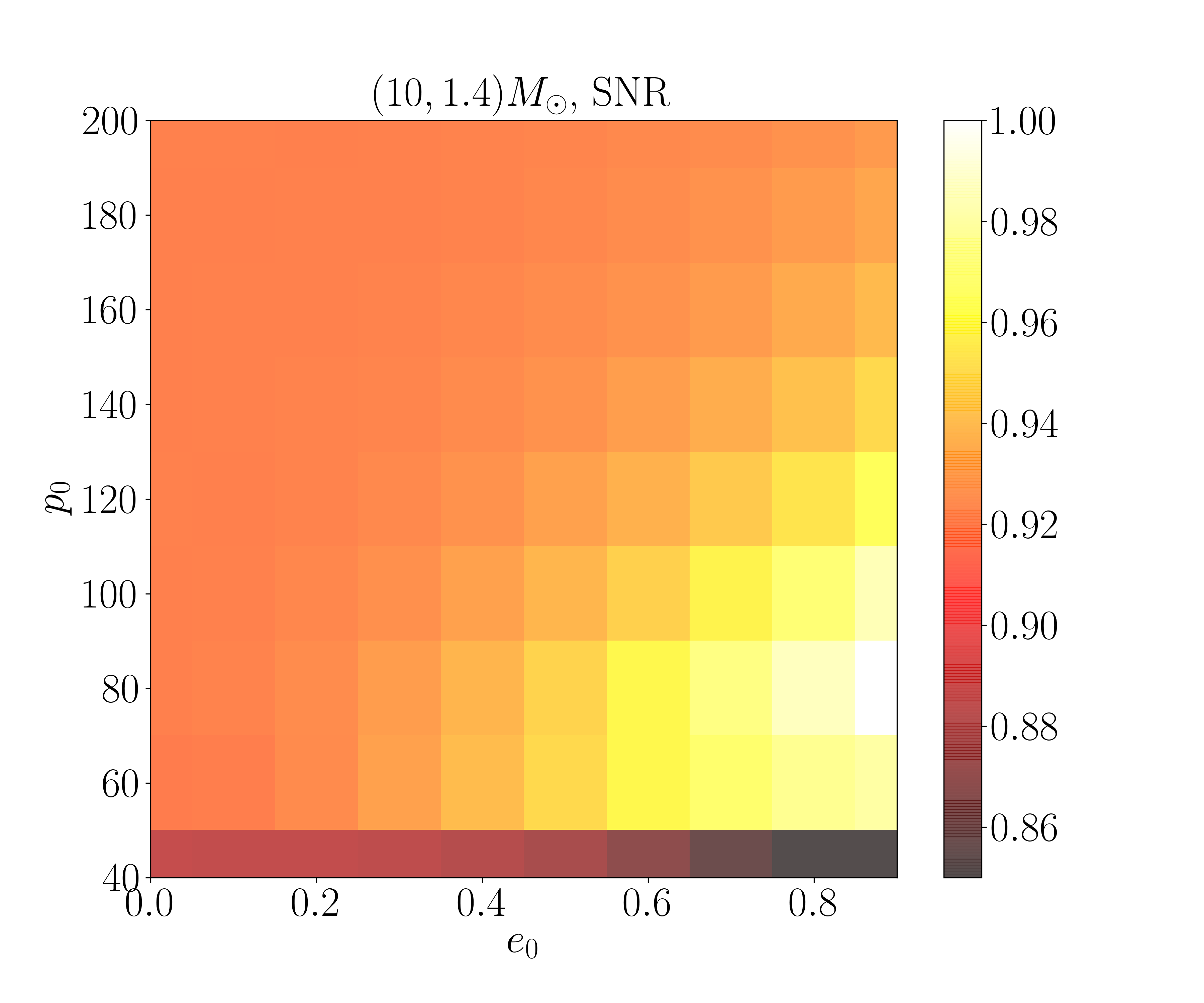}
\includegraphics[clip=true,angle=0,width=0.4\textwidth]{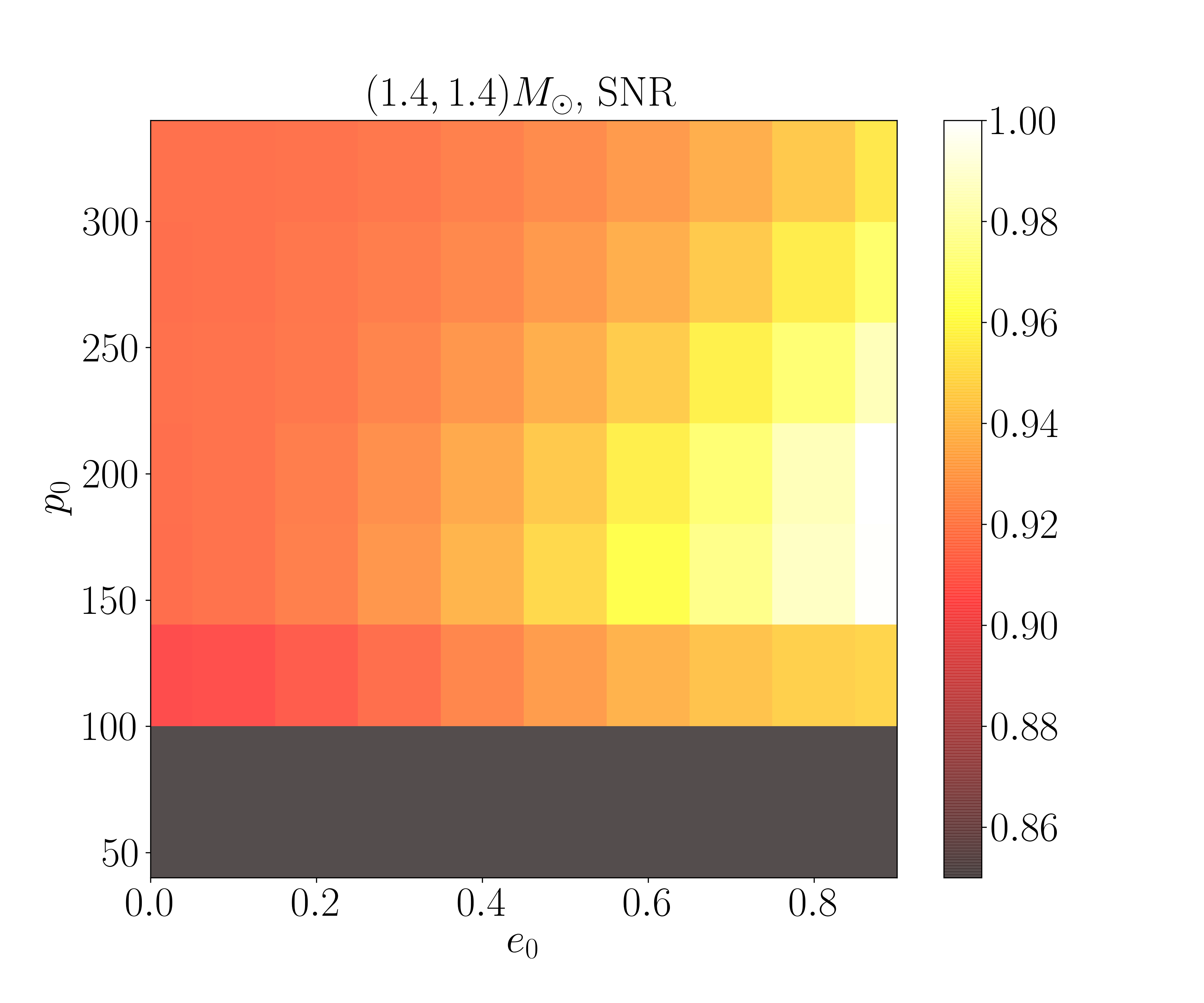}
\caption{\label{fig:SNR} The SNR (normalized by the largest SNR present) as a function of mass and initial orbital eccentricity and semi-latus rectum. In the less massive case there is still significant SNR even for initial semi-latus recta greater than 300, a region where our model works well.}
\end{figure*}

The matches appearing in Fig.~\ref{fig:match} misleadingly suggest that the model is unfaithful for a significant amount of the parameter space, but this is only an artifact of the particular parameter space region we chose to calculate the match over. For initial semi-latus recta that are larger than those considered here, the model is very faithful with matches larger than 0.98. Let us then study if eccentric corrections to the model matter at larger initial separations than those considered in Fig.~\ref{fig:match}. In order to roughly determine which combinations of mass, initial eccentricity, and initial semi-latus recta lead to significant eccentric corrections, we employ a signal-to-noise ratio (SNR) measure, defined by
\begin{equation}
\text{SNR}^2 = (h_1|h_1) \, .
\end{equation}
Figure \ref{fig:SNR} shows the SNR normalized by the largest SNR present in the parameter space explored for the three systems considered in Fig.~\ref{fig:match}. We see that particularly in the less massive case there is significant SNR for systems with larger initial semi-latus recta than those considered in Fig.~\ref{fig:match}. This suggests that our TaylorF2e model is useful for systems in a larger parameter space than considered in Fig.~\ref{fig:match}. We leave a more thorough investigation of the region of parameter space in initial eccentricity and semi-latus recta where eccentric effects matter to future work.

\subsection{Error Analysis in Different Components of TaylorF2e}
\label{sec:err_spa}
In this section we wish to investigate which of the major analytic components which enter the TaylorF2e contribute to the most error in match. Namely we wish to know which of the functions which appear in the phase ($\lambda(e)$, $l(e)$, and $t(e)$) give the most loss in match, so that if one wished to improve the model they would have some guidance as to where to start. In order to do so we obtain numerical solutions for the phase functions by numerically solving $\dot{\lambda}$, $\dot{l}$, $\dot{y}$, and $\dot{e}$. With these solutions in hand we invert $e(t)$ to obtain $t(e)$ and use this to map $\lambda$ and $l$ to the eccentricity domain which allows them to be used in our TaylorF2e model. 

\begin{figure*}[htp]
\includegraphics[clip=true,angle=0,width=0.4\textwidth]{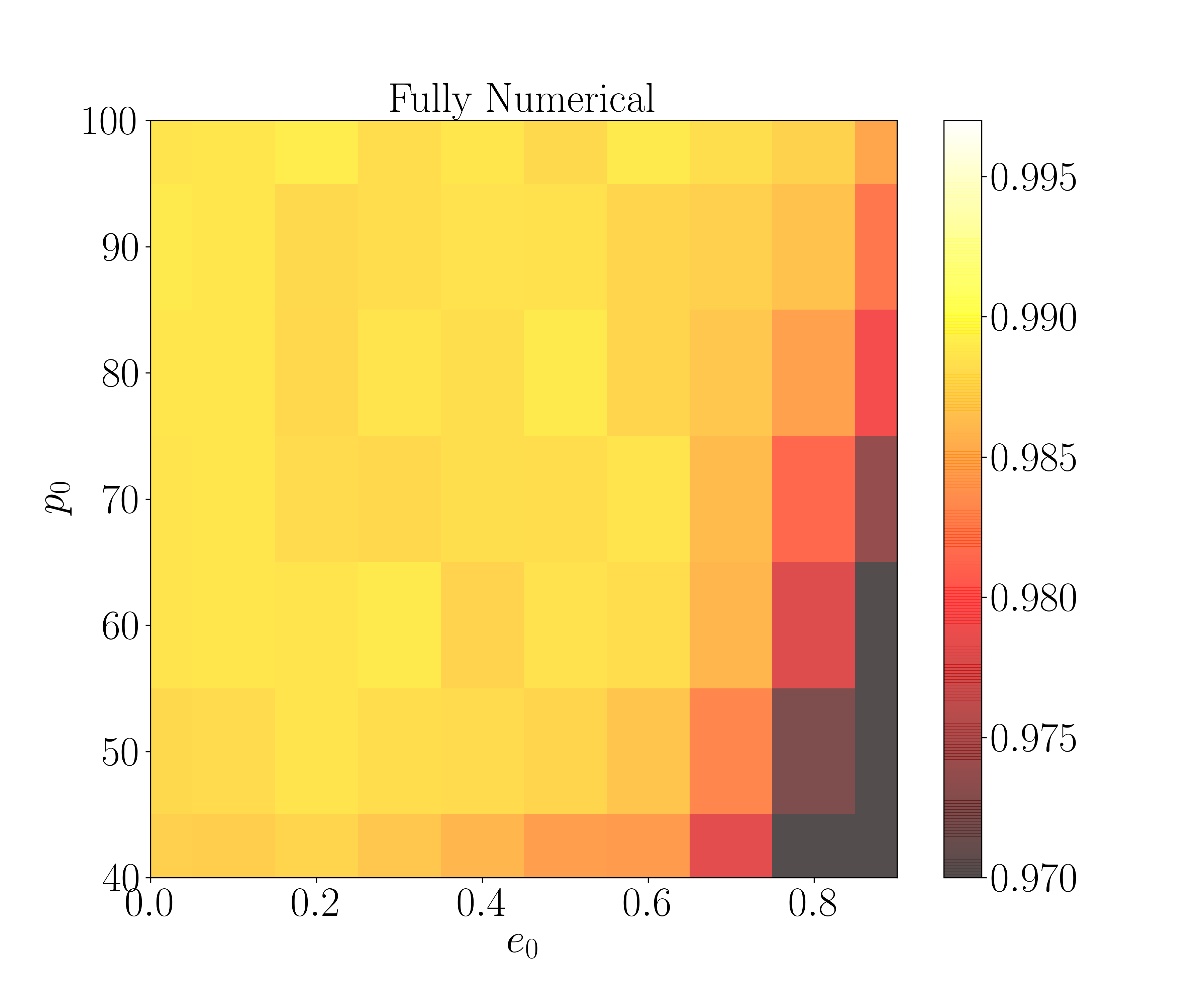}
\includegraphics[clip=true,angle=0,width=0.4\textwidth]{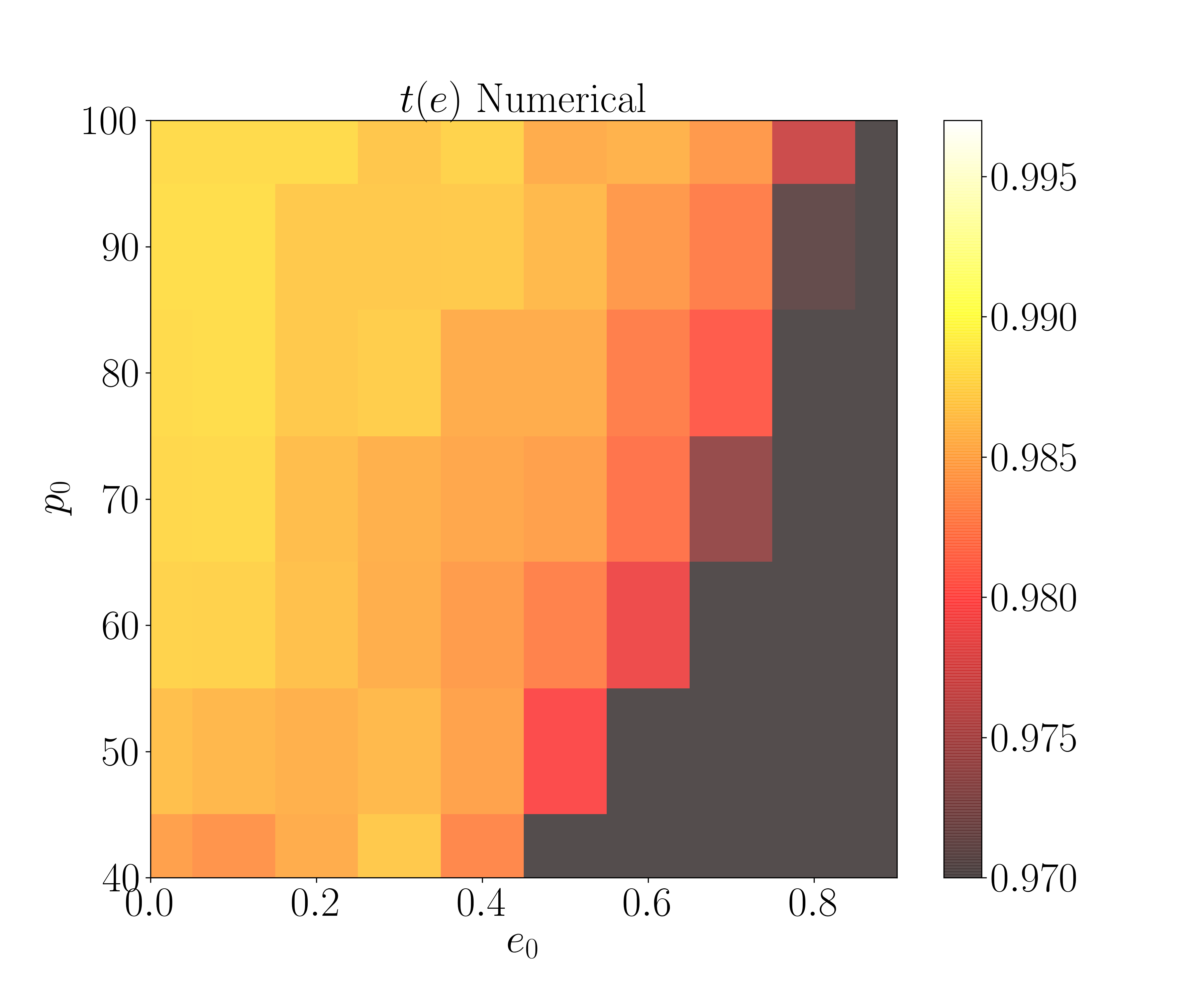}
\includegraphics[clip=true,angle=0,width=0.4\textwidth]{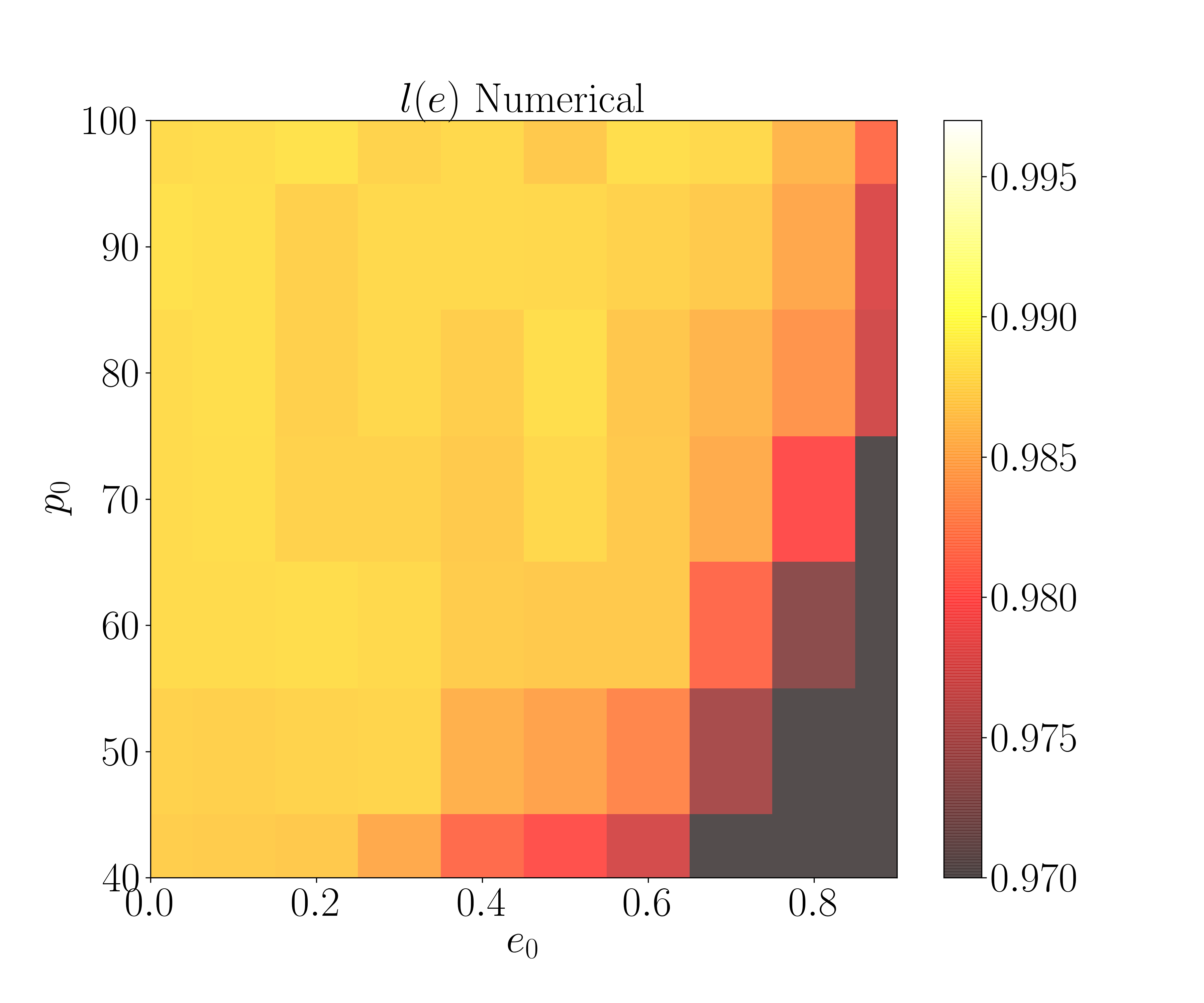}
\includegraphics[clip=true,angle=0,width=0.4\textwidth]{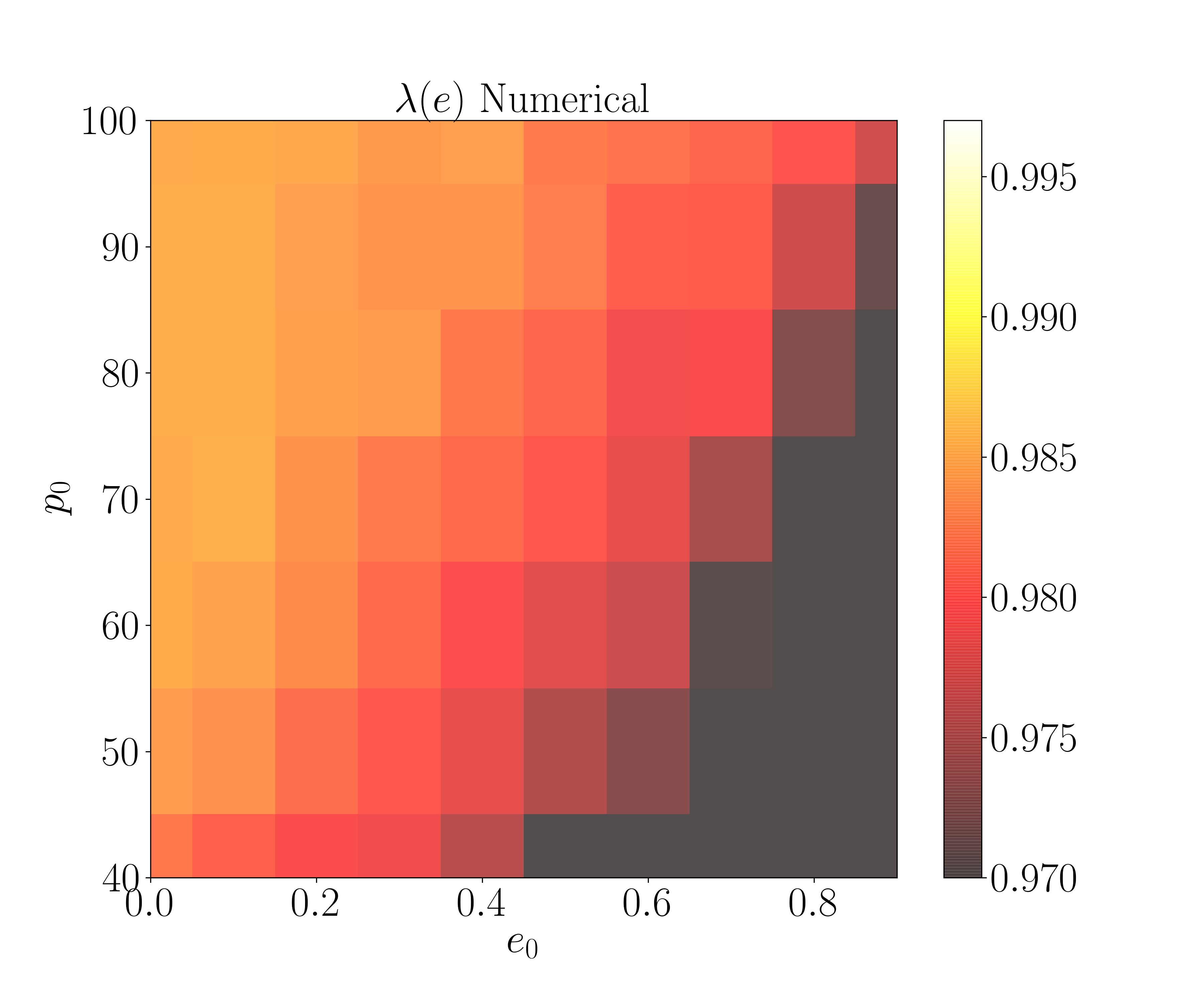}
\caption{\label{fig:err_num} The match between TaylorF2e and TaylorT4t where we have kept all phase functions ($\lambda(e)$, $l(e)$, and $t(e)$) numerical (top left), just $t(e)$ numerical (top right), just $l(e)$ numerical (bottom left), and just $\lambda(e)$ numerical (bottom right). The largest improvement (besides using all numerical phase functions) in match occurs when $l(e)$ is kept numerical.}
\end{figure*}

In Figure \ref{fig:err_num} we show the value of the match for a $(10, 10)M_{\odot}$ system where we have kept the phase functions all numerical, or just one of any of them numerical. We see, as expected, when all of the phase functions are numerical we recover the result of Figure \ref{fig:harmdecomp_study}. The error in match is due to the truncation of harmonics in Eq.\eqref{eq:spa}. The best increase in match (besides keeping all phase functions numerical) occurs when $l(e)$ is computed numerically. Interestingly, the match when $\lambda(e)$ is kept numerical is slightly worse than had the analytic expression for $\lambda(e)$ been used. This is because there is cancellation in error between the analytic expression for $t(e)$ and $\lambda(e)$. 

\section{Conclusions \& Future Work}
\label{sec:conc}

We have extended the eccentric Fourier-domain Newtonian model of paper 1 to 3PN order. This model combines a harmonic decomposition of the time domain signal, the stationary phase approximation, and PN theory. We have thoroughly validated the model and shown that it is faithful to very high initial orbital eccentricities ($e_0 \sim 0.7$), but of course this depends on the particular system and the initial separation at which the initial eccentricity is defined. We expect that the model is valid to even higher eccentricities in relevant regions of ($e_0$, $p_0$) parameter space. 

In addition to this main result, we have validated the harmonic decomposition of the eccentric time-domain signal in terms of a sum of harmonics of the two orbital frequencies with amplitudes that are expanded in small eccentricity. We also investigated the importance of higher PN corrections in the overall amplitude of the time domain model, and we investigated the agreement of different PN consistent time domain approximants. We found that the harmonic decomposition presented here is valid for very large initial eccentricities ($e_0 \sim 0.9$), and that in the equal mass case the effects of higher PN corrections in the amplitude are negligible. Interestingly, we found that different PN consistent ways of solving for the orbital dynamics lead to time-domain approximants that disagree at moderate initial eccentricities and relatively close initial separation -- an effect which is enhanced in the unequal mass limit. However, we expect that this result only excludes faithful modeling in a small region of the relevant parameter space.

Given our model's 3PN accuracy, we believe it is sufficient for preliminary data analysis applications. However, first we must address the efficiency of the model. In this work we sought to validate the model, but we did not implement it in the fastest possible way. We expect that through a careful evaluation of the Fourier phase, the inversion of the stationary phase condition, and neglecting certain low SNR harmonics, the model can retain its accuracy and become very efficient.

With such an efficient and 3PN accurate eccentric model we could seek to answer some important questions about eccentric inspirals using Bayesian techniques. We plan to investigate the region of $(M, \eta, p_0, e_0)$ parameter space where eccentric corrections to the signal are important (i.e.~where in this parameter space we could differentiate between a quasi-circular signal and an eccentric one). We also plan to investigate the ability to measure parameters and any biases of the model. 

We also expect that, using our model, we can constrain certain modified theories of gravity. If such a modified theory of gravity modifies the eccentricity evolution of a binary, these corrections could be incorporated into the model presented here. With these corrections due to a modified theory present in the model, we could investigate the ability of detectors to constrain these corrections and test General relativity perhaps more stringently than possible with a quasi-circular model. 

Lastly, in order to fully model an eccentric signal we require the incorporation of effects due to the merger and the ringdown. The model presented here could be used to create an inspiral-merger-ringdown (IMR) type hybrid model. The only IMR model which incorporates orbital eccentricity is that of~\cite{2018PhRvD..97b4031H}. It will be useful to have several IMR type models with which to extract signals in the future for validation purposes, as is the case for quasi-circular coalescences.

\section*{Acknowledgments} 
B. M. was supported by the Joan L. Dalton Memorial Fellowship in Astronomy from Montclair State University. B.~M.~and N.~Y.~ also acknowledge support from NSF Grant No. PHY-1759615, as well as 
NASA grants NNX16AB98G and 80NSSC17M0041. We thank Antoine Klein for useful conversations and providing us with the expressions for the tail terms. We also thank Travis Robson and Alejandro C\'ardenas-Avenda\~no for useful conversations. Computational efforts were performed on the Hyalite High Performance Computing system, which is supported by University Information Technology at Montana State University.

\appendix
\section{2PN Polarizations in Modified Harmonic Gauge}
\label{app:MH_polarizations}

Here we provide the necessary corrections in order to obtain Eq.\eqref{eq:pn_amp} using the expressions in \cite{2002PhRvD..65h4011G}. In the latter reference, the authors list the amplitude corrections as 
\begin{align}
\label{eq:hp-hc}
h_{+,\times} &= \frac{m \eta}{R}\xi^{2/3}\left(\bar{H}^{(0)}_{+,\times } + \xi^{1/3} \bar{H}^{(1/2)}_{+,\times } 
\right. 
\nonumber \\
& \left. +  \xi^{2/3} \bar{H}^{(1)}_{+,\times } +
\xi \bar{H}^{(3/2)}_{+,\times } + \xi^{4/3} \bar{H}^{(2), \mbox{\tiny ADM}}_{+,\times }\right)
\end{align}  
where $\xi = mn$ and we have introduced an overbar to differentiate these amplitudes from those listed in Eq.\eqref{eq:pn_amp}. Because gauge differences arise at 2PN order, only the 2PN amplitude contains differences due to gauge. First, we must re-express the above in terms of $y$. At 2PN order, $y$ and $\xi$ are related by 
\begin{align}
\xi &= y^3(1-e^2)^{3/2}\left\lbrace 1 - 3y^2 
\right.
\nonumber \\ 
 & \left. + \frac{1}{4}y^4 \left[ -18 + 28\eta + e^2(-51 + 26 \eta) \right] \right\rbrace
\end{align}
Inserting the above into the expressions for $h_{+,\times}$ in Eq.~\eqref{eq:hp-hc}, expanding to 2PN order in $y$, and collecting like terms in $y$ such that it takes the form 
\begin{align}
h_{+,\times} &= \frac{m \eta}{R}y^2(1-e^2)\left(H^{(0)}_{+,\times } + y H^{(1/2)}_{+,\times } 
\right. 
\nonumber \\
& \left. +  y^2 H^{(1)}_{+,\times } +
y^3 H^{(3/2)}_{+,\times } + y^4 H^{(2), \mbox{\tiny ADM}}_{+,\times }\right)
\end{align}
gives the unbarred amplitudes as a function of the barred ones
\begin{subequations}
\begin{align}
H^{(0)}_{+,\times } &= \bar{H}^{(0)}_{+,\times } \\ 
H^{(1/2)}_{+,\times } &= (1-e^2)^{1/2} \bar{H}^{(1/2)}_{+,\times } \\
H^{(1)}_{+,\times } &= (1-e^2)\bar{H}^{(1)}_{+,\times } -2\bar{H}^{(0)}_{+,\times } \\
H^{(3/2)}_{+,\times } &= (1-e^2)^{3/2}\bar{H}^{(3/2)}_{+,\times } - 3 (1-e^2)^{1/2} \bar{H}^{(1/2)}_{+,\times } \\
H^{(2), \mbox{\tiny ADM}}_{+,\times } &= (1-e)^4\bar{H}^{(2), \mbox{\tiny ADM}}_{+,\times } - 4(1-e^2)\bar{H}^{(1)}_{+,\times }
\nonumber \\
& + \frac{1}{6}\left[-24+28\eta + e^2(-51+26\eta) \right]\bar{H}^{(0)}_{+,\times }
\end{align}
\end{subequations}

In order to express these results in MH gauge we must substitute Eq.\eqref{eq:ADMtoMH} into the expression for $\bar{H}^{(0)}_{+,\times }$. This will produce 2PN corrections that when added to $\bar{H}^{(2), \mbox{\tiny ADM}}_{+,\times }$ yields $\bar{H}^{(2), \mbox{\tiny MH}}_{+,\times }$. We have then 
\allowdisplaybreaks[4]
\begin{subequations}
\begin{align}
\bar{H}^{(2), \mbox{\tiny MH}}_{+} &=  \bar{H}^{(2), \mbox{\tiny ADM}}_{+} + \Lambda_{+} \\
\bar{H}^{(2), \mbox{\tiny MH}}_{\times} &= \bar{H}^{(2), \mbox{\tiny ADM}}_{\times} + \Lambda_{\times}
\end{align}
\end{subequations}
where the $\Lambda_{+, \times}$ are given by 
\begin{widetext}
\begin{subequations}
\begin{align}
\Lambda_{+} &= \frac{1+17\eta}{4(e \cos u - 1)^3}e \left[-(1 + C^2)(3\cos u - 4e + \cos^2u)\cos(2\bar{\lambda} + 2W) - S^2\cos u (1-e\cos u)
\right.
\nonumber \\
& \left. -2\sin u (1+C^2)\frac{(1+e\cos u - 2 e^2)}{\sqrt{1-e^2}}\sin(2\bar{\lambda} + 2W) \right] \\
\Lambda_{\times} &= C\frac{1+17\eta}{4(e \cos u - 1)^3}e \left[ \frac{\sin u}{\sqrt{1-e^2}}(1 + e \cos u - 2 e^2)\cos(2\bar{\lambda} + 2W) - (3\cos u -4e + e\cos^2 u)\sin(2\bar{\lambda} + 2W)  \right]
\end{align}
\end{subequations}
\end{widetext}
We these expressions, we now have all that we need to express the polarization in MH gauge and in terms of the parameter $y$. 
\bibliography{master}
\end{document}